\documentclass[sigplan,10pt,screen]{acmart}
\renewcommand\footnotetextcopyrightpermission[1]{}

\usepackage{amsthm}
\usepackage{multirow}
\usepackage{xurl}

\newtheorem{challenge}{Challenge}
\newtheorem{invariant}{Invariant}

\AtBeginDocument{%
  }

\setcopyright{acmlicensed}
\copyrightyear{2023}
\acmYear{2023}
\acmDOI{XXXXXXX.XXXXXXX}

\acmConference[Conference acronym 'XX]{Make sure to enter the correct
  conference title from your rights confirmation emai}{June 03--05,
  2018}{Woodstock, NY}
\acmPrice{15.00}
\acmISBN{978-1-4503-XXXX-X/18/06}

\def\System{InversOS}
\def\Technique{Privilege Inversion}
\def\elevated{elevated}

\def\XOR{\texorpdfstring{$\oplus$}{\^{}}}

\def\Comment#1{}

\begin{document}

\title{{\System}: Efficient Control-Flow Protection for AArch64 Applications with {\Technique}}

\author{Zhuojia Shen}
\orcid{0000-0002-8052-8643}
\affiliation{%
  \institution{University of Rochester}
  \city{Rochester}
  \state{NY}
  \country{USA}
  \postcode{14627}
}
\email{zshen10@cs.rochester.edu}

\author{John Criswell}
\orcid{0000-0003-2176-3659}
\affiliation{%
  \institution{University of Rochester}
  \city{Rochester}
  \state{NY}
  \country{USA}
  \postcode{14627}
}
\email{criswell@cs.rochester.edu}

\renewcommand{\shortauthors}{Shen and Criswell}

\begin{abstract}
  With the increasing popularity of AArch64 processors in
  general-purpose computing,
  securing software running on AArch64 systems against
  control-flow hijacking attacks has become a critical part toward
  secure computation.
  Shadow stacks keep shadow copies of function return addresses and,
  when protected from illegal modifications and
  coupled with forward-edge control-flow integrity,
  form an effective and proven defense against such attacks.
  However, AArch64 lacks native support for write-protected
  shadow stacks,
  while software alternatives either incur prohibitive performance
  overhead or provide weak security guarantees.

  We present \emph{\System},
  the first hardware-assisted write-protected shadow stacks for
  AArch64 user-space applications,
  utilizing commonly available features of AArch64 to achieve
  efficient intra-address space isolation (called \emph{\Technique})
  required to protect shadow stacks.
  {\Technique} adopts unconventional design choices that
  run protected applications in the kernel mode
  and mark operating system (OS) kernel memory as user-accessible;
  {\System} therefore uses a novel combination of
  OS kernel modifications,
  compiler transformations,
  and another AArch64 feature
  to ensure the safety of doing so and to support legacy applications.
  We show that {\System} is
  secure by design,
  effective against various control-flow hijacking attacks,
  and performant on selected benchmarks and applications
  (incurring overhead of 7.0\% on LMBench, 7.1\% on SPEC CPU 2017,
  and 3.0\% on Nginx web server).
\end{abstract}


%

\settopmatter{printfolios=true}
\settopmatter{printacmref=false}

\maketitle

\pagestyle{plain}

\section{Introduction}
\label{sec:intro}

%
%
%
%
%
%
%

AArch64 (64-bit ARM) processors are becoming increasingly popular,
not only in embedded and mobile platforms but also in
personal computers~\cite{M1:Apple} and
high-performance servers and
data centers~\cite{EC2-A1:AWS,Arm:MicrosoftAzure,Arm:GoogleCloud,%
AmpereA1:OCI}.
Given the popularity of AArch64 processors used in production and in
our daily lives, securing software on such systems is critical.
In particular, a large portion of AArch64 application code is written
in memory-unsafe programming languages (e.g., C and C++)
and is vulnerable to
control-flow hijacking attacks~\cite{Ret2Libc:RAID11,ROP:TISSEC12}
that exploit memory safety errors.
While basic code injection attacks are prevented by the wide deployment
of the W{\XOR}X policy~\cite{NoExec:PaX00},
which disallows memory to be writable and executable at the same time,
advanced code-reuse attacks like
return-oriented programming (ROP)~\cite{ROP:CCS07,ROP:TISSEC12}
and jump-oriented programming (JOP)~\cite{JOP:ASIACCS11}
are still possible.
These attacks hijack a program's control flow by corrupting
code pointers (e.g., return addresses and function pointers)
to point to reusable code of the attacker's choosing.
Worse yet, recent research~\cite{RiscyROP:RAID22} has demonstrated
automation of ROP attacks on AArch64,
necessitating effective and practical defenses to be deployed.

Control-flow integrity (CFI)~\cite{CFI:CCS05,CFI:TISSEC09},
a seminal mitigation to control-flow hijacking attacks,
restricts a program's control flow to follow its intended
control-flow graph.
While ineffective by itself~\cite{OutOfCtrl:Oakland14,%
ROPDanger:UsenixSec14,StitchGadgets:UsenixSec14,LoseCtrl:CCS15},
CFI necessitates a mechanism that protects the integrity of
return addresses, such as
write-protected shadow stacks~\cite{RAD:ICDCS01,SoK:SS:Oakland19},
to form an effective defense~\cite{CFBending:UsenixSec15}.
However, software approaches to protecting return address integrity
either
suffer from high performance overhead
(e.g., software-based shadow stacks~\cite{RAD:ICDCS01,%
StackGhost:UsenixSec01,DISE:WASSA04,RECFISH:ECRTS19,%
Silhouette:UsenixSec20})
or only provide probabilistic guarantees
(e.g., information hiding~\cite{Lockdown:DIMVA15,Zieris:ASIACCS18,%
SoK:SS:Oakland19,Randezvous:ACSAC22}).
Hardware-assisted shadow stack protection,
such as Control-flow Enforcement Technology
(CET)~\cite{CET:HASP19} on x86,
offers the best security and performance
but is not natively available on AArch64.

In this paper, we present \emph{\System}, a system that provides
AArch64 user-space applications with hardware-assisted write-protected
shadow stacks.
{\System} does so without requiring the most recent hardware security
features on AArch64 or modifying hardware.
Instead, {\System} uses two widely available AArch64 features~\cite{ARM:Manual},
namely \emph{unprivileged load/store instructions} and
\emph{Privileged Access Never},
in a novel way to create an efficient domain-based instruction-level
intra-address space isolation technique which we call \emph{\Technique}.
With {\Technique}, {\System} runs protected applications in
the same privilege mode as an operating system (OS) kernel,
sets up incorruptible shadow stack memory accessible only by
unprivileged load/store instructions,
and ensures the safety of running privileged user-space code via
a combination of OS kernel modifications and compiler transformations.
To keep compatibility with legacy untransformed application binaries,
{\System} repurposes another AArch64 feature to support coexistence of
legacy and protected applications securely and efficiently.

We built a prototype implementation of {\System} based on
the Linux kernel v4.19.219~\cite{linux-4-19-219-src} and
the LLVM/Clang compiler v13.0.1~\cite{LLVM:CGO04}.
We analyzed the security of {\System} and assessed the strength of
its defense against different types of control-flow hijacking attacks.
Our evaluation of {\System} on a real AArch64 system
and a comprehensive set of benchmarks and applications
(LMBench~\cite{lmbench:ATC96}, SPEC CPU 2017~\cite{CPU2017:SPEC},
and Nginx~\cite{Nginx})
shows low performance overhead
(7.0\% on LMBench, 7.1\% on SPEC CPU 2017, and 3.0\% on Nginx),
indicating that {\System} is practical for deployment.
We open-sourced {\System} at \url{https://github.com/URSec/InversOS}.

To summarize, we make the following contributions:
\begin{itemize}
  \item
    We present {\Technique},
    the first domain-based intra-address space isolation technique
    for AArch64 user-space applications,
    using only widely available features on commodity hardware.

  \item
    We designed and implemented {\System},
    an OS-kernel-compiler co-design that provides the first
    hardware-assisted protected shadow stacks on AArch64
    utilizing {\Technique} and is compatible with existing binaries.

  \item
    We evaluated the security and performance of {\System}
    and showed that {\System} is both efficacious and efficient.
\end{itemize}

The rest of the paper is organized as follows.
Section~\ref{sec:bg} provides background information.
Section~\ref{sec:threat} defines our threat model.
Sections~\ref{sec:design} and~\ref{sec:impl} describe the design and
implementation of {\System}, respectively.
Section~\ref{sec:security} analyzes the security of {\System}.
Section~\ref{sec:eval} presents the performance evaluation of {\System},
Section~\ref{sec:related} discusses related work,
and Section~\ref{sec:conc} concludes and discusses future work.

\section{Background}
\label{sec:bg}

%
%
%
%

In this section, we provide background information on protected shadow
stacks.
We also briefly introduce features of AArch64 instruction set
architecture (ISA) that are relevant to the design and implementation of
{\System}.

\subsection{Protected Shadow Stacks}
\label{sec:bg:ss}

Control-flow hijacking attacks like ROP~\cite{ROP:CCS07,ROP:TISSEC12}
corrupt
saved return addresses on the stack.
One way to mitigate such attacks is to use
shadow stacks~\cite{SoK:SS:Oakland19},
which keep copies of return addresses in separate memory regions.
When calling a function, a return address is pushed onto both the
regular stack and the shadow stack; on return, the program loads the
return address
from the shadow stack and either compares it to the one on the regular
stack to ensure its
validity~\cite{SS:ASIACCS15,ROPdefender:ASIACCS11,RAD:ICDCS01}
or jumps to the
value loaded from the shadow stack
directly~\cite{Randezvous:ACSAC22,IskiOS:RAID21,Silhouette:UsenixSec20,%
Zieris:ASIACCS18,CFI:TISSEC09}.
To enforce return address integrity, however, shadow stacks themselves
require protection that disallows illegal modifications.
Prior approaches to protecting shadow stack integrity rely on
system calls~\cite{RECFISH:ECRTS19,StackGhost:UsenixSec01,RAD:ICDCS01},
software fault isolation
(SFI)~\cite{Silhouette:UsenixSec20,DISE:WASSA04},
information hiding~\cite{Randezvous:ACSAC22,SoK:SS:Oakland19,%
Zieris:ASIACCS18,Lockdown:DIMVA15},
or special hardware such as segmentation~\cite{CFI:TISSEC09},
Memory Protection Extensions (MPX)~\cite{SoK:SS:Oakland19,LMP:ACSAC16,%
Ombro:ATC22},
Memory Protection Keys (MPK)~\cite{IskiOS:RAID21,SoK:SS:Oakland19},
and CET~\cite{CET:HASP19}).
To the best of our knowledge,
no hardware-assisted shadow stack protection exists on AArch64.

\subsection{AArch64 Architecture}
\label{sec:bg:arch}

\paragraph{Exception Levels}

AArch64~\cite{ARM:Manual} provides four Exception Levels from EL0 to
EL3, with increasing execution privileges.
Typically user-space software executes in EL0 and OS
kernels execute in EL1.
EL2 and EL3 are for hypervisors and a secure monitor, respectively.
A processor core enters from a lower Exception Level to a same or higher
non-EL0 Exception Level via taking synchronous exceptions (e.g., traps,
system calls) or asynchronous exceptions (e.g., interrupts) and returns
via executing an {\tt ERET} instruction.
Each Exception Level EL$x$ has a dedicated stack
pointer register {\tt SP\_EL$x$}.
Software running in EL$x$ ($x\ge1$) can select {\tt SP\_EL0} or
{\tt SP\_EL$x$} as the current stack pointer, referred to as running in
EL$x$t or EL$x$h (i.e., thread or handler mode).
The two modes are different only in the stack pointer register in use,
which also determines the set of exception vectors to use when an
exception occurs that targets the same Exception Level.
The Linux kernel, as of v4.19.219, executes in EL1h and leaves EL1t
(and thus the corresponding set of exception vectors)
unused~\cite{linux-4-19-219-src}.
Unless otherwise noted,
hereafter we only focus on EL0 and EL1(t/h) and refer to them as
unprivileged and privileged (thread/handler) modes, respectively.

\paragraph{Address Space and Page Tables}

\begin{figure}[tb]
  \centering
  \resizebox{1.0\linewidth}{!}{%
    \includegraphics{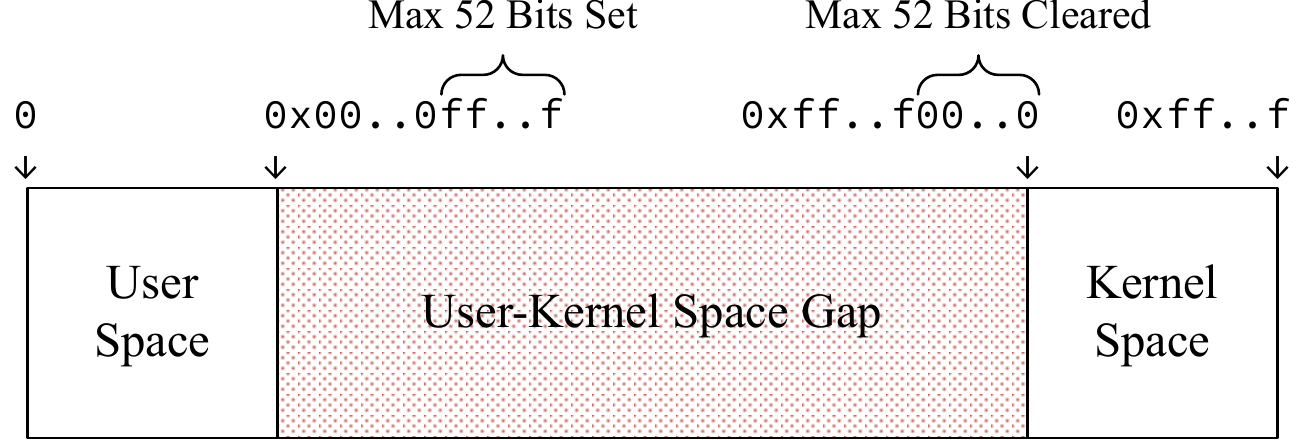}
  }
  \caption{AArch64 Virtual Address Space}
  \Description{AArch64 Virtual Address Space}
  \label{fig:addr-space}
\end{figure}

AArch64~\cite{ARM:Manual} uses hierarchical page tables
and a hardware memory management unit (MMU)
to provide
virtual memory, with two Translation Table Base Registers
{\tt TTBR0\_EL1} and {\tt TTBR1\_EL1} holding the root page table
addresses.
{\tt TTBR0\_EL1} is for the lower half of the virtual address space
(which typically corresponds to the user space), while {\tt TTBR1\_EL1}
is for the upper half (which typically corresponds to the kernel space).
Not all 64~bits of an virtual address are used in address translation;
AArch64 supports a virtual address space up to 52~bits,
thus leaving a gap between the two halves,
as Figure~\ref{fig:addr-space} shows.

AArch64~\cite{ARM:Manual} supports page-level access permissions,
controlled by the {\tt UXN} (Unprivileged eXecution Never) bit, the {\tt PXN}
(Privileged eXecution Never) bit, and two {\tt AP[2:1]} (Access
Permission) bits in last-level page table entries (PTEs).
As the names imply, {\tt UXN} and {\tt PXN}, when set, disable
unprivileged and privileged instruction access of the corresponding
page, respectively.
{\tt AP[1]} disables unprivileged data access when cleared, and
{\tt AP[2]} disables write access when set.

In addition to the above PTE bits, AArch64~\cite{ARM:Manual} also
supports hierarchical access permission control via
the {\tt UXNTable} bit,
the {\tt PXNTable} bit,
and
two {\tt APTable[1:0]} bits in top- and mid-level PTEs
(PTEs that point to a next-level page table rather than a page).
Unlike their last-level PTE counterparts, these bits can apply access
restrictions to the whole corresponding address range on top of the
permission of subsequent levels.
When set, {\tt UXNTable} and {\tt PXNTable} disallow unprivileged and
privileged instruction access, respectively.
{\tt APTable[0]} disallows unprivileged data access when set, and
{\tt APTable[1]} disallows write access when set.
The Linux kernel, as of v4.19.219,
always keeps these bits cleared and instead only controls
access permissions at page level~\cite{linux-4-19-219-src}.

\paragraph{Unprivileged Load/Store Instructions}

A special feature of AArch64~\cite{ARM:Manual}
(and many other ARM ISAs such as ARMv7-M~\cite{ARMv7-M:Manual})
is unprivileged load and
store (LSU) instructions.
These instructions, with mnemonics starting with {\tt LDTR} or
{\tt STTR} on AArch64,
check unprivileged memory access permissions even when executed in
the privileged mode.
This makes LSU instructions useful in accessing user-space memory
inside the OS kernel (e.g., Linux's {\tt get\_user()} and
{\tt put\_user()} functions~\cite{Bovet:Linux:3}).

\paragraph{Architecture Extensions}

AArch64~\cite{ARM:Manual} has architecture extensions; the initial ISA
is called ARMv8.0-A, and subsequent releases (e.g., ARMv8.1-A) are based
on the previous ISA with new hardware features.
Specifically, we focus on the following hardware features:
Privileged Access Never (PAN),
User Access Override (UAO),
Hierarchical Permission Disable (HPDS), and E0PD.

PAN~\cite{ARM:Manual} is an ARMv8.1-A feature which prevents privileged
code from accessing unprivileged-accessible data memory, similar to
x86's Supervisor Mode Access Prevention
(SMAP)~\cite{X86:Intel:Manual,X86:AMD:Manual}.
When PAN is enabled via setting the {\tt PAN} bit in the processor state
{\tt PSTATE}, all loads and
stores (except LSU instructions) executed in the privileged mode
that try to access
memory accessible in the unprivileged mode will generate a permission fault.

UAO~\cite{ARM:Manual} is an ARMv8.2-A feature which, when enabled via
setting the {\tt PSTATE.UAO} bit, allows LSU instructions
executed in the privileged mode
to act as
regular loads/stores.

HPDS~\cite{ARM:Manual}, introduced in ARMv8.1-A, allows disabling
hierarchical access permission bits ({\tt UXNTable}, {\tt PXNTable}, and
{\tt APTable[1:0]}) during page table lookups.
Software running in the privileged mode
can set the {\tt HPD\{0,1\}} bits in Translation
Control Register {\tt TCR\_EL1} to disable hierarchical access
permission checks in address translation from {\tt TTBR\{0,1\}\_EL1}.
However, as AArch64 allows caching {\tt TCR\_EL1.HPD\{0,1\}} in
translation lookaside buffers (TLBs), flipping either bit may require a
local TLB flush to take effect.

E0PD~\cite{ARM:Manual}, introduced in ARMv8.5-A as a hardware mitigation
to side-channel attacks that leverage fault timing (e.g.,
Meltdown~\cite{Meltdown:UsenixSec18}), prevents code running in the
unprivileged mode from
accessing (lower or upper or both)
halves of the virtual address space and generates faults in
constant time.
Similar to HPDS, there are two bits {\tt TCR\_EL1.E0PD\{0,1\}} that
privileged software can use to
control whether unprivileged access to which half of the address space is
disabled.

\section{Threat Model}
\label{sec:threat}

%
%
%
%

We assume a powerful attacker trying to achieve arbitrary code
execution on a benign but potentially buggy application by exploiting
arbitrary memory read/write vulnerabilities to hijack the control flow.
We assume that the underlying OS kernel and hardware are trusted and
unexploitable, providing
the user space with the basic W{\XOR}X protection~\cite{NoExec:PaX00}.
Non-control data attacks~\cite{NonCtrlDataAttack:UsenixSec05} (such as
data-oriented programming~\cite{DOP:Oakland16} and block-oriented
programming~\cite{BOP:CCS18}), side-channel attacks, and physical
attacks are out of scope.
This threat model is in line with recent work on user-space
control-flow hijacking attacks~\cite{RiscyROP:RAID22,LoseCtrl:CCS15} and
defenses~\cite{RetTag:EuroSec22,PACStack:UsenixSec21,%
ZipperStack:ESORICS20,SoK:SS:Oakland19}.

\section{Design}
\label{sec:design}

%
%
%
%
%
%

In this section, we present the design of {\System}.
The goal of {\System} is to provide low-cost return address integrity to
user-space applications running on commodity AArch64 systems, which may
or may not come with the most recent hardware security features such as
Pointer Authentication (PAuth),
Branch Target Identification (BTI), and
Memory Tagging Extension (MTE)~\cite{ARM:Manual}.
To do so, {\System} must only rely on AArch64 features from the early
ISA versions.
We therefore require {\System}'s target platform to support at least PAN
and HPDS
(i.e., conforming to ARMv8.1-A~\cite{ARM:Manual}); this allows {\System}
to be deployed on most of AArch64 systems released since
2017~\cite{ARMCoreList:Wikipedia}.

Overall, we devise {\System} as a co-design between an OS kernel and a
compiler.
The {\System}-compliant OS kernel utilizes \emph{\Technique}, a novel
intra-address space
isolation technique we invented, to provide user-space applications an
extra protection domain accessible only by LSU instructions.
The {\System}-compliant compiler then instruments user-space code to
leverage the protection
domain for efficient protected shadow stacks as well as to enforce
forward-edge CFI~\cite{CFI:CCS05,CFI:TISSEC09},
allowing {\System} to protect user-space applications without modifying
their source code.
The nature of {\Technique} dictates running user-space applications in
the privileged mode; we therefore combine
CFI,
a compile-time bit-masking compiler pass,
a load-time code scanner in the OS kernel, and
a set of kernel modifications
to together ensure the safety and security of
doing so.
Lastly, {\System} supports running legacy untransformed applications to
keep compatibility with existing binaries
via a novel use of HPDS or E0PD (if available).

\subsection{\Technique}
\label{sec:design:isolate}

LSU instructions in AArch64, as described in Section~\ref{sec:bg:arch},
show a great potential in implementing efficient intra-address space
isolation; previous work~\cite{ILDI:DAC17} has explored their usage in
kernel-level data isolation.
However, using these instructions to compartmentalize user-space
applications poses challenges as they act like regular loads/stores
when executed in the unprivileged mode.
Essentially the underlying hardware only supports one protection domain
for unprivileged software.

\begin{figure}[tb]
  \centering
  \resizebox{1.0\linewidth}{!}{%
    \includegraphics{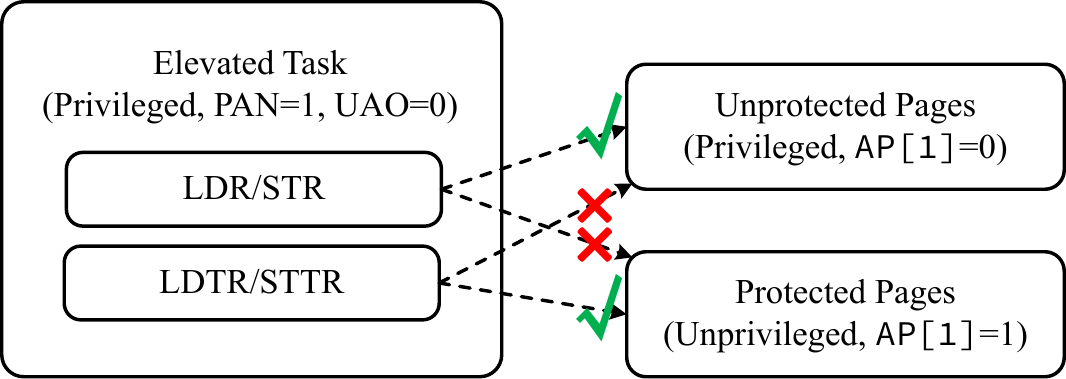}
  }
  \caption{Compartmentalization by {\Technique}}
  \Description{Compartmentalization by {\Technique}}
  \label{fig:compartment}
\end{figure}

We devise {\Technique}, a novel intra-address space isolation technique
that creates a separate protection domain for AArch64 user-space
applications.
With {\Technique}, the OS kernel runs a user-space application needing
an extra protection domain in the privileged mode.
We dub such an application as an \emph{{\elevated} task}.
When launching an {\elevated} task, the OS kernel configures its memory
pages as unprivileged-inaccessible (i.e., with {\tt AP[1]} cleared in
PTEs),
marks its code pages as privileged-executable
(i.e., with {\tt PXN} cleared and {\tt UXN} set in PTEs),
and enables PAN during its execution.
Then, pages that the {\elevated} task wants to place in the separate
protection domain are marked as unprivileged-accessible (i.e., with
{\tt AP[1]} set in PTEs).
Note that the {\elevated} task's pages are still mapped to the user
space (translated by {\tt TTBR0\_EL1}); the above changes only apply to
their access permission bits in the PTEs.
This configuration allows LSU instructions in {\elevated} task code to
access the protected pages but forbids accesses to them made by all
regular loads/stores due to PAN.
In the meanwhile, it leaves all other unprotected pages
in the {\elevated} task accessible by regular loads/stores but
inaccessible by LSU instructions, effectively compartmentalizing
the {\elevated} task into two separate protection domains (one for
regular loads/stores and the other for LSU instructions),
as Figure~\ref{fig:compartment} shows.
Note that in systems with UAO support, UAO has to be turned off during
{\elevated} task execution; otherwise LSU instructions would act just
like regular loads/stores.

However, in order to make {\Technique} safe and useful, we need to
address the following challenges:

\begin{challenge}
  As {\elevated} tasks run in the privileged mode, kernel memory becomes
  accessible by their regular loads/stores.
  \label{challenge:kmem}
\end{challenge}

\begin{challenge}
  As {\elevated} tasks run in the privileged mode, their control-flow
  transfer instructions can jump to the kernel space to execute
  arbitrary kernel code (i.e., kernel memory with {\tt PXN} cleared).
  \label{challenge:ctrl-flow}
\end{challenge}

\begin{challenge}
  As {\elevated} tasks run in the privileged mode, they may contain and
  execute special privileged instructions that would only be allowed to
  execute in kernel code (e.g., instructions that flip
  {\tt PSTATE.PAN}).
  \label{challenge:priv-instr}
\end{challenge}

To address Challenge~\ref{challenge:kmem}, we incorporate a set of
kernel modifications that mark all kernel memory as
unprivileged-accessible and disable PAN during kernel execution.
Such modifications, while radical in idea, effectively stop regular
loads/stores in {\elevated} tasks from accessing kernel memory
and still keep the OS kernel functional.
The ramifications of modifying the OS kernel in this way are two folds.
First, LSU instructions in {\elevated} tasks can now access kernel
memory.
We therefore require that {\elevated} tasks not contain LSU instructions
by themselves
(which is the case in C/C++ code compiled by GCC or LLVM/Clang)
and use a compiler pass to insert vetted LSU instructions for enforcing
the desired protection policies.
Our shadow stack pass described in Section~\ref{sec:design:ss-cfi}
provides a good example.
Second, if we are to support running legacy untransformed applications
in the unprivileged mode still, they can access kernel memory as well;
Section~\ref{sec:design:compat} discusses how we tackle this problem.

To address Challenge~\ref{challenge:ctrl-flow}, we use a bit-masking
compiler pass, which instruments all indirect control-flow transfer
instructions (i.e., indirect calls, indirect jumps, and returns) in
{\elevated} tasks by preceding them with a bit-masking instruction that
clears the top bit of the target register.%
\footnote{AArch64 returns via the {\tt RET} instruction,
which uses the link register {\tt LR} (by default)
or another explicitly specified register
as the return address~\cite{ARM:Manual}.}
This limits the control-flow transfer target to be within the user space
or to become an invalid pointer pointing to the user-kernel space gap.
Such instrumentation alone, however, can be bypassed by
attacker-manipulated control flow that jumps over the bit-masking
instruction; we therefore combine it with CFI to ensure its execution,
which we discuss in Section~\ref{sec:design:ss-cfi}.
Note that direct control-flow transfer instructions (i.e., direct calls
and jumps) do not need such instrumentation; their target is PC-relative
and always points to a known location within the user space.

To address Challenge~\ref{challenge:priv-instr}, we add to the OS kernel
a load-time code scanner which scans for privileged instructions that
unprivileged software should never execute.
Whenever a page in an {\elevated} task is being marked as executable,
the OS kernel invokes our code scanner to scan the whole page;
if the page contains any forbidden privileged instruction,
the execution permission of the whole page is denied.
As AArch64 instructions are 4-byte sized and aligned~\cite{ARM:Manual},
a linear non-overlapping scan should suffice.

\subsection{Protected Shadow Stacks and Forward-Edge CFI}
\label{sec:design:ss-cfi}

With {\Technique} creating an extra protection domain, we can now
leverage the protection domain to enforce efficient shadow stack
protection for the user space.
Specifically, the OS kernel allocates unprivileged memory for a shadow
stack when a new {\elevated} task is launched via {\tt exec()} or when
a new thread in an {\elevated} task is created via {\tt clone()}.
The compiler utilizes a shadow stack pass to instrument
{\elevated} task code; a copy of the return address is saved onto
a shadow stack via an {\tt STTR} instruction inserted into the prologue
of functions that save the return address to the regular stack,
and the return address is loaded from the shadow stack via an
{\tt LDTR} instruction inserted into the epilogue(s) of these functions.
A special case for shadow stacks to handle is irregular control flow
such as {\tt setjmp()}/{\tt longjmp()} in C and exception handling in
C++.
Since support for such irregular control flow depends on the specific
shadow stack scheme used~\cite{SoK:SS:Oakland19},
we discuss how our {\System} prototype supports such code constructs in
Section~\ref{sec:impl:compiler}.

To form a complete control-flow protection, we couple our shadow stacks
with forward-edge CFI~\cite{CFI:CCS05,CFI:TISSEC09},
which ensures that the target
of indirect calls and jumps is within a set of allowed code locations.
Specifically, we use a label-based CFI pass in the compiler.
For each indirect call or tail-call indirect jump in {\elevated} task
code, the pass
inserts a CFI label at the beginning of every function that might be
the call target and inserts a CFI check before the call.
Similarly, for each non-tail-call indirect jump in {\elevated} task
code, the pass inserts
a CFI label at the beginning of every successor basic block and inserts
a CFI check before the jump.
The CFI check ensures that a proper CFI label is present
at the control-flow target; otherwise it generates a fault and
traps the execution.

\subsection{Compatibility}
\label{sec:design:compat}

Not all AArch64 user-space applications need a separate protection
domain, nor can all of them be recompiled.
{\System} must therefore allow existing application and library binaries
that are not compiled by the {\System}-compliant compiler to run without
compromising its security.

We propose two methods to allow safe execution of legacy applications
in the unprivileged mode (dubbed as \emph{legacy tasks}),
depending on hardware feature availability.
In systems with E0PD support (ARMv8.5-A and onward), the OS kernel can
directly enable E0PD via setting {\tt TCR\_EL1.E0PD1}
during legacy task execution.
This way, even though kernel memory is marked unprivileged-accessible,
legacy tasks running in the unprivileged mode still cannot access
kernel memory translated by {\tt TTBR1\_EL1}.

\begin{figure}[tb]
  \centering
  \resizebox{1.0\linewidth}{!}{%
    \includegraphics{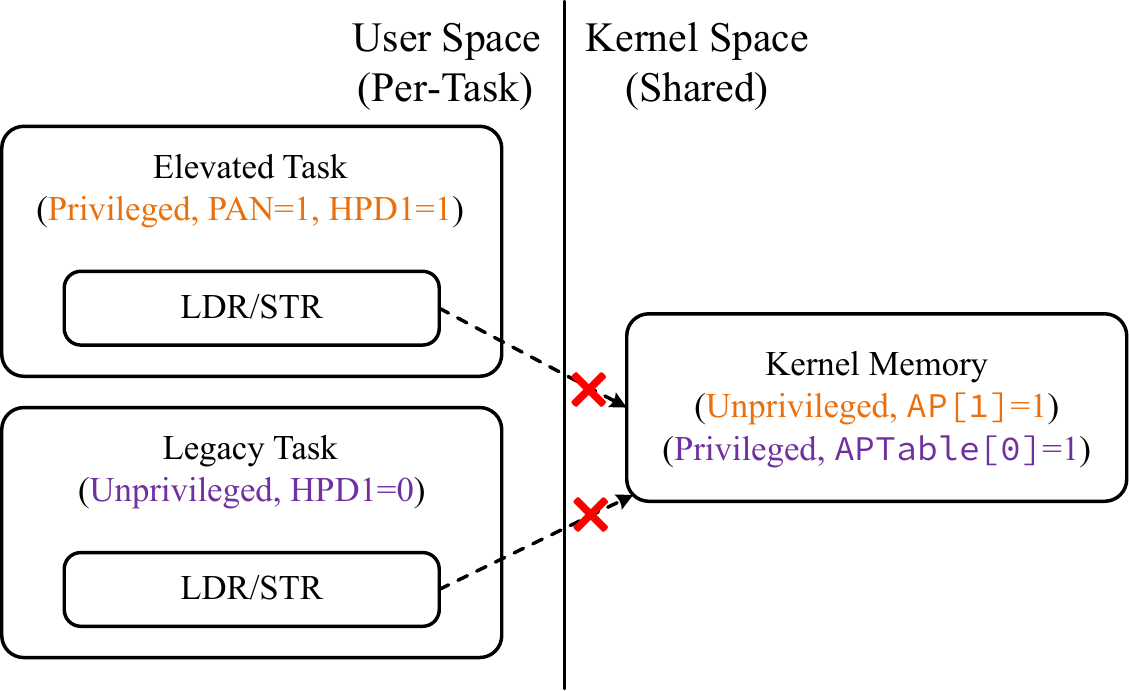}
  }
  \caption{Different ``Views'' of Kernel Memory Due to HPDS}
  \Description{Different View of Kernel Memory Due to HPDS}
  \label{fig:hpds}
\end{figure}

In pre-ARMv8.5-A systems without E0PD support, however, we rely on HPDS
to provide a less-efficient solution.
Specifically, the OS kernel first sets {\tt APTable[0]} in all
top- and mid-level PTEs of kernel memory when establishing page tables
for the kernel space.
This effectively marks all kernel pages as unprivileged-inaccessible
even if {\tt AP[1]} in their last-level PTEs is set.
Then, the OS kernel
enables HPDS via setting {\tt TCR\_EL1.HPD1} before running an
{\elevated} task,
disables HPDS via clearing {\tt TCR\_EL1.HPD1} before running a legacy
task,
and flushes the local TLBs every time after flipping
{\tt TCR\_EL1.HPD1}.
This way, legacy and {\elevated} tasks will possess different ``views''
of kernel memory, as Figure~\ref{fig:hpds} depicts.
Specifically, legacy tasks see kernel memory as
unprivileged-inaccessible due to {\tt APTable[0]} being set,
while {\elevated} tasks see kernel memory as unprivileged-accessible
because HPDS disables {\tt APTable[0]} in top- and mid-level PTEs and
{\tt AP[1]} in last-level PTEs takes effect.
As a result, both types of tasks cannot access kernel memory.

Note that relying on HPDS prevents the OS kernel from mapping kernel
memory with the largest huge pages on certain systems
(e.g., 1~GB huge pages with a page size of 4~KB and a 39-bit virtual
address space),
because such pages have no top- or mid-level PTEs for setting
{\tt APTable[0]}.
However, we believe this has no practical impact on the OS kernel's
address translation and memory usage;
the use of the largest huge pages is rare and infrequent.

\section{Implementation}
\label{sec:impl}

%
%
%
%
%
%
%

We implemented a prototype of {\System} on the Linux kernel
v4.19.219~\cite{linux-4-19-219-src}
and the LLVM/Clang compiler v13.0.1~\cite{LLVM:CGO04}.
Using Tokei v12.1.2~\cite{Tokei:GitHub},
our kernel modifications include
1,815~lines of C code and
207~lines of assembly code,
and our changes to LLVM contain
1,003~lines of C++ code.
To provide complete and transparent {\System} support for user-space
applications, we also modified the musl libc v1.2.2~\cite{musl} and
LLVM's LLD linker~\cite{LLD:LLVM},
compiler-rt builtin runtime library~\cite{compiler-rt:LLVM}, and
libunwind~\cite{libunwind:LLVM},
totalling
27~lines of C code,
131~lines of C++ code, and
299~lines of assembly code.

\subsection{OS Kernel Modifications}
\label{sec:impl:kernel}

{\Technique} requires running {\elevated} tasks in the privileged mode.
As Linux does not use the privileged thread mode
(as Section~\ref{sec:bg:arch} describes),
our prototype therefore utilizes it to run {\elevated} tasks.
This way, the Linux kernel can keep using the privileged handler mode
for its own operations without interference from {\elevated} tasks.
It also greatly simplifies our implementation.
To enable the privileged thread mode, our prototype enables an unused
set of exception vectors that are responsible for taking exceptions
from the privileged thread mode to the privileged handler mode.
Changes were also made to Linux's existing AArch64 exception handler
code so that our prototype can reuse most of the code to handle
exceptions from the privileged thread mode and to resume {\elevated}
task execution properly.
Note that {\elevated} tasks in our prototype still use the {\tt SVC}
instruction for system calls, which is unnecessary because {\elevated}
tasks are already privileged; we leave system call optimizations as
future work.

Apart from the architectural usage of {\tt AP[1]},
Linux also uses {\tt AP[1]} to distinguish whether a page
is kernel or user memory.
As {\System} marks kernel memory unprivileged-accessible,
{\tt AP[1]} can no longer serve for that purpose.
Our prototype therefore utilizes an unused bit (bit 63)
in last-level PTEs to differentiate between kernel and user memory;
the hardware MMU ignores this bit automatically~\cite{ARM:Manual}.

When launching a new task, {\System} must decide whether it should be
run as a legacy or {\elevated} task.
For simplicity and ease of implementation,
our prototype checks the presence of an environment variable
{\tt INVERSOS=1} to make such a decision; if it is present, the task is
started as an {\elevated} task.
Production systems can use a more enhanced mechanism
(e.g., checking the presence of a code signature generated by an
{\System}-compliant compiler) to qualify an {\elevated} task.

%
%
\begin{table}[tb]
\caption{Forbidden Privileged Instructions by Code Scanner}
\label{tbl:priv-instr}
\centering
{\sffamily
\footnotesize{
\begin{tabular}{@{}l|l@{}}
\toprule
  {\bf Instruction} & {\bf Description} \\
\midrule
  {\tt MRS}$^*$/{\tt MSR}$^*$       & Read/Write System Register \\
  {\tt IC}$^*$/{\tt DC}$^*$         & Invalidate Instruction/Data Cache \\
  {\tt TLBI}                        & Invalidate Translation Lookaside Buffer \\
  {\tt HVC}                         & Hypervisor Call \\
  {\tt SMC}                         & Secure Monitor Call \\
  {\tt AT}                          & Address Translation \\
  {\tt ERET}                        & Exception Return \\
  {\tt CFP}/{\tt CPP}/{\tt DVP}     & Prediction Restriction \\
  {\tt LDGM}/{\tt STGM}/{\tt STZGM} & Load/Store Tag Multiple (MTE) \\
  {\tt BRB}                         & Branch Record Buffer \\
  {\tt SYS}/{\tt SYSL}              & Other System Instructions \\
\midrule
\multicolumn{2}{@{}l@{}}{$^*$ Instructions with Certain Operands Allowed} \\
\bottomrule
\end{tabular}
}}
\end{table}

The load-time code scanner, as part of our kernel modifications,
scans for illegal privileged instructions in {\elevated} task code.
Instead of directly scanning a user-space code page,
our prototype maps the page to the kernel space for scanning
in order to avoid frequently calling {\tt get\_user()}.
Table~\ref{tbl:priv-instr} lists all types of privileged instructions
that our prototype forbids, which roughly correspond to instructions
that would generate a fault when executed in the unprivileged mode but
might not when executed in the privileged mode~\cite{ARM:Manual}.
In particular, {\tt MRS}/{\tt MSR}/{\tt IC}/{\tt DC} instructions with
certain operands
(e.g., reading the unprivileged thread ID register {\tt TPIDR\_EL0}
via {\tt MRS})
are allowed in unprivileged software, so these instructions are also
permitted in {\elevated} tasks.

Our kernel modifications take responsibility of setting up and
tearing down memory for protected shadow stacks in {\elevated} tasks,
as Section~\ref{sec:design:ss-cfi} describes.
Each shadow stack region in an {\elevated} task can grow as much as a
regular stack can grow, supporting both parallel and compact
shadow stack schemes~\cite{SoK:SS:Oakland19}.
To prevent shadow stack overflow and underflow, each shadow stack region
is surrounded by two guard regions inaccessible by both regular
loads/stores and LSU instructions.
Mappings of shadow stack and guard regions are unmodifiable by
{\tt munmap()}, {\tt mremap()}, and {\tt mprotect()} requests from the
user space.

Lastly, our prototype implements the HPDS support for running legacy
tasks, as described in Section~\ref{sec:design:compat}.
We omitted implementing the E0PD alternative due to the lack of hardware
that supports E0PD.
As Linux has introduced support for E0PD since
v5.6~\cite{linux-5-6-src} (which is enabled by default),
a simple backport of the relevant changes would suffice.

\subsection{Compiler, Linker, and Library Modifications}
\label{sec:impl:compiler}

We implemented the shadow stack, forward-edge CFI, and bit-masking
compiler passes in a single LLVM pass that transforms
LLVM machine intermediate representation (IR).

\begin{figure}[tb]
  \centering
  \resizebox{1.0\linewidth}{!}{%
    \includegraphics{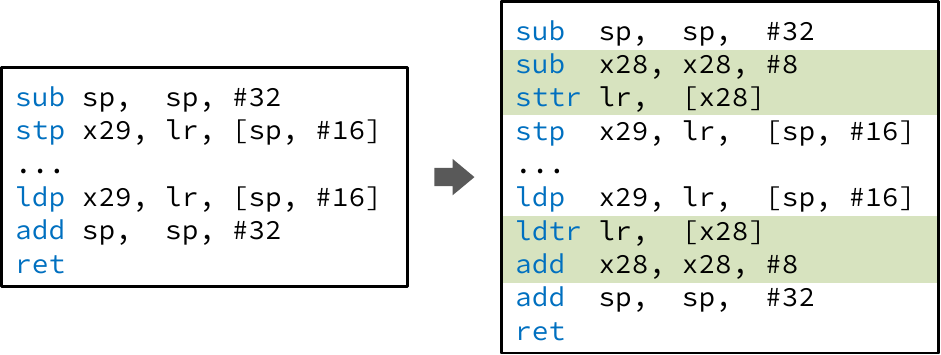}
  }
  \caption{Shadow Stack Transformations}
  \Description{Shadow Stack Transformations}
  \label{fig:ss}
\end{figure}

Our shadow stack transformations adopt the compact shadow stack
scheme~\cite{SoK:SS:Oakland19} and reserve the {\tt X28} register
(a callee-saved register) as the shadow stack pointer register.
Figure~\ref{fig:ss} demonstrates our shadow stack transformations
performed on a function's prologue and epilogue.
Our prototype supports  C's {\tt setjmp()}/{\tt longjmp()} functions
and C++ exception handling via modifications to the musl libc and
LLVM's libunwind, respectively.
Instead of directly guaranteeing the integrity of return address saved
by {\tt setjmp()} or {\tt \_\_unw\_getcontext()}, our prototype provides
shadow stack pointer integrity when restoring the saved context
in {\tt longjmp()} or {\tt \_\_libunwind\_Registers\_arm64\_jumpto()}.
Specifically, rather than overriding {\tt X28} with the saved value,
we unwind {\tt X28} step by step until
a matched return address is found or
it reaches a guard region to cause shadow stack underflow.

\begin{figure}[tb]
  \centering
  \resizebox{1.0\linewidth}{!}{%
    \includegraphics{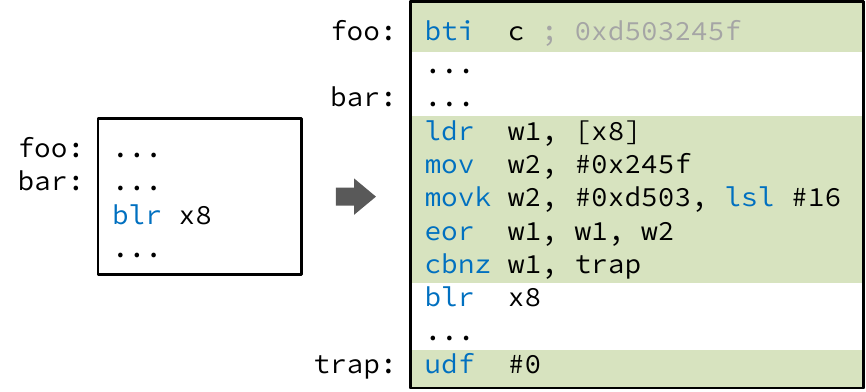}
  }
  \caption{Forward-Edge CFI Transformations}
  \Description{Forward-Edge CFI Transformations}
  \label{fig:cfi}
\end{figure}

Our forward-edge CFI transformations use the {\tt BTI} instructions
as CFI labels to keep forward compatibility with ARMv8.5-A's
BTI~\cite{ARM:Manual}, a hardware-assisted forward-edge CFI mechanism
rolling out to new AArch64 processors.
Processors not supporting BTI execute a {\tt BTI} instruction as
a no-operation.
An appropriate CFI check is inserted before every indirect call or jump
to ensure that the target contains a correct CFI label
({\tt BTI C} for indirect calls and tail-call indirect jumps
and {\tt BTI\ J} for non-tail-call indirect jumps).
Figure~\ref{fig:cfi} illustrates our forward-edge CFI transformations
performed on an indirect call and one of its target functions.
On AArch64, a non-tail-call indirect jump can only be generated from a
{\tt switch} or computed {\tt goto} statement;
the former is bounds-checked against a read-only jump table,
and our prototype restricts the latter by transforming it to a
{\tt switch} statement using the
{\tt IndirectBrExpandPass}~\cite{IndirectBrExpand:LLVM}.
Consequently, a non-tail-call indirect jump is limited to jump within
its function and cannot branch to other functions.

Our bit-masking transformation inserts an {\tt AND} instruction before
every indirect call, indirect jump, or return to clear the top bit of
control-flow transfer target.
For indirect calls and jumps, the instruction is placed after the CFI
check.

While our all-in-one LLVM machine IR pass transforms most of
{\elevated} task code, it fails to cover certain pieces of code in
the user space when compiling the application.
One piece of untransformed code is the procedure linkage table (PLT)
generated by the linker.
We therefore also modified LLD to be able to generate
CFI-checked and bit-masked PLT code.
Another piece of untransformed code is Linux's virtual dynamic shared
object (vDSO); it is compiled with the Linux kernel and stored within
the kernel's read-only data.
We therefore applied our compiler transformations to the vDSO as well
during kernel compilation.
The last case is assembly code (including assembly files and inline
assembly statements).
We manually instrumented assembly code in the musl libc and
compiler-rt builtin runtime library.

\subsection{Discussion}
\label{sec:impl:discuss}

\paragraph{Virtualization Host Support}

ARMv8.1-A adds Virtualization Host Extensions (VHE)~\cite{ARM:Manual}
to accelerate hosted (Type 2) hypervisors
such as Linux's KVM~\cite{KVM/ARM:ASPLOS14}
and FreeBSD's bhyve~\cite{bhyvearm64:BSDCan19}.
In pre-VHE systems,
a host OS kernel (running in EL1) needs to partition its hypervisor
into a ``high-visor'' (running in EL1)
and a ``low-visor'' (running in EL2)
and thus incurs heavy overhead when context-switching between the two
parts.
VHE allows the host OS kernel to run entirely in EL2 to reduce the cost.
The Linux kernel, as of v4.19.219~\cite{linux-4-19-219-src},
stays in EL2h for execution
when having detected VHE support during early boot.
Our prototype therefore transparently supports
running {\elevated} tasks in EL2t in such a case.

\paragraph{AArch32 Support}

Quite a few AArch64 processors still allow running AArch32 (32-bit ARM)
applications for compatibility.
While there are no technical difficulties to support an {\elevated} task
running in the AArch32 state
(i.e., LSU instructions and PAN are also available on AArch32),
we opted not to implement AArch32 support for the sake of time.

\section{Security Analysis}
\label{sec:security}

%
%
%
%

In this section, we analyze the security of {\System} by providing
answers to the following security questions:
\begin{enumerate}
  \item[\bf SQ1]
    Why is {\System} secure
    (to run \emph{instrumented} {\elevated} tasks in the privileged mode
    and \emph{arbitrary} legacy tasks in the unprivileged mode)?

  \item[\bf SQ2]
    How well does {\System} mitigate control-flow hijacking attacks
    on {\elevated} tasks?
\end{enumerate}

\subsection{Security by Design}
\label{sec:security:design}

To answer {\bf SQ1}, we examine \emph{all} potential ways to
compromise {\System} from a legacy or {\elevated} task:
\begin{enumerate}
  \item
    A task may try to read from/write to memory of other tasks to break
    their confidentiality/integrity.

  \item
    A task may try to read from/write to kernel memory to break
    the confidentiality/integrity of the OS kernel.

  \item
    A task may try to allocate an excessive amount of resources
    (e.g., time, memory) to break the availability of {\System}.

  \item
    A task may try to execute detrimental instructions that could
    undermine the security of {\System}.

  \item
    A task may try to jump to kernel code and use kernel code as a
    ``confused deputy'' for the above goals.
\end{enumerate}
As each task's memory (sans shared memory) is mapped exclusively to
the task's own address space,
reading and writing other tasks' memory can only be carried out by
accessing kernel memory or jumping to kernel code.
Since kernel memory has {\tt AP[1]}
(and {\tt APTable[0]}, if using HPDS) set,
accessing kernel memory is disabled via PAN for {\elevated} tasks
and via HPDS or E0PD for legacy tasks.
Jumping to kernel code is also impossible;
having {\tt UXN} set for kernel code prevents legacy tasks from
executing kernel code,
while {\System}'s CFI and bit-masking instrumentation ensures that
control-flow transfers in {\elevated} tasks never reach
the kernel space.
As for attacks on availability,
we argue that {\System} does not introduce new availability problems;
running an {\elevated} task in the privileged mode does not prioritize
it on resource allocation over all other legacy or {\elevated} tasks
and the OS kernel.
The remaining case is privileged instructions,
the execution of which is restricted by hardware automatically for
legacy tasks
and by {\System}'s load-time code scanner for {\elevated} tasks.
Conclusively, {\System} does not introduce new security flaws and
is secure by design.

\subsection{Efficacy against Control-Flow Hijacking}
\label{sec:security:efficacy}

To answer {\bf SQ2}, we first define and explain a list of invariants
that {\System} maintains for guaranteeing return address integrity
of {\elevated} tasks and then reason about why return address integrity
significantly reduces the control-flow hijacking attack surface.
Specifically, {\System} maintains the following invariants for
{\elevated} tasks:

\begin{invariant}
  A function in an {\elevated} task either pushes its return address
  in {\tt LR} to a shadow stack,
  or never spills the return address to memory.
  \label{inv:save}
\end{invariant}

\begin{invariant}
  If a function in an {\elevated} task pushed its return address to
  a shadow stack,
  its epilogue will always load the return address from the shadow
  stack location in which its prologue saved the return address.
  \label{inv:match}
\end{invariant}

\begin{invariant}
  An {\elevated} task cannot corrupt shadow stacks by itself or by
  using a system call as a ``confused deputy''
  (e.g., calling {\tt read(fd, buf, size)} where {\tt buf} points to
  shadow stack memory~\cite{SEIMI:Oakland20}).
  \label{inv:protect}
\end{invariant}

Invariant~\ref{inv:save} is easily upheld by our shadow stack pass,
which instruments {\tt LR}-saving function prologues to push {\tt LR}
to the shadow stack.
With the counterpart instrumentation on epilogue(s) of these functions
to pop {\tt LR} from the shadow stack,
our shadow stack pass guarantees that only a function's prologue and
epilogue(s) can update the shadow stack pointer with a matched
decrement/increment,
contributing to Invariant~\ref{inv:match}.
Since our forward-edge CFI pass ensures that all indirect calls and
tail-call indirect jumps target the beginning of a function and
all non-tail-call indirect jumps are restricted within their
containing function,
shadow stack pointer decrements and increments are guaranteed to occur
in a matched order, sustaining Invariant~\ref{inv:match}.
Finally, Invariant~\ref{inv:protect} is maintained because
the shadow stacks are unprivileged and no existing/new
LSU instructions can be
exploited/introduced to corrupt the shadow stacks (due to CFI/W{\XOR}X),
and because of the benign nature of {\elevated} tasks assumed by
our threat model in Section~\ref{sec:threat}.

With return address integrity, control-flow hijacking attacks that
require corrupting return addresses
(such as return-into-libc~\cite{Ret2Libc:RAID11} and
ROP~\cite{ROP:CCS07,ROP:TISSEC12})
are effectively prevented.
Furthermore, as non-tail-call indirect jumps cannot break the ``jail''
of their containing function, attacks that exploit indirect jumps
(such as JOP~\cite{JOP:ASIACCS11}) no longer work.
The remaining attack surface requires attackers to do purely
\emph{call-oriented programming}
(i.e., using only corrupted function pointers);
while such attacks are possible~\cite{COOP:Oakland15,CtrlJujutsu:CCS15},
they are limited by forward-edge CFI and can be further restrained if
{\System} refines CFI's granularity.
In short, {\System} greatly reduces the control-flow hijacking attack
surface for {\elevated} tasks.

\section{Performance Evaluation}
\label{sec:eval}

%
%
%
%
%

We evaluated the performance of {\System} on a Station P2 mini-PC which
has an RK3568 quad-core Cortex-A55 processor implementing the
ARMv8.2-A architecture
that can run up to 2.0~GHz.
The mini-PC comes with 8~GB of LPDDR4 DRAM up to 1,600~MHz,
64~GB of internal eMMC storage (unused),
and 1~TB of SATA SSD.
It runs Ubuntu 20.04 LTS modified by the manufacturer.

We ran all our experiments using two configurations: Baseline and
{\System}.
In Baseline, we compiled program and library code using LLVM/Clang
v13.0.1~\cite{LLVM:CGO04} without the {\System} compiler transformations
and ran the generated binary executables on a Linux v4.19.219
kernel~\cite{linux-4-19-219-src}
without our kernel modifications.
In {\System}, all program and library code was compiled with the
{\System} compiler transformations (i.e., shadow stack, forward-edge
CFI, and bit-masking transformations)
and executed on the same version of the Linux kernel modified
with our kernel changes.
When running an {\System} executable, we set an environment variable
{\tt INVERSOS=1} to inform the OS kernel that the program should be started
as an {\elevated} task, as Section~\ref{sec:impl:kernel} describes.
As the processor lacks E0PD support, we rely on HPDS to prevent
legacy tasks from accessing kernel memory.
Both configurations used {\tt -O2} optimizations and performed static
linking against the musl libc v1.2.2~\cite{musl} and LLVM's compiler-rt
builtin runtime library v13.0.1~\cite{compiler-rt:LLVM}.
C++ code in our experiments was compiled with and statically linked
against libc++~\cite{libc++:LLVM}, libc++abi~\cite{libc++abi:LLVM}, and
libunwind~\cite{libunwind:LLVM} from LLVM v13.0.1.
Libraries for Baseline and {\System} are compiled without and with
our modifications described in Section~\ref{sec:impl:compiler},
respectively.

\subsection{Microbenchmarks}
\label{sec:eval:micro}

To understand the performance impact of the {\System} Linux
kernel modifications, we used LMBench v3.0-alpha9~\cite{lmbench:ATC96},
a microbenchmark suite that measures the latency and bandwidth of
various OS services.
For each microbenchmark that supports parallelism, we ran four parallel
workloads to reduce variance.
We report an average and a standard deviation of 10~rounds of execution
for each microbenchmark.

%
%
\begin{table}[tb]
\caption{LMBench Latency (Lower is Better)}
\label{tbl:lat-lmbench}
\centering
{\sffamily
\footnotesize{
\resizebox{\linewidth}{!}{
\begin{tabular}{@{}l|rrrr@{}}
\toprule
  {\bf Microbenchmark} & {\bf Baseline} {($\mu$s)} & {\bf stdev} {($\mu$s)}
  & {\bf {\System}} {($\times$)} &  {\bf stdev} {($\times$)} \\
\midrule
  null syscall     &     0.148 &  0.000 & 1.047 & 0.007 \\
  read             &     0.482 &  0.001 & 1.054 & 0.004 \\
  write            &     0.351 &  0.002 & 0.991 & 0.003 \\
  stat             &     4.928 &  0.023 & 1.066 & 0.003 \\
  fstat            &     0.422 &  0.003 & 1.052 & 0.005 \\
  open/close       &     9.744 &  0.017 & 0.989 & 0.003 \\
  select 500 fd    &    24.365 &  0.017 & 1.002 & 0.001 \\
  signal install   &     0.375 &  0.001 & 1.059 & 0.003 \\
  signal catch     &     3.801 &  0.009 & 1.493 & 0.002 \\
  protection fault &     0.408 &  0.005 & 0.980 & 0.029 \\
  pipe             &    16.115 &  0.067 & 0.948 & 0.004 \\
  AF\_UNIX stream  &    27.314 &  0.618 & 1.051 & 0.008 \\
  AF\_UNIX connect &    99.329 &  0.733 & 1.012 & 0.009 \\
  fork+exit        &   266.767 &  6.945 & 1.256 & 0.012 \\
  fork+exec        &   562.585 &  7.046 & 1.188 & 0.009 \\
  fork+shell       & 2,878.983 & 12.869 & 4.007 & 0.015 \\
  page fault       &     0.910 &  0.016 & 1.038 & 0.009 \\
  mmap 1~MB        &    42.700 &  3.318 & 1.019 & 0.007 \\
  udp              &    76.490 &  0.214 & 1.018 & 0.005 \\
  tcp              &    63.472 &  0.200 & 1.011 & 0.002 \\
  connect          &   102.196 &  0.503 & 1.004 & 0.006 \\
  context switch   &    59.318 &  0.880 & 0.993 & 0.014 \\
  fcntl            &     8.772 &  1.643 & 0.992 & 0.219 \\
  semaphore        &     3.083 &  0.515 & 0.954 & 0.162 \\
  usleep           &    78.661 &  1.579 & 0.995 & 0.020 \\
\midrule
  {\bf Geomean}    &       --- &    --- & 1.103 &   --- \\
\bottomrule
\end{tabular}
}}}
\end{table}

%
%
\begin{table}[tb]
\caption{LMBench Bandwidth (Higher is Better)}
\label{tbl:bw-lmbench}
\centering
{\sffamily
\footnotesize{
\begin{tabular}{@{}l|rrrr@{}}
\toprule
  \multirow{2}{*}{\bf Microbenchmark}
                       & {\bf Baseline} & {\bf stdev} & {\bf {\System}} &  {\bf stdev} \\
                       &       {(MB/s)} &    {(MB/s)} &    {($\times$)} & {($\times$)} \\
\midrule
  pipe                 &  1,096.147 & 72.703 & 0.991 & 0.049 \\
  AF\_UNIX stream      &    931.933 &  6.753 & 1.003 & 0.011 \\
  read 1~MB            &  3,706.665 & 65.823 & 0.978 & 0.013 \\
  read 1~MB open2close &  3,474.633 & 45.699 & 0.990 & 0.015 \\
  mmap 1~MB            & 10,689.636 & 36.243 & 1.006 & 0.001 \\
  mmap 1~MB open2close &  6,365.563 & 43.215 & 0.972 & 0.008 \\
  tcp                  &    720.056 & 48.645 & 0.987 & 0.013 \\
\midrule
  {\bf Geomean}        &        --- &    --- & 0.989 &   --- \\
\bottomrule
\end{tabular}
}}
\end{table}

\begin{figure}[tb]
  \centering
  \resizebox{1.0\linewidth}{!}{%
    \includegraphics{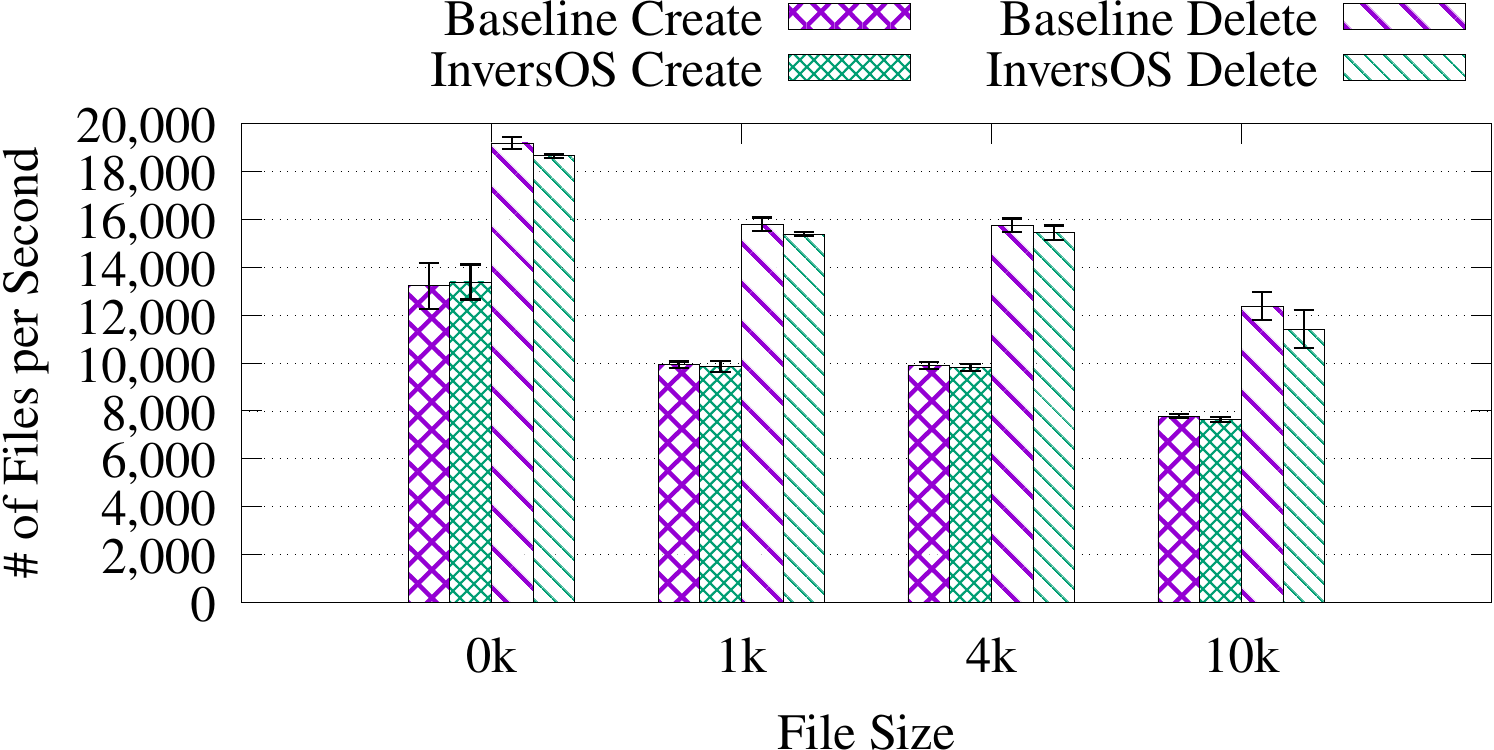}
  }
  \caption{LMBench File Operation Rate (Higher is Better)}
  \Description{LMBench File Operation Rate (Higher is Better)}
  \label{fig:fs-lmbench}
\end{figure}

Tables~\ref{tbl:lat-lmbench} and~\ref{tbl:bw-lmbench} and
Figure~\ref{fig:fs-lmbench} show LMBench performance of both Baseline
and {\System}.
Overall, {\System} incurred a geometric mean of 7.0\% overhead:
10.3\% on
latency, 1.1\% on bandwidth, and 2.2\% on file operation rate.
In most microbenchmarks the overhead is miniscule.
Most notably, {\tt fork+shell} exhibited a 4$\times$ slowdown because
{\System} had to scan every code page of a newly executed shell.
The same goes with {\tt fork+exec}, in which the executed program is
much smaller than the shell and thus incurred much less overhead
(18.8\%).
In {\tt fork+exit}, the 25.6\% overhead comes from an optimization of
copying code page
PTEs upfront; Linux by default only sets up shared page table mappings
of a child process at page faults (i.e., when the child first accesses
the page), which, however, would cause redundant code scanning in
{\System} as {\System} invokes the code scanner whenever a page in an
{\elevated} task is marked executable.
We therefore optimized {\System} to avoid redundant code scanning by
copying an {\elevated} task's code page PTEs during {\tt fork()}
and enabled this optimization in all {\System} experiments.
{\System} incurred 49.3\% overhead in signal catching because of
additional flipping of {\tt PSTATE.UAO} (due to PAN being disabled)
when setting up and tearing down a signal
frame;
this could be optimized away by simply disabling UAO support
in the Linux kernel,
which we opted not to in order to avoid introducing less relevant
changes.

\subsection{Macrobenchmarks and Applications}
\label{sec:eval:macro}

To see how {\System} performs on real workloads, we used
SPEC CPU 2017 v1.1.9~\cite{CPU2017:SPEC}
and Nginx v1.23.3~\cite{Nginx}.
SPEC CPU 2017 is a comprehensive benchmark suite containing CPU- and
memory-intensive programs written in C, C++, and/or Fortran that stress
a computer system's performance.
Nginx is a high performance web server written in C that has been widely
used in the real world.

For SPEC CPU 2017, we evaluated 28 (out of 43) benchmark programs
in C/C++ as LLVM/Clang cannot compile Fortran code.
We used the {\tt train} (instead of the larger {\tt ref}) input set
because {\tt train} yielded execution time of at least 20~seconds
in each benchmark already.
We report average execution time
with 10~rounds of execution for each benchmark; standard deviations are
negligible (less than 1\%).

For Nginx, we used Nginx to host randomly generated static files ranging
from 1~KB to 512~MB with one worker process listening to port 8080 for
HTTP requests.
We then ran ApacheBench ({\tt ab})~\cite{ApacheBench} on the same
machine to measure Nginx's bandwidth of transferring files within a
period of 10~seconds.
We report an average and a standard deviation over 10~rounds of
execution for each file size.

%
%
\begin{table}[tb]
\caption{SPEC CPU 2017 Execution Time (Lower is Better)}
\label{tbl:perf-cpu2017-baseline}
\centering
{\sffamily
\footnotesize{
\begin{tabular}{@{}l|r||l|r@{}}
\toprule
  \multirow{2}{*}{\bf Benchmark (Rate)}  & {\bf Baseline} &
  \multirow{2}{*}{\bf Benchmark (Speed)} & {\bf Baseline} \\
  & {(s)} & & {(s)} \\
\midrule
  500.perlbench\_r & 135.795 & 600.perlbench\_s &   135.289 \\
  502.gcc\_r       & 268.035 & 602.gcc\_s       &   268.294 \\
  505.mcf\_r       & 431.810 & 605.mcf\_s       &   428.423 \\
  520.omnetpp\_r   & 354.081 & 620.omnetpp\_s   &   353.981 \\
  523.xalancbmk\_r & 242.465 & 623.xalancbmk\_s &   242.501 \\
  525.x264\_r      &  96.540 & 625.x264\_s      &    96.527 \\
  531.deepsjeng\_r & 203.713 & 631.deepsjeng\_s &   227.060 \\
  541.leela\_r     & 216.941 & 641.leela\_s     &   217.306 \\
  557.xz\_r        & 128.610 & 657.xz\_s        &   127.926 \\
  508.namd\_r      & 157.894 &                  &           \\
  510.parest\_r    & 330.373 &                  &           \\
  511.povray\_r    &  25.722 &                  &           \\
  519.lbm\_r       & 231.428 & 619.lbm\_s       & 1,718.814 \\
  526.blender\_r   & 533.649 &                  &           \\
  538.imagick\_r   & 167.810 & 638.imagick\_s   &   168.136 \\
  544.nab\_r       & 396.789 & 644.nab\_s       &   397.586 \\
\bottomrule
\end{tabular}
}}
\end{table}

\begin{figure}[tb]
  \centering
  \resizebox{1.0\linewidth}{!}{%
    \includegraphics{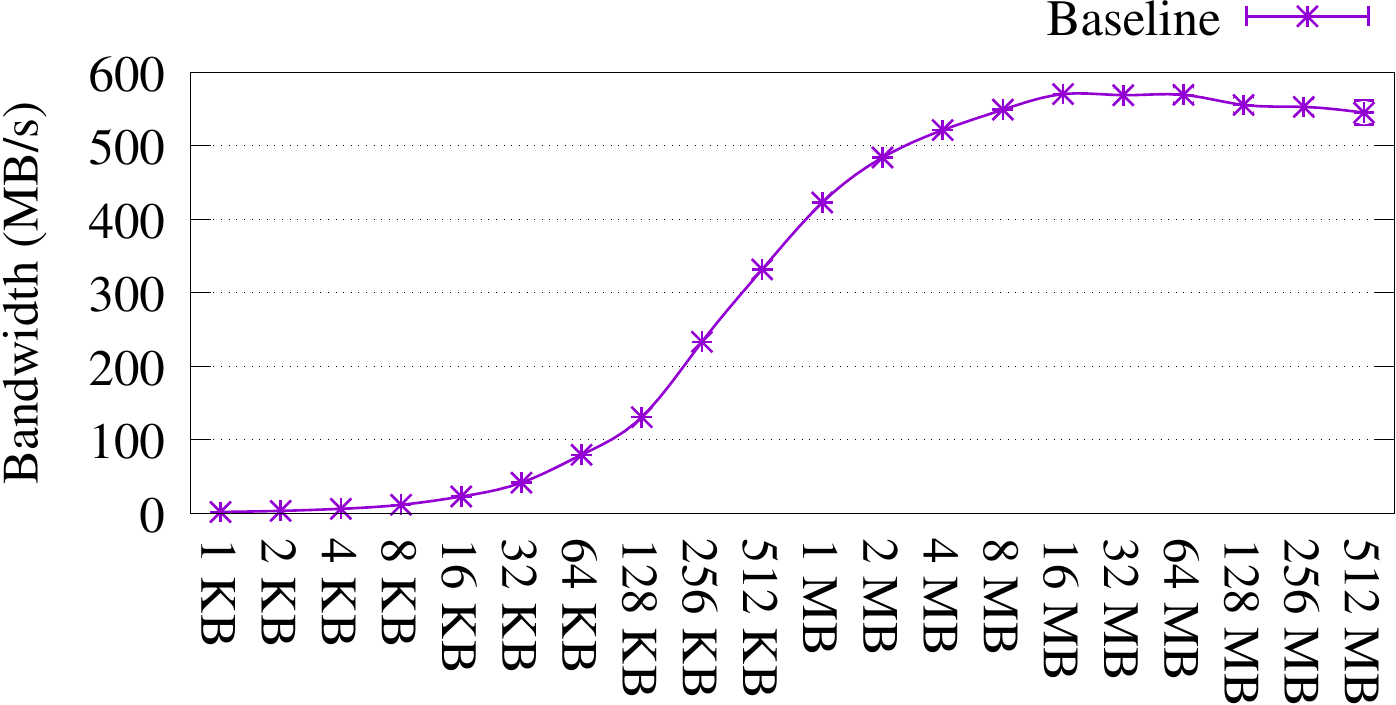}
  }
  \caption{Nginx Bandwidth (Higher is Better)}
  \Description{Nginx Bandwidth (Higher is Better)}
  \label{fig:bw-nginx-baseline}
\end{figure}

\begin{figure*}[tb]
  \centering
  \resizebox{1.0\linewidth}{!}{%
    \includegraphics{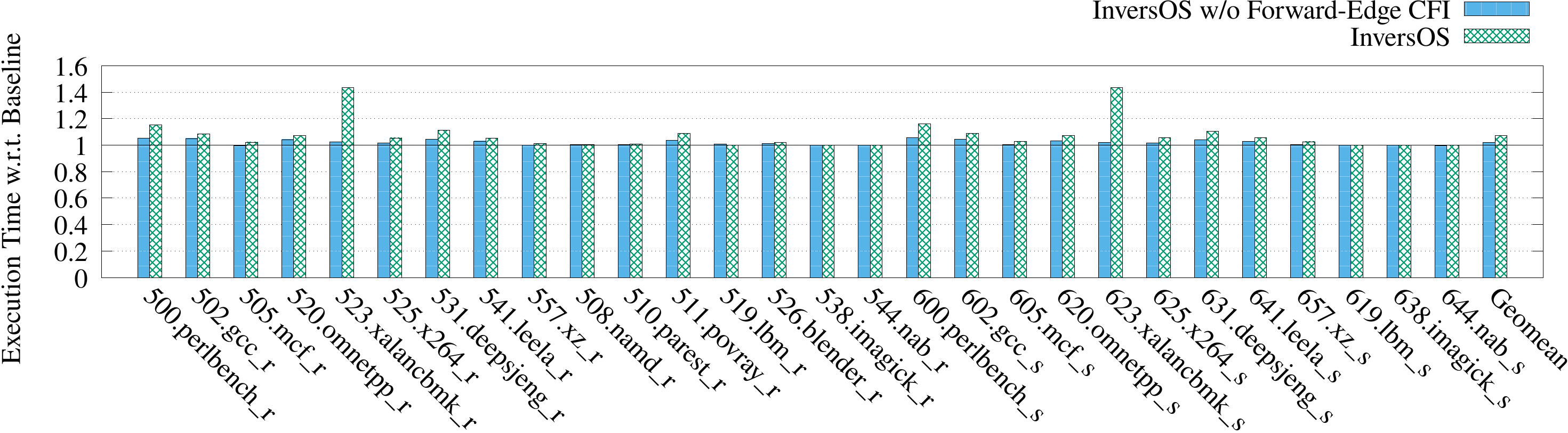}
  }
  \caption{SPEC CPU 2017 Execution Time (Normalized, Lower is Better)}
  \Description{SPEC CPU 2017 Execution Time (Normalized, Lower is Better)}
  \label{fig:perf-cpu2017}
\end{figure*}

\begin{figure}[tb]
  \centering
  \resizebox{1.0\linewidth}{!}{%
    \includegraphics{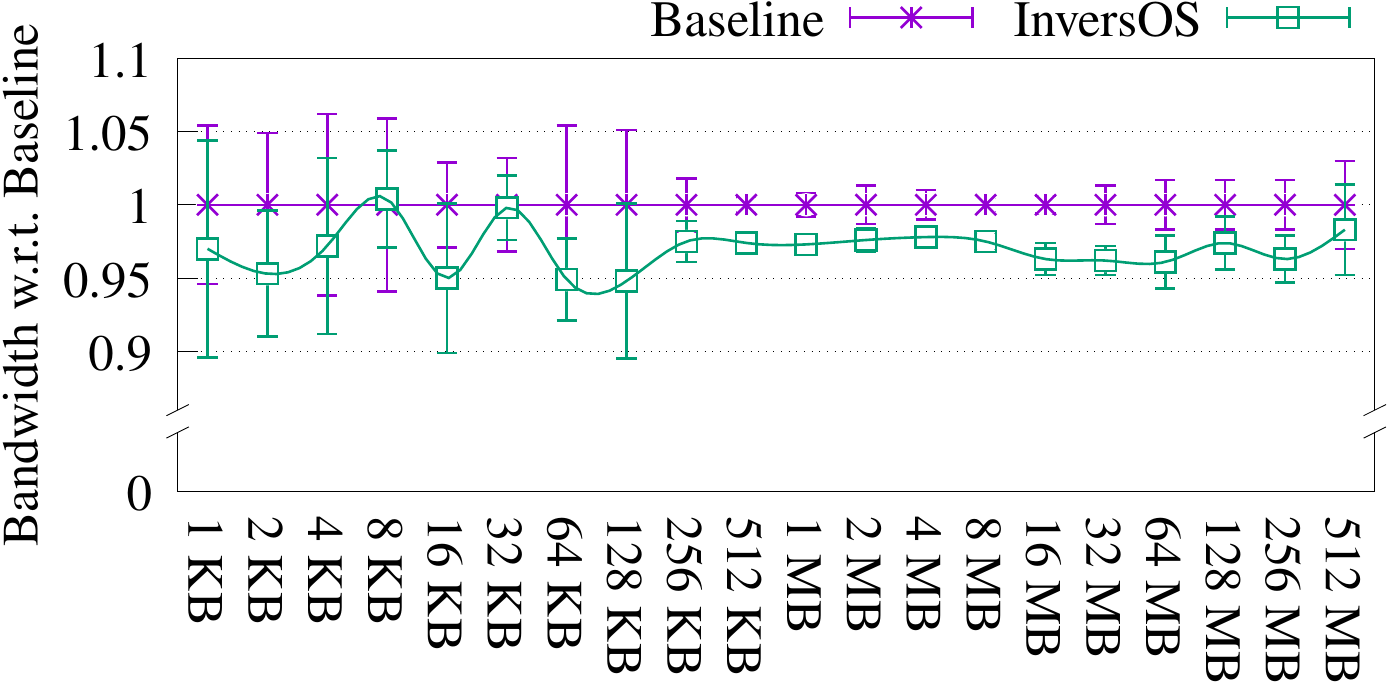}
  }
  \caption{Nginx Bandwidth (Normalized, Higher is Better)}
  \Description{Nginx Bandwidth (Normalized, Higher is Better)}
  \label{fig:bw-nginx}
\end{figure}

Table~\ref{tbl:perf-cpu2017-baseline} and
Figure~\ref{fig:bw-nginx-baseline} present the Baseline performance of
SPEC CPU 2017 and Nginx, respectively.
Figures~\ref{fig:perf-cpu2017} and~\ref{fig:bw-nginx} show the
performance overhead {\System} incurred on SPEC CPU 2017 and Nginx,
respectively.
Overall, {\System} increased the execution time of SPEC CPU 2017 by a
geometric mean of 7.1\% and degraded the bandwidth of Nginx by a
geometric mean of 3.0\%.
We studied the overhead on SPEC CPU 2017 and discovered that our
software-based forward-edge CFI caused most of the overhead;
with that disabled, the overhead decreased to a geometric mean of 1.9\%
(in particular, {\tt xalancbmk}'s overhead dropped down from more than
40\% to less than 3\%).
This indicates that {\System}'s shadow stack and bit-masking
transformations and kernel
modifications have minimal performance impact on SPEC CPU 2017,
compared with software-based forward-edge CFI.
Incorporating BTI~\cite{ARM:Manual},
we expect {\System}'s performance
overhead to be greatly reduced; with BTI, no explicit CFI checks
(as shown in Figure~\ref{fig:cfi})
are needed.
However, as BTI does not provide protected shadow stacks by itself,
\mbox{(post-)ARMv8.5-A} systems can still leverage {\System}'s {\Technique}
to protect the integrity of shadow stacks.
Nginx saw significant variance especially on file sizes $\le$ 128~KB.
We suspect that the cause of high variance is caching and file system
behaviors.

\section{Related Work}
\label{sec:related}
\subsection{Control-Flow Integrity}
\label{sec:related:cfi}

Since the introduction of the original CFI
work~\cite{CFI:CCS05,CFI:TISSEC09},
a long line of research has been proposed to improve its precision,
performance, and/or applicability~\cite{XFI:OSDI06,BGI:SOSP09,%
HyperSafe:Oakland10,ROPdefender:ASIACCS11,CFL:ACSAC11,Zeng:CCS11,%
MoCFI:NDSS12,CFIMon:DSN12,CCFIR:Oakland13,Strato:UsenixSec13,%
Bin-CFI:UsenixSec13,kBouncer:UsenixSec13,MIP:CCS13,CFR:ACSAC13,%
ROPecker:NDSS14,SafeDispatch:NDSS14,KCoFI:Oakland14,MCFI:PLDI14,%
IFCC:UsenixSec14,O-CFI:NDSS15,SS:ASIACCS15,Lockdown:DIMVA15,%
HAFIX:DAC15,PICFI:CCS15,PathArmor:CCS15,CCFI:CCS15,CFIGuard:RAID15,%
BinCC:ACSAC15,Bounov:NDSS16,HCFI:CODASPY16,KernelCFI:EuroSP16,%
TypeArmor:Oakland16,LMP:ACSAC16,FlowGuard:HPCA17,PT-CFI:CODASPY17,%
GRIFFIN:ASPLOS17,PittyPat:UsenixSec17,CaRE:RAID17,CFIXX:NDSS18,%
Zieris:ASIACCS18,uCFI:CCS18,SoK:SS:Oakland19,CFI-LB:EuroSP19,%
CET:HASP19,RECFISH:ECRTS19,PARTS:UsenixSec19,OS-CFI:UsenixSec19,%
uRAI:NDSS20,Silhouette:UsenixSec20,ZipperStack:ESORICS20,%
PACStack:UsenixSec21,IskiOS:RAID21,RetTag:EuroSec22,Ombro:ATC22,%
PAL:UsenixSec22,%
Kage:UsenixSec22,PACtight:UsenixSec22,CFI:CSUR17,TZmCFI:IJPP21}.
As {\System} leverages label-based CFI for forward edges and
protected shadow stacks for backward edges,
we compare {\System} with various types of CFI schemes.

\paragraph{Stateless CFI}

The original CFI~\cite{CFI:CCS05,CFI:TISSEC09} restricts forward-edge
indirect control-flow targets via a coarse-grained context-insensitive
analysis, which statically assigns a distinct label to allowed targets
(an equivalence class or EC)
of each indirect call or jump and inserts checks for a matched label
at indirect call and jump sites.
Subsequent research on stateless forward-edge CFI makes trade-offs
between granularity and performance~\cite{CFL:ACSAC11,CCFIR:Oakland13,%
Strato:UsenixSec13,Bin-CFI:UsenixSec13,MIP:CCS13,MCFI:PLDI14,%
IFCC:UsenixSec14,Lockdown:DIMVA15,PICFI:CCS15,BinCC:ACSAC15,%
TypeArmor:Oakland16},
strengthens other security policies~\cite{XFI:OSDI06,BGI:SOSP09,%
Zeng:CCS11,O-CFI:NDSS15},
or applies to new platforms~\cite{HyperSafe:Oakland10,MoCFI:NDSS12,%
CFR:ACSAC13,SafeDispatch:NDSS14,KCoFI:Oakland14,Bounov:NDSS16,%
KernelCFI:EuroSP16,CaRE:RAID17,CFIXX:NDSS18,RECFISH:ECRTS19,%
uRAI:NDSS20,Silhouette:UsenixSec20,TZmCFI:IJPP21,Ombro:ATC22,%
Kage:UsenixSec22}.
Hardware support for stateless forward-edge CFI
(such as HAFIX~\cite{HAFIX:DAC15}, HCFI~\cite{HCFI:CODASPY16},
Intel CET~\cite{CET:HASP19}, and ARM BTI~\cite{ARM:Manual})
has been proposed, which further lowers the performance overhead but
only provides coarse-grained protection similar to the original CFI.
{\System}'s forward-edge CFI,
while currently prototyped with two labels,
can seamlessly adopt any of the above available finer-grained schemes
for better security.
It can also utilize BTI on newer processors for better performance.

\paragraph{Stateful CFI}

Due to imprecision of context-insensitive CFI,
researchers have focused on context-sensitive CFI policies that
take previous execution history into account.
Using a runtime monitor (inlined or as a separate process),
these systems track executed
branches~\cite{CFIMon:DSN12,kBouncer:UsenixSec13,ROPecker:NDSS14,%
CFIGuard:RAID15,PT-CFI:CODASPY17},
paths~\cite{PathArmor:CCS15,PittyPat:UsenixSec17,uCFI:CCS18},
call-sites~\cite{CFI-LB:EuroSP19,OS-CFI:UsenixSec19},
code pointer origins~\cite{OS-CFI:UsenixSec19},
or complete control flows~\cite{FlowGuard:HPCA17,GRIFFIN:ASPLOS17}
to reduce the size of ECs.
However, such dynamic CFI schemes require hardware features only found
on x86 processors, such as
Branch Trace Store (BTS)~\cite{CFIMon:DSN12},
Last Branch Record (LBR)~\cite{kBouncer:UsenixSec13,ROPecker:NDSS14,%
PathArmor:CCS15,CFIGuard:RAID15},
Performance Monitoring Unit (PMU)~\cite{CFIGuard:RAID15},
Processor Trace (PT)~\cite{FlowGuard:HPCA17,PT-CFI:CODASPY17,%
GRIFFIN:ASPLOS17,PittyPat:UsenixSec17,uCFI:CCS18},
Transactional Synchronization Extensions (TSX)~\cite{CFI-LB:EuroSP19,%
OS-CFI:UsenixSec19},
and MPX~\cite{OS-CFI:UsenixSec19},
limiting their applicability on AArch64.
Compared with stateful CFI, {\System} offers a weaker protection on
forward edges but provides the strongest security on backward edges
with better performance and less resource consumption.

\paragraph{Shadow Stacks}

The original CFI~\cite{CFI:CCS05,CFI:TISSEC09} uses shadow stacks for
backward-edge protection; their debut dates back to
RAD~\cite{RAD:ICDCS01} and StackGhost~\cite{StackGhost:UsenixSec01},
which all used the compact shadow stack design.
Dang et al.~\cite{SS:ASIACCS15} proposed the parallel shadow stack
design, improving the performance but wasting more memory.
As described in Section~\ref{sec:bg:ss},
in order to guarantee return address integrity,
shadow stacks need a protection mechanism that forbids unauthorized
tampering.
A few systems~\cite{ROPdefender:ASIACCS11,SS:ASIACCS15}
simply leave shadow stacks unprotected,
while some rely on
system calls~\cite{RAD:ICDCS01,StackGhost:UsenixSec01,RECFISH:ECRTS19}
or
SFI~\cite{DISE:WASSA04,Silhouette:UsenixSec20}
for protection but incur prohibitive overhead.
More commonly used is information hiding (i.e., ASLR~\cite{ASLR:PaX01}),
which places shadow stacks at a random location in the address
space to increase the difficulty for attackers to locate the shadow
stacks~\cite{Lockdown:DIMVA15,Zieris:ASIACCS18,SoK:SS:Oakland19,%
Randezvous:ACSAC22}.
Though achieving the best performance among software-only solutions,
information hiding provides the weakest guarantee and is vulnerable to
information disclosure
attacks~\cite{Overread:EuroSec09,JIT-ROP:Oakland13,BROP:Oakland14,%
CROP:NDSS16,APM:UsenixSec16,AllocOracle:UsenixSec16}.
Hardware-assisted shadow stack protection significantly lowers
the performance cost and can be fulfilled differently on different ISAs.
On x86\_32, segmentation~\cite{CFI:CCS05,CFI:TISSEC09} provides the most
efficient implementation.
CET~\cite{CET:HASP19} offers native support for protected shadow stacks
on x86\_64 but is only available on most recent
processors~\cite{X86:Intel:Manual,X86:AMD:Manual};
a few solutions repurposed
MPX~\cite{LMP:ACSAC16,SoK:SS:Oakland19,Ombro:ATC22} or
MPK~\cite{SoK:SS:Oakland19,IskiOS:RAID21} for non-CET-equipped
Intel processors
but reported vastly different overhead numbers.
HCFI~\cite{HCFI:CODASPY16} implements an in-chip non-memory-mapped
shadow stack on SPARC via a custom ISA extension.
In the microcontroller world,
Silhouette~\cite{Silhouette:UsenixSec20} and
Kage~\cite{Kage:UsenixSec22} transform regular store instructions into
LSU stores on ARMv7-M~\cite{ARMv7-M:Manual},
while CaRE~\cite{CaRE:RAID17} and TZmCFI~\cite{TZmCFI:IJPP21} leverage
TrustZone-M on ARMv8-M~\cite{ARMv8-M:Manual}.
To the best of our knowledge,
{\System} is the first to provide hardware-assisted protected shadow
stacks on AArch64; our {\Technique} technique is inspired by
Silhouette-Invert~\cite{Silhouette:UsenixSec20}.

\paragraph{Cryptographic CFI}

Mashtizadeh et al.~\cite{CCFI:CCS15} created Cryptographic CFI (CCFI),
which uses message authentication codes (MACs) to sign and verify code
pointers and leverages x86's AES-NI instructions to accelerate MAC
calculation.
ARMv8.3-A's PAuth~\cite{ARM:Manual} adds hardware support for pointer
authentication codes (PACs) and places PACs in unused upper bits of
pointers.
Qualcomm has adopted PAuth to enforce CFI~\cite{PAuth:Qualcomm-WP17}.
However, CCFI and plain PAuth suffer from pointer reuse attacks,
in which attackers use buffer overread
vulnerabilities~\cite{Overread:EuroSec09} to harvest signed pointers
for later reuse.
Utilizing PAuth,
PARTS~\cite{PARTS:UsenixSec19} signs code pointers with type IDs;
this limits reuse of signed return addresses within the same functions
and signed function pointers within the same types.
PACStack~\cite{PACStack:UsenixSec21} and
PACtight~\cite{PACtight:UsenixSec22} are also based on PAuth;
both solutions sign a return address with the PAC of the previous
return address, creating an authenticated stack.
PACtight further signs a function pointer with its address and a random
tag.
Studies on type-ID-based PACs~\cite{RetTag:EuroSec22} and
authenticated chain of return addresses~\cite{ZipperStack:ESORICS20}
have also been explored on RISC-V as custom ISA extensions.
PAL~\cite{PAL:UsenixSec22} uses PAuth to provide CFI for OS kernels.

As PACStack~\cite{PACStack:UsenixSec21} and
PACtight~\cite{PACtight:UsenixSec22} share the most similar
threat model, assumptions, and security guarantees with {\System},
we compare {\System} with them in more detail.
PACStack claims that its authenticated stack ``achieves security
comparable to hardware-assisted shadow stacks
\emph{without requiring dedicated hardware}'';
we show that {\System} achieves hardware-assisted shadow stacks
\emph{with even less hardware requirements}
(ARMv8.1-A's PAN and HPDS vs. ARMv8.3-A's PAuth).
Furthermore, PACStack requires forward-edge CFI but reported
performance numbers without accounting its overhead.
For an apples-to-apples comparison,
{\System} without forward-edge CFI outperforms PACStack
(1.9\% vs. $\approx$3.0\% on SPEC CPU 2017 and
$\le$3.0\% vs. 6--13\% on Nginx).
PACtight enforces finer-grained forward-edge CFI than {\System}
and its performance (4.0\% on Nginx) is roughly on par with {\System}.
However, PACtight maintains an in-memory metadata storage
for the random tags at runtime and relies on
ASLR~\cite{ASLR:PaX01} to hide its location.
Essentially, PAC-based systems only offer probabilistic security
even if the entropy they provide is large.
In contrast, {\System}'s shadow stacks are integrity-enforced,
providing the strongest guarantees.

\paragraph{Other Approaches}

Kuznetsov et al.~\cite{CPI:OSDI14} developed
code-pointer integrity (CPI),
an approach to ensuring memory safety of all code pointers and
data related to code pointers.
CPI identifies such data via static analysis and instrumentation
and places the data in isolated safe regions.
Again, segmentation~\cite{X86:Intel:Manual,X86:AMD:Manual} and
ASLR~\cite{ASLR:PaX01} were used to protect the safe regions
on x86\_32 and x86\_64, respectively.
PACtight-CPI~\cite{PACtight:UsenixSec22} implements CPI using PAuth,
incurring 4.07\% performance overhead on average.
{\System}'s {\Technique} provides an alternative option to protect
CPI's safe regions with potentially less overhead.
$\mu$RAI~\cite{uRAI:NDSS20} enforces return address integrity
on microcontrollers by encoding return addresses in
a reserved register and ensuring that the register value is never
corrupted;
it relies on system calls to spill the register value to protected
memory when needing to fold a call chain longer than
what a single register can hold.
While $\mu$RAI is in theory applicable to general-purpose systems
like x86 and AArch64,
we believe such an approach provides poor scalability
and may incur high performance overhead
due to more nested function calls than on microcontrollers.

\subsection{Intra-Address Space Isolation}
\label{sec:related:isolate}

{\System} uses {\Technique} for efficient intra-address space isolation.
We omit discussing custom hardware modifications that compartmentalize
software (e.g., CODOMs~\cite{CODOMs:ISCA14} and
Mondrian~\cite{Mondrian:ASPLOS02,Mondrix:SOSP05})
and limit our discussion on related work utilizing recent commodity
hardware.
Approaches used to enforce CFI are also not repeated here.

SFI~\cite{SFI:SOSP93,PittSFIeld:UsenixSec06} instruments program loads
and stores to prevent them from accessing certain memory regions and
has been used to sandbox
untrusted code~\cite{NaCl:Oakland09,PNaCl:UsenixSec10,Privbox:ATC22}.
While some systems~\cite{MemSentry:EuroSys17,Apparition:UsenixSec18}
accelerate SFI checks using MPX on x86,
the overhead of SFI is still considered high
(on both performance~\cite{SEIMI:Oakland20} and
memory usage~\cite{SoK:SS:Oakland19})
and grows as the number of isolated regions increases.
Furthermore, SFI often requires CFI to ensure that SFI checks are not
bypassed by attacker-manipulated control flow.
Another address-based isolation technique is hardware-enforced
address range monitoring.
PicoXOM~\cite{PicoXOM:SecDev20} enforces execute-only memory (XOM) by
configuring ARM debug registers to watch over a code segment against
read accesses.
Such approaches are limited by hardware resources available and
cannot scale up.

Recent defenses enforce domain-based isolation;
memory regions are associated with a protection domain,
and different mechanisms are used to allow or disallow accesses
to the protection domain at runtime.
On x86, researchers have explored domain-based memory access control
using hardware features such as
Virtual Machine Extensions
(VMX)~\cite{SeCage:CCS15,MemSentry:EuroSys17,SkyBridge:EuroSys19,%
Hodor:ATC19,LVDs:VEE20,xMP:Oakland20,SEIMI:Oakland20,EPK:ATC22},
MPK~\cite{libmpk:ATC19,Hodor:ATC19,ERIM:UsenixSec19,libhermitMPK:VEE20,%
MonGuard:EuroSec20,UnderBridge:ATC20,Donky:UsenixSec20,%
CubicleOS:ASPLOS21,Cerberus:EuroSys22,EPK:ATC22},
SMAP~\cite{SEIMI:Oakland20}, and CET~\cite{CETIS:CCS22}.
ARMlock~\cite{ARMlock:CCS14} and Shreds~\cite{Shreds:Oakland16}
use ARM domains, which are only available on AArch32~\cite{ARM:Manual}.
Previous work has also used LSU instructions for isolation.
ILDI~\cite{ILDI:DAC17} utilizes LSU instructions and PAN to protect
a safe region inside the OS kernel; it relies on a more privileged
hypervisor to moderate sensitive kernel operations.
uXOM~\cite{uXOM:UsenixSec19} transforms regular loads/stores to LSU
instructions to enforce XOM on microcontrollers,
where application code typically executes in the privileged mode
already.
{\System}, employing {\Technique}, is the first to extend domain-based
isolation to AArch64 user space.

We notice that Privbox~\cite{Privbox:ATC22} and
SEIMI~\cite{SEIMI:Oakland20}, like {\System}, also proposed executing
user-space code in the privileged mode (x86's ring 0).
Privbox does so to accelerate system call invocation and uses SFI to
safely run elevated code.
The overhead of its heavy instrumentation, however, may outweigh its
speedup from faster system calls on certain programs.
{\System} can benefit from the idea of system call acceleration for
{\elevated} tasks,
which we leave as future work.
SEIMI flips SMAP (x86 equivalence to PAN) to create a safe region for
trusted user-space code;
its OS kernel is then elevated to run in ring -1 via VMX.
Compared with SEIMI, {\System}'s {\Technique} provides
instruction-level isolation and requires no frequent domain switching.

\section{Conclusions and Future Work}
\label{sec:conc}

In conclusion, we presented {\System},
a hardware-assisted protected shadow stack implementation for AArch64,
which utilizes common hardware features to create novel and efficient
intra-address space isolation and safely executes user-space code
in the privileged mode via OS kernel and compiler restraints.
{\System} is backward-compatible with existing application binaries
by a novel use of another AArch64 feature.
Our analysis shows that {\System} is secure and effective in
mitigating attacks,
and our performance evaluation demonstrates the low costs of {\System}
on real-world benchmarks and applications.
Our prototype of {\System} is open-sourced at
\url{https://github.com/URSec/InversOS}.

We see several directions for future work.
First, we can explore system call optimizations
(such as Privbox~\cite{Privbox:ATC22})
for {\elevated} tasks;
these tasks already run in the privileged mode and can accelerate
system call invocation by avoiding the costly {\tt SVC} instructions.
Second, we can leverage {\Technique} to enforce other security policies
such as CPI~\cite{CPI:OSDI14} and
full memory safety~\cite{SAFECode:PLDI06,SoftBound:PLDI09,CETS:ISMM10,%
BOGO:ASPLOS19},
reducing their overheads significantly.
Finally, we intend to investigate potential performance improvements
to {\System} by using more recent ISA features
(e.g., BTI and E0PD)~\cite{ARM:Manual}
on real hardware.


\begin{acks}

This work was supported by ONR Award N00014-17-1-2996 and NSF Awards
CNS-1955498 and CNS-2154322.

\end{acks}

\bibliographystyle{ACM-Reference-Format}
\bibliography{inversos}


\begin{thebibliography}{155}


\ifx \showCODEN    \undefined \def \showCODEN     #1{\unskip}     \fi
\ifx \showDOI      \undefined \def \showDOI       #1{#1}\fi
\ifx \showISBNx    \undefined \def \showISBNx     #1{\unskip}     \fi
\ifx \showISBNxiii \undefined \def \showISBNxiii  #1{\unskip}     \fi
\ifx \showISSN     \undefined \def \showISSN      #1{\unskip}     \fi
\ifx \showLCCN     \undefined \def \showLCCN      #1{\unskip}     \fi
\ifx \shownote     \undefined \def \shownote      #1{#1}          \fi
\ifx \showarticletitle \undefined \def \showarticletitle #1{#1}   \fi
\ifx \showURL      \undefined \def \showURL       {\relax}        \fi
\providecommand\bibfield[2]{#2}
\providecommand\bibinfo[2]{#2}
\providecommand\natexlab[1]{#1}
\providecommand\showeprint[2][]{arXiv:#2}

\bibitem[Abadi et~al\mbox{.}(2005)]%
        {CFI:CCS05}
\bibfield{author}{\bibinfo{person}{Mart\'{\i}n Abadi}, \bibinfo{person}{Mihai
  Budiu}, \bibinfo{person}{\'{U}lfar Erlingsson}, {and} \bibinfo{person}{Jay
  Ligatti}.} \bibinfo{year}{2005}\natexlab{}.
\newblock \showarticletitle{Control-Flow Integrity}. In
  \bibinfo{booktitle}{\emph{Proceedings of the 12th ACM Conference on Computer
  and Communications Security}} \emph{(\bibinfo{series}{CCS '05})}.
  \bibinfo{publisher}{ACM}, \bibinfo{address}{Alexandria, VA, USA},
  \bibinfo{pages}{340--353}.
\newblock
\showISBNx{1-59593-226-7}
\urldef\tempurl%
\url{https://doi.org/10.1145/1102120.1102165}
\showDOI{\tempurl}


\bibitem[Abadi et~al\mbox{.}(2009)]%
        {CFI:TISSEC09}
\bibfield{author}{\bibinfo{person}{Mart\'{\i}n Abadi}, \bibinfo{person}{Mihai
  Budiu}, \bibinfo{person}{\'{U}lfar Erlingsson}, {and} \bibinfo{person}{Jay
  Ligatti}.} \bibinfo{year}{2009}\natexlab{}.
\newblock \showarticletitle{Control-Flow Integrity Principles, Implementations,
  and Applications}.
\newblock \bibinfo{journal}{\emph{ACM Transactions on Information and System
  Security}} \bibinfo{volume}{13}, \bibinfo{number}{1}, Article
  \bibinfo{articleno}{4} (\bibinfo{date}{Nov.} \bibinfo{year}{2009}),
  \bibinfo{numpages}{40}~pages.
\newblock
\showISSN{1094-9224}
\urldef\tempurl%
\url{https://doi.org/10.1145/1609956.1609960}
\showDOI{\tempurl}


\bibitem[Advanced Micro Devices Inc.(2023)]%
        {X86:AMD:Manual}
Advanced Micro Devices Inc. \bibinfo{year}{2023}\natexlab{}.
\newblock \bibinfo{booktitle}{\emph{AMD64 Architecture Programmer's Manual}}.
\newblock Advanced Micro Devices Inc.
\newblock
\urldef\tempurl%
\url{https://www.amd.com/en/support/tech-docs/amd64-architecture-programmers-manual-volumes-1-5}
\showURL{%
\tempurl}
\newblock
\shownote{40332 Rev 4.06}.


\bibitem[Almakhdhub et~al\mbox{.}(2020)]%
        {uRAI:NDSS20}
\bibfield{author}{\bibinfo{person}{Naif~Saleh Almakhdhub},
  \bibinfo{person}{Abraham~A. Clements}, \bibinfo{person}{Saurabh Bagchi},
  {and} \bibinfo{person}{Mathias Payer}.} \bibinfo{year}{2020}\natexlab{}.
\newblock \showarticletitle{{$\mu$RAI}: Securing Embedded Systems with Return
  Address Integrity}. In \bibinfo{booktitle}{\emph{Proceedings of the 2020
  Network and Distributed System Security Symposium}}
  \emph{(\bibinfo{series}{NDSS '20})}. \bibinfo{publisher}{Internet Society},
  \bibinfo{address}{San Diego, CA, USA}, \bibinfo{numpages}{18}~pages.
\newblock
\showISBNx{1-891562-61-4}
\urldef\tempurl%
\url{https://doi.org/10.14722/ndss.2020.24016}
\showDOI{\tempurl}


\bibitem[Amazon Web Services(2023)]%
        {EC2-A1:AWS}
Amazon Web Services \bibinfo{year}{2023}\natexlab{}.
\newblock \bibinfo{booktitle}{\emph{Amazon EC2 A1 Instances: Optimized cost and
  performance for scale-out workloads}}.
\newblock
\urldef\tempurl%
\url{https://aws.amazon.com/ec2/instance-types/a1}
\showURL{%
\tempurl}


\bibitem[Apache(2023)]%
        {ApacheBench}
Apache \bibinfo{year}{2023}\natexlab{}.
\newblock \bibinfo{booktitle}{\emph{{ab} - {Apache} {HTTP} server benchmarking
  tool}}.
\newblock
\urldef\tempurl%
\url{https://httpd.apache.org/docs/current/programs/ab.html}
\showURL{%
\tempurl}


\bibitem[Apple(2020)]%
        {M1:Apple}
Apple \bibinfo{year}{2020}\natexlab{}.
\newblock \bibinfo{booktitle}{\emph{Apple unleashes M1}}.
\newblock
\urldef\tempurl%
\url{https://www.apple.com/newsroom/2020/11/apple-unleashes-m1}
\showURL{%
\tempurl}


\bibitem[Arm Holdings(2021)]%
        {ARMv7-M:Manual}
Arm Holdings \bibinfo{year}{2021}\natexlab{}.
\newblock \bibinfo{booktitle}{\emph{Arm\textsuperscript{\textregistered}v7-M
  Architecture Reference Manual}}.
\newblock Arm Holdings.
\newblock
\urldef\tempurl%
\url{https://developer.arm.com/documentation/ddi0403/ee}
\showURL{%
\tempurl}
\newblock
\shownote{{DDI} 0403E.e}.


\bibitem[Arm Holdings(2022a)]%
        {ARM:Manual}
Arm Holdings \bibinfo{year}{2022}\natexlab{a}.
\newblock \bibinfo{booktitle}{\emph{Arm\textsuperscript{\textregistered}
  Architecture Reference Manual: for A-profile architecture}}.
\newblock Arm Holdings.
\newblock
\urldef\tempurl%
\url{https://developer.arm.com/documentation/ddi0487/ia}
\showURL{%
\tempurl}
\newblock
\shownote{{DDI} 0487I.a}.


\bibitem[Arm Holdings(2022b)]%
        {ARMv8-M:Manual}
Arm Holdings \bibinfo{year}{2022}\natexlab{b}.
\newblock \bibinfo{booktitle}{\emph{Arm\textsuperscript{\textregistered}v8-M
  Architecture Reference Manual}}.
\newblock Arm Holdings.
\newblock
\urldef\tempurl%
\url{https://developer.arm.com/documentation/ddi0553/bv}
\showURL{%
\tempurl}
\newblock
\shownote{{DDI} 0553B.v}.


\bibitem[Bittau et~al\mbox{.}(2014)]%
        {BROP:Oakland14}
\bibfield{author}{\bibinfo{person}{Andrea Bittau}, \bibinfo{person}{Adam
  Belay}, \bibinfo{person}{Ali Mashtizadeh}, \bibinfo{person}{David
  Mazi\`{e}res}, {and} \bibinfo{person}{Dan Boneh}.}
  \bibinfo{year}{2014}\natexlab{}.
\newblock \showarticletitle{Hacking Blind}. In
  \bibinfo{booktitle}{\emph{Proceedings of the 2014 IEEE Symposium on Security
  and Privacy}} \emph{(\bibinfo{series}{SP '14})}. \bibinfo{publisher}{IEEE
  Computer Society}, \bibinfo{address}{San Jose, CA, USA},
  \bibinfo{pages}{227--242}.
\newblock
\showISBNx{978-1-4799-4686-0}
\urldef\tempurl%
\url{https://doi.org/10.1109/SP.2014.22}
\showDOI{\tempurl}


\bibitem[Bletsch et~al\mbox{.}(2011a)]%
        {CFL:ACSAC11}
\bibfield{author}{\bibinfo{person}{Tyler Bletsch}, \bibinfo{person}{Xuxian
  Jiang}, {and} \bibinfo{person}{Vince Freeh}.}
  \bibinfo{year}{2011}\natexlab{a}.
\newblock \showarticletitle{Mitigating Code-Reuse Attacks with Control-Flow
  Locking}. In \bibinfo{booktitle}{\emph{Proceedings of the 27th Annual
  Computer Security Applications Conference}} \emph{(\bibinfo{series}{ACSAC
  '11})}. \bibinfo{publisher}{ACM}, \bibinfo{address}{Orlando, FL, USA},
  \bibinfo{pages}{353--362}.
\newblock
\showISBNx{978-1-4503-0672-0}
\urldef\tempurl%
\url{https://doi.org/10.1145/2076732.2076783}
\showDOI{\tempurl}


\bibitem[Bletsch et~al\mbox{.}(2011b)]%
        {JOP:ASIACCS11}
\bibfield{author}{\bibinfo{person}{Tyler Bletsch}, \bibinfo{person}{Xuxian
  Jiang}, \bibinfo{person}{Vince~W. Freeh}, {and} \bibinfo{person}{Zhenkai
  Liang}.} \bibinfo{year}{2011}\natexlab{b}.
\newblock \showarticletitle{Jump-Oriented Programming: A New Class of
  Code-Reuse Attack}. In \bibinfo{booktitle}{\emph{Proceedings of the 6th ACM
  Symposium on Information, Computer and Communications Security}}
  \emph{(\bibinfo{series}{ASIACCS '11})}. \bibinfo{publisher}{ACM},
  \bibinfo{address}{Hong Kong, China}, \bibinfo{pages}{30--40}.
\newblock
\showISBNx{978-1-4503-0564-8}
\urldef\tempurl%
\url{https://doi.org/10.1145/1966913.1966919}
\showDOI{\tempurl}


\bibitem[Bounov et~al\mbox{.}(2016)]%
        {Bounov:NDSS16}
\bibfield{author}{\bibinfo{person}{Dimitar Bounov},
  \bibinfo{person}{Rami~G\"{o}khan K\i{}c\i{}}, {and} \bibinfo{person}{Sorin
  Lerner}.} \bibinfo{year}{2016}\natexlab{}.
\newblock \showarticletitle{Protecting {C++} Dynamic Dispatch Through {VTable}
  Interleaving}. In \bibinfo{booktitle}{\emph{Proceedings of the 2016 Network
  and Distributed System Security Symposium}} \emph{(\bibinfo{series}{NDSS
  '16})}. \bibinfo{publisher}{Internet Society}, \bibinfo{address}{San Diego,
  CA, USA}, \bibinfo{numpages}{15}~pages.
\newblock
\showISBNx{1-891562-41-X}
\urldef\tempurl%
\url{https://doi.org/10.14722/ndss.2016.23421}
\showDOI{\tempurl}


\bibitem[Bovet and Cesati(2005)]%
        {Bovet:Linux:3}
\bibfield{author}{\bibinfo{person}{Daniel~P. Bovet} {and}
  \bibinfo{person}{Marco Cesati}.} \bibinfo{year}{2005}\natexlab{}.
\newblock \bibinfo{booktitle}{\emph{Understanding the Linux Kernel}
  (\bibinfo{edition}{3rd} ed.)}.
\newblock \bibinfo{publisher}{O'Reilly \& Associates Inc},
  \bibinfo{address}{Sebastopol, CA, USA}.
\newblock
\showISBNx{0-5960-0565-2}


\bibitem[Burow et~al\mbox{.}(2017)]%
        {CFI:CSUR17}
\bibfield{author}{\bibinfo{person}{Nathan Burow}, \bibinfo{person}{Scott~A.
  Carr}, \bibinfo{person}{Joseph Nash}, \bibinfo{person}{Per Larsen},
  \bibinfo{person}{Michael Franz}, \bibinfo{person}{Stefan Brunthaler}, {and}
  \bibinfo{person}{Mathias Payer}.} \bibinfo{year}{2017}\natexlab{}.
\newblock \showarticletitle{Control-Flow Integrity: Precision, Security, and
  Performance}.
\newblock \bibinfo{journal}{\emph{Comput. Surveys}} \bibinfo{volume}{50},
  \bibinfo{number}{1}, Article \bibinfo{articleno}{16} (\bibinfo{date}{April}
  \bibinfo{year}{2017}), \bibinfo{numpages}{33}~pages.
\newblock
\showISSN{0360-0300}
\urldef\tempurl%
\url{https://doi.org/10.1145/3054924}
\showDOI{\tempurl}


\bibitem[Burow et~al\mbox{.}(2018)]%
        {CFIXX:NDSS18}
\bibfield{author}{\bibinfo{person}{Nathan Burow}, \bibinfo{person}{Derrick
  McKee}, \bibinfo{person}{Scott~A. Carr}, {and} \bibinfo{person}{Mathias
  Payer}.} \bibinfo{year}{2018}\natexlab{}.
\newblock \showarticletitle{{CFIXX}: Object Type Integrity for {C++}}. In
  \bibinfo{booktitle}{\emph{Proceedings of the 2018 Network and Distributed
  System Security Symposium}} \emph{(\bibinfo{series}{NDSS '18})}.
  \bibinfo{publisher}{Internet Society}, \bibinfo{address}{San Diego, CA, USA},
  \bibinfo{numpages}{14}~pages.
\newblock
\showISBNx{1-891562-49-5}
\urldef\tempurl%
\url{https://doi.org/10.14722/ndss.2018.23279}
\showDOI{\tempurl}


\bibitem[Burow et~al\mbox{.}(2019)]%
        {SoK:SS:Oakland19}
\bibfield{author}{\bibinfo{person}{Nathan Burow}, \bibinfo{person}{Xinping
  Zhang}, {and} \bibinfo{person}{Mathias Payer}.}
  \bibinfo{year}{2019}\natexlab{}.
\newblock \showarticletitle{{SoK}: Shining Light on Shadow Stacks}. In
  \bibinfo{booktitle}{\emph{Proceedings of the 2019 IEEE Symposium on Security
  and Privacy}} \emph{(\bibinfo{series}{SP '19})}. \bibinfo{publisher}{IEEE
  Computer Society}, \bibinfo{address}{San Francisco, CA, USA},
  \bibinfo{pages}{985--999}.
\newblock
\showISBNx{978-1-5386-6660-9}
\urldef\tempurl%
\url{https://doi.org/10.1109/SP.2019.00076}
\showDOI{\tempurl}


\bibitem[Carlini et~al\mbox{.}(2015)]%
        {CFBending:UsenixSec15}
\bibfield{author}{\bibinfo{person}{Nicolas Carlini}, \bibinfo{person}{Antonio
  Barresi}, \bibinfo{person}{Mathias Payer}, \bibinfo{person}{David Wagner},
  {and} \bibinfo{person}{Thomas~R. Gross}.} \bibinfo{year}{2015}\natexlab{}.
\newblock \showarticletitle{Control-Flow Bending: On the Effectiveness of
  Control-flow Integrity}. In \bibinfo{booktitle}{\emph{Proceedings of the 24th
  USENIX Security Symposium}} \emph{(\bibinfo{series}{Security '15})}.
  \bibinfo{publisher}{USENIX Association}, \bibinfo{address}{Washington, DC,
  USA}, \bibinfo{pages}{161--176}.
\newblock
\showISBNx{978-1-931971-232}
\urldef\tempurl%
\url{https://www.usenix.org/conference/usenixsecurity15/technical-sessions/presentation/carlini}
\showURL{%
\tempurl}


\bibitem[Carlini and Wagner(2014)]%
        {ROPDanger:UsenixSec14}
\bibfield{author}{\bibinfo{person}{Nicholas Carlini} {and}
  \bibinfo{person}{David Wagner}.} \bibinfo{year}{2014}\natexlab{}.
\newblock \showarticletitle{{ROP} is Still Dangerous: Breaking Modern
  Defenses}. In \bibinfo{booktitle}{\emph{Proceedings of the 23rd USENIX
  Security Symposium}} \emph{(\bibinfo{series}{Security '14})}.
  \bibinfo{publisher}{USENIX Association}, \bibinfo{address}{San Diego, CA,
  USA}, \bibinfo{pages}{385--399}.
\newblock
\showISBNx{978-1-931971-15-7}
\urldef\tempurl%
\url{https://www.usenix.org/conference/usenixsecurity14/technical-sessions/presentation/carlini}
\showURL{%
\tempurl}


\bibitem[Castro et~al\mbox{.}(2009)]%
        {BGI:SOSP09}
\bibfield{author}{\bibinfo{person}{Miguel Castro}, \bibinfo{person}{Manuel
  Costa}, \bibinfo{person}{Jean-Philippe Martin}, \bibinfo{person}{Marcus
  Peinado}, \bibinfo{person}{Periklis Akritidis}, \bibinfo{person}{Austin
  Donnelly}, \bibinfo{person}{Paul Barham}, {and} \bibinfo{person}{Richard
  Black}.} \bibinfo{year}{2009}\natexlab{}.
\newblock \showarticletitle{Fast Byte-Granularity Software Fault Isolation}. In
  \bibinfo{booktitle}{\emph{Proceedings of the 22nd ACM SIGOPS Symposium on
  Operating Systems Principles}} \emph{(\bibinfo{series}{SOSP '09})}.
  \bibinfo{publisher}{ACM}, \bibinfo{address}{Big Sky, MT, USA},
  \bibinfo{pages}{45--58}.
\newblock
\showISBNx{978-1-6055-8752-3}
\urldef\tempurl%
\url{https://doi.org/10.1145/1629575.1629581}
\showDOI{\tempurl}


\bibitem[Chen et~al\mbox{.}(2005)]%
        {NonCtrlDataAttack:UsenixSec05}
\bibfield{author}{\bibinfo{person}{Shuo Chen}, \bibinfo{person}{Jun Xu},
  \bibinfo{person}{Emre~C. Sezer}, \bibinfo{person}{Prachi Gauriar}, {and}
  \bibinfo{person}{Ravishankar~K. Iyer}.} \bibinfo{year}{2005}\natexlab{}.
\newblock \showarticletitle{Non-Control-Data Attacks Are Realistic Threats}. In
  \bibinfo{booktitle}{\emph{Proceedings of the 14th USENIX Security Symposium}}
  \emph{(\bibinfo{series}{Security '05})}. \bibinfo{publisher}{USENIX
  Association}, \bibinfo{address}{Baltimore, MD, USA},
  \bibinfo{pages}{177--191}.
\newblock
\urldef\tempurl%
\url{https://www.usenix.org/conference/14th-usenix-security-symposium/non-control-data-attacks-are-realistic-threats}
\showURL{%
\tempurl}


\bibitem[Chen et~al\mbox{.}(2016)]%
        {Shreds:Oakland16}
\bibfield{author}{\bibinfo{person}{Yaohui Chen}, \bibinfo{person}{Sebassujeen
  Reymondjohnson}, \bibinfo{person}{Zhichuang Sun}, {and} \bibinfo{person}{Long
  Lu}.} \bibinfo{year}{2016}\natexlab{}.
\newblock \showarticletitle{Shreds: Fine-Grained Execution Units with Private
  Memory}. In \bibinfo{booktitle}{\emph{Proceedings of the 2016 IEEE Symposium
  on Security and Privacy}} \emph{(\bibinfo{series}{SP '16})}.
  \bibinfo{publisher}{IEEE Computer Society}, \bibinfo{address}{San Jose, CA,
  USA}, \bibinfo{pages}{56--71}.
\newblock
\showISBNx{978-1-5090-0824-7}
\urldef\tempurl%
\url{https://doi.org/10.1109/SP.2016.12}
\showDOI{\tempurl}


\bibitem[Cheng et~al\mbox{.}(2014)]%
        {ROPecker:NDSS14}
\bibfield{author}{\bibinfo{person}{Yueqiang Cheng}, \bibinfo{person}{Zongwei
  Zhou}, \bibinfo{person}{Miao Yu}, \bibinfo{person}{Xuhua Ding}, {and}
  \bibinfo{person}{Robert~H. Deng}.} \bibinfo{year}{2014}\natexlab{}.
\newblock \showarticletitle{{ROPecker}: A Generic and Practical Approach for
  Defending Against {ROP} Attacks}. In \bibinfo{booktitle}{\emph{Proceedings of
  the 2014 Network and Distributed System Security Symposium}}
  \emph{(\bibinfo{series}{NDSS '14})}. \bibinfo{publisher}{Internet Society},
  \bibinfo{address}{San Diego, CA, USA}, \bibinfo{numpages}{14}~pages.
\newblock
\showISBNx{1-891562-35-5}
\urldef\tempurl%
\url{https://doi.org/10.14722/ndss.2014.23156}
\showDOI{\tempurl}


\bibitem[Chiueh and Hsu(2001)]%
        {RAD:ICDCS01}
\bibfield{author}{\bibinfo{person}{Tzi-Cker Chiueh} {and}
  \bibinfo{person}{Fu-Hau Hsu}.} \bibinfo{year}{2001}\natexlab{}.
\newblock \showarticletitle{{RAD}: A Compile-Time Solution to Buffer Overflow
  Attacks}. In \bibinfo{booktitle}{\emph{Proceedings of the 21st International
  Conference on Distributed Computing Systems}} \emph{(\bibinfo{series}{ICDCS
  '01})}. \bibinfo{publisher}{IEEE Computer Society}, \bibinfo{address}{Mesa,
  AZ, USA}, \bibinfo{pages}{409--417}.
\newblock
\showISBNx{0-7695-1077-9}
\urldef\tempurl%
\url{https://doi.org/10.1109/ICDSC.2001.918971}
\showDOI{\tempurl}


\bibitem[Cho et~al\mbox{.}(2017)]%
        {ILDI:DAC17}
\bibfield{author}{\bibinfo{person}{Yeongpil Cho}, \bibinfo{person}{Donghyun
  Kwon}, {and} \bibinfo{person}{Yunheung Paek}.}
  \bibinfo{year}{2017}\natexlab{}.
\newblock \showarticletitle{Instruction-Level Data Isolation for the Kernel on
  {ARM}}. In \bibinfo{booktitle}{\emph{Proceedings of the 54th ACM/EDAC/IEEE
  Annual Design Automation Conference}} \emph{(\bibinfo{series}{DAC '17})}.
  \bibinfo{publisher}{ACM}, \bibinfo{address}{Austin, TX, USA}, Article
  \bibinfo{articleno}{26}, \bibinfo{numpages}{6}~pages.
\newblock
\showISBNx{978-1-4503-4927-7}
\urldef\tempurl%
\url{https://doi.org/10.1145/3061639.3062267}
\showDOI{\tempurl}


\bibitem[Christoulakis et~al\mbox{.}(2016)]%
        {HCFI:CODASPY16}
\bibfield{author}{\bibinfo{person}{Nick Christoulakis}, \bibinfo{person}{George
  Christou}, \bibinfo{person}{Elias Athanasopoulos}, {and}
  \bibinfo{person}{Sotiris Ioannidis}.} \bibinfo{year}{2016}\natexlab{}.
\newblock \showarticletitle{{HCFI}: Hardware-Enforced Control-Flow Integrity}.
  In \bibinfo{booktitle}{\emph{Proceedings of the 6th ACM Conference on Data
  and Application Security and Privacy}} \emph{(\bibinfo{series}{CODASPY
  '16})}. \bibinfo{publisher}{ACM}, \bibinfo{address}{New Orleans, LA, USA},
  \bibinfo{pages}{38--49}.
\newblock
\showISBNx{978-1-4503-3935-3}
\urldef\tempurl%
\url{https://doi.org/10.1145/2857705.2857722}
\showDOI{\tempurl}


\bibitem[Cloosters et~al\mbox{.}(2022)]%
        {RiscyROP:RAID22}
\bibfield{author}{\bibinfo{person}{Tobias Cloosters}, \bibinfo{person}{David
  Paa\ss{}en}, \bibinfo{person}{Jianqiang Wang}, \bibinfo{person}{Oussama
  Draissi}, \bibinfo{person}{Patrick Jauernig}, \bibinfo{person}{Emmanuel
  Stapf}, \bibinfo{person}{Lucas Davi}, {and} \bibinfo{person}{Ahmad-Reza
  Sadeghi}.} \bibinfo{year}{2022}\natexlab{}.
\newblock \showarticletitle{{RiscyROP}: Automated Return-Oriented Programming
  Attacks on {RISC-V} and {ARM64}}. In \bibinfo{booktitle}{\emph{Proceedings of
  the 25th International Symposium on Research in Attacks, Intrusions and
  Defenses}} \emph{(\bibinfo{series}{RAID '22})}. \bibinfo{publisher}{ACM},
  \bibinfo{address}{Limassol, Cyprus}, \bibinfo{pages}{30--42}.
\newblock
\showISBNx{978-1-4503-9704-9}
\urldef\tempurl%
\url{https://doi.org/10.1145/3545948.3545997}
\showDOI{\tempurl}


\bibitem[Conti et~al\mbox{.}(2015)]%
        {LoseCtrl:CCS15}
\bibfield{author}{\bibinfo{person}{Mauro Conti}, \bibinfo{person}{Stephen
  Crane}, \bibinfo{person}{Lucas Davi}, \bibinfo{person}{Michael Franz},
  \bibinfo{person}{Per Larsen}, \bibinfo{person}{Marco Negro},
  \bibinfo{person}{Christopher Liebchen}, \bibinfo{person}{Mohaned Qunaibit},
  {and} \bibinfo{person}{Ahmad-Reza Sadeghi}.} \bibinfo{year}{2015}\natexlab{}.
\newblock \showarticletitle{Losing Control: On the Effectiveness of
  Control-Flow Integrity under Stack Attacks}. In
  \bibinfo{booktitle}{\emph{Proceedings of the 22nd ACM SIGSAC Conference on
  Computer and Communications Security}} \emph{(\bibinfo{series}{CCS '15})}.
  \bibinfo{publisher}{ACM}, \bibinfo{address}{Denver, CO, USA},
  \bibinfo{pages}{952--963}.
\newblock
\showISBNx{978-1-4503-3832-5}
\urldef\tempurl%
\url{https://doi.org/10.1145/2810103.2813671}
\showDOI{\tempurl}


\bibitem[Corliss et~al\mbox{.}(2005)]%
        {DISE:WASSA04}
\bibfield{author}{\bibinfo{person}{Marc~L. Corliss},
  \bibinfo{person}{E.~Christopher Lewis}, {and} \bibinfo{person}{Amir Roth}.}
  \bibinfo{year}{2005}\natexlab{}.
\newblock \showarticletitle{Using {DISE} to Protect Return Addresses from
  Attack}.
\newblock \bibinfo{journal}{\emph{SIGARCH Computer Architecture News}}
  \bibinfo{volume}{33}, \bibinfo{number}{1} (\bibinfo{date}{March}
  \bibinfo{year}{2005}), \bibinfo{pages}{65--72}.
\newblock
\showISSN{0163-5964}
\urldef\tempurl%
\url{https://doi.org/10.1145/1055626.1055636}
\showDOI{\tempurl}


\bibitem[Criswell et~al\mbox{.}(2014)]%
        {KCoFI:Oakland14}
\bibfield{author}{\bibinfo{person}{John Criswell}, \bibinfo{person}{Nathan
  Dautenhahn}, {and} \bibinfo{person}{Vikram Adve}.}
  \bibinfo{year}{2014}\natexlab{}.
\newblock \showarticletitle{{KCoFI}: Complete Control-Flow Integrity for
  Commodity Operating System Kernels}. In \bibinfo{booktitle}{\emph{Proceedings
  of the 2014 IEEE Symposium on Security and Privacy}}
  \emph{(\bibinfo{series}{SP '14})}. \bibinfo{publisher}{IEEE Computer
  Society}, \bibinfo{address}{San Jose, CA, USA}, \bibinfo{pages}{292--307}.
\newblock
\showISBNx{978-1-4799-4686-0}
\urldef\tempurl%
\url{https://doi.org/10.1109/SP.2014.26}
\showDOI{\tempurl}


\bibitem[Dall and Nieh(2014)]%
        {KVM/ARM:ASPLOS14}
\bibfield{author}{\bibinfo{person}{Christoffer Dall} {and}
  \bibinfo{person}{Jason Nieh}.} \bibinfo{year}{2014}\natexlab{}.
\newblock \showarticletitle{{KVM/ARM}: The Design and Implementation of the
  {Linux} {ARM} Hypervisor}. In \bibinfo{booktitle}{\emph{Proceedings of the
  19th International Conference on Architectural Support for Programming
  Languages and Operating Systems}} \emph{(\bibinfo{series}{ASPLOS '14})}.
  \bibinfo{publisher}{ACM}, \bibinfo{address}{Salt Lake City, UT, USA},
  \bibinfo{pages}{333--348}.
\newblock
\showISBNx{978-1-4503-2305-5}
\urldef\tempurl%
\url{https://doi.org/10.1145/2541940.2541946}
\showDOI{\tempurl}


\bibitem[Dang et~al\mbox{.}(2015)]%
        {SS:ASIACCS15}
\bibfield{author}{\bibinfo{person}{Thurston~H.Y. Dang}, \bibinfo{person}{Petros
  Maniatis}, {and} \bibinfo{person}{David Wagner}.}
  \bibinfo{year}{2015}\natexlab{}.
\newblock \showarticletitle{The Performance Cost of Shadow Stacks and Stack
  Canaries}. In \bibinfo{booktitle}{\emph{Proceedings of the 10th ACM Symposium
  on Information, Computer and Communications Security}}
  \emph{(\bibinfo{series}{ASIACCS '15})}. \bibinfo{publisher}{ACM},
  \bibinfo{address}{Singapore, Republic of Singapore},
  \bibinfo{pages}{555--566}.
\newblock
\showISBNx{978-1-4503-3245-3}
\urldef\tempurl%
\url{https://doi.org/10.1145/2714576.2714635}
\showDOI{\tempurl}


\bibitem[Davi et~al\mbox{.}(2012)]%
        {MoCFI:NDSS12}
\bibfield{author}{\bibinfo{person}{Lucas Davi}, \bibinfo{person}{Alexandra
  Dmitrienko}, \bibinfo{person}{Manuel Egele}, \bibinfo{person}{Thomas
  Fischer}, \bibinfo{person}{Thorsten Holz}, \bibinfo{person}{Ralf Hund},
  \bibinfo{person}{Stefan N{\"u}rnberger}, {and} \bibinfo{person}{Ahmad-Reza
  Sadeghi}.} \bibinfo{year}{2012}\natexlab{}.
\newblock \showarticletitle{{MoCFI}: A Framework to Mitigate Control-Flow
  Attacks on Smartphones}. In \bibinfo{booktitle}{\emph{Proceedings of the 2012
  Network and Distributed System Security Symposium}}
  \emph{(\bibinfo{series}{NDSS '12})}. \bibinfo{publisher}{Internet Society},
  \bibinfo{address}{San Diego, CA, USA}, \bibinfo{numpages}{17}~pages.
\newblock
\urldef\tempurl%
\url{https://www.ndss-symposium.org/ndss2012/ndss-2012-programme/mocfi-framework-mitigate-control-flow-attacks-smartphonesoverlay-contextmocfi-framework-mitigate-control-flow-attacks-smartphones}
\showURL{%
\tempurl}


\bibitem[Davi et~al\mbox{.}(2015)]%
        {HAFIX:DAC15}
\bibfield{author}{\bibinfo{person}{Lucas Davi}, \bibinfo{person}{Matthias
  Hanreich}, \bibinfo{person}{Debayan Paul}, \bibinfo{person}{Ahmad-Reza
  Sadeghi}, \bibinfo{person}{Patrick Koeberl}, \bibinfo{person}{Dean Sullivan},
  \bibinfo{person}{Orlando Arias}, {and} \bibinfo{person}{Yier Jin}.}
  \bibinfo{year}{2015}\natexlab{}.
\newblock \showarticletitle{{HAFIX}: Hardware-Assisted Flow Integrity
  Extension}. In \bibinfo{booktitle}{\emph{Proceedings of the 52nd
  ACM/EDAC/IEEE Annual Design Automation Conference}}
  \emph{(\bibinfo{series}{DAC '15})}. \bibinfo{publisher}{ACM},
  \bibinfo{address}{San Francisco, CA, USA}, Article \bibinfo{articleno}{74},
  \bibinfo{numpages}{6}~pages.
\newblock
\showISBNx{978-1-4503-3520-1}
\urldef\tempurl%
\url{https://doi.org/10.1145/2744769.2744847}
\showDOI{\tempurl}


\bibitem[Davi et~al\mbox{.}(2014)]%
        {StitchGadgets:UsenixSec14}
\bibfield{author}{\bibinfo{person}{Lucas Davi}, \bibinfo{person}{Ahmad-Reza
  Sadeghi}, \bibinfo{person}{Daniel Lehmann}, {and} \bibinfo{person}{Fabian
  Monrose}.} \bibinfo{year}{2014}\natexlab{}.
\newblock \showarticletitle{Stitching the Gadgets: On the Ineffectiveness of
  Coarse-Grained Control-Flow Integrity Protection}. In
  \bibinfo{booktitle}{\emph{Proceedings of the 23rd USENIX Security Symposium}}
  \emph{(\bibinfo{series}{Security '14})}. \bibinfo{publisher}{USENIX
  Association}, \bibinfo{address}{San Diego, CA, USA},
  \bibinfo{pages}{401--416}.
\newblock
\showISBNx{978-1-931971-15-7}
\urldef\tempurl%
\url{https://www.usenix.org/conference/usenixsecurity14/technical-sessions/presentation/davi}
\showURL{%
\tempurl}


\bibitem[Davi et~al\mbox{.}(2011)]%
        {ROPdefender:ASIACCS11}
\bibfield{author}{\bibinfo{person}{Lucas Davi}, \bibinfo{person}{Ahmad-Reza
  Sadeghi}, {and} \bibinfo{person}{Marcel Winandy}.}
  \bibinfo{year}{2011}\natexlab{}.
\newblock \showarticletitle{{ROPdefender}: A Detection Tool to Defend against
  Return-Oriented Programming Attacks}. In
  \bibinfo{booktitle}{\emph{Proceedings of the 6th ACM Symposium on
  Information, Computer and Communications Security}}
  \emph{(\bibinfo{series}{ASIACCS '11})}. \bibinfo{publisher}{ACM},
  \bibinfo{address}{Hong Kong, China}, \bibinfo{pages}{40--51}.
\newblock
\showISBNx{978-1-4503-0564-8}
\urldef\tempurl%
\url{https://doi.org/10.1145/1966913.1966920}
\showDOI{\tempurl}


\bibitem[Dhurjati et~al\mbox{.}(2006)]%
        {SAFECode:PLDI06}
\bibfield{author}{\bibinfo{person}{Dinakar Dhurjati}, \bibinfo{person}{Sumant
  Kowshik}, {and} \bibinfo{person}{Vikram Adve}.}
  \bibinfo{year}{2006}\natexlab{}.
\newblock \showarticletitle{{SAFECode}: Enforcing Alias Analysis for Weakly
  Typed Languages}. In \bibinfo{booktitle}{\emph{Proceedings of the 2006 ACM
  SIGPLAN Conference on Programming Language Design and Implementation}}
  \emph{(\bibinfo{series}{PLDI '06})}. \bibinfo{publisher}{ACM},
  \bibinfo{address}{Ottawa, ON, Canada}, \bibinfo{pages}{144--157}.
\newblock
\showISBNx{1-59593-320-4}
\urldef\tempurl%
\url{https://doi.org/10.1145/1133981.1133999}
\showDOI{\tempurl}


\bibitem[Ding et~al\mbox{.}(2017)]%
        {PittyPat:UsenixSec17}
\bibfield{author}{\bibinfo{person}{Ren Ding}, \bibinfo{person}{Chenxiong Qian},
  \bibinfo{person}{Chengyu Song}, \bibinfo{person}{William Harris},
  \bibinfo{person}{Taesoo Kim}, {and} \bibinfo{person}{Wenke Lee}.}
  \bibinfo{year}{2017}\natexlab{}.
\newblock \showarticletitle{Efficient Protection of Path-Sensitive Control
  Security}. In \bibinfo{booktitle}{\emph{Proceedings of the 26th USENIX
  Security Symposium}} \emph{(\bibinfo{series}{Security '17})}.
  \bibinfo{publisher}{USENIX Association}, \bibinfo{address}{Vancouver, BC,
  Canada}, \bibinfo{pages}{131--148}.
\newblock
\showISBNx{978-1-931971-40-9}
\urldef\tempurl%
\url{https://www.usenix.org/conference/usenixsecurity17/technical-sessions/presentation/ding}
\showURL{%
\tempurl}


\bibitem[Dong et~al\mbox{.}(2018)]%
        {Apparition:UsenixSec18}
\bibfield{author}{\bibinfo{person}{Xiaowan Dong}, \bibinfo{person}{Zhuojia
  Shen}, \bibinfo{person}{John Criswell}, \bibinfo{person}{Alan~L. Cox}, {and}
  \bibinfo{person}{Sandhya Dwarkadas}.} \bibinfo{year}{2018}\natexlab{}.
\newblock \showarticletitle{Shielding Software from Privileged Side-Channel
  Attacks}. In \bibinfo{booktitle}{\emph{Proceedings of the 27th USENIX
  Security Symposium}} \emph{(\bibinfo{series}{Security '18})}.
  \bibinfo{publisher}{USENIX Association}, \bibinfo{address}{Baltimore, MD,
  USA}, \bibinfo{pages}{1441--1458}.
\newblock
\showISBNx{978-1-939133-04-5}
\urldef\tempurl%
\url{https://www.usenix.org/conference/usenixsecurity18/presentation/dong}
\showURL{%
\tempurl}


\bibitem[Du et~al\mbox{.}(2022)]%
        {Kage:UsenixSec22}
\bibfield{author}{\bibinfo{person}{Yufei Du}, \bibinfo{person}{Zhuojia Shen},
  \bibinfo{person}{Komail Dharsee}, \bibinfo{person}{Jie Zhou},
  \bibinfo{person}{Robert~J. Walls}, {and} \bibinfo{person}{John Criswell}.}
  \bibinfo{year}{2022}\natexlab{}.
\newblock \showarticletitle{Holistic Control-Flow Protection on Real-Time
  Embedded Systems with {Kage}}. In \bibinfo{booktitle}{\emph{Proceedings of
  the 31st USENIX Security Symposium}} \emph{(\bibinfo{series}{Security '22})}.
  \bibinfo{publisher}{USENIX Association}, \bibinfo{address}{Boston, MA, USA},
  \bibinfo{pages}{2281--2298}.
\newblock
\showISBNx{978-1-939133-31-1}
\urldef\tempurl%
\url{https://www.usenix.org/conference/usenixsecurity22/presentation/du}
\showURL{%
\tempurl}


\bibitem[Elisei(2019)]%
        {bhyvearm64:BSDCan19}
\bibfield{author}{\bibinfo{person}{Alexandru Elisei}.}
  \bibinfo{year}{2019}\natexlab{}.
\newblock \showarticletitle{{bhyvearm64}: {CPU} and Memory Virtualization on
  {Armv8.0-A}}.
\newblock In \bibinfo{booktitle}{\emph{The BSDCan Conference}}.
  \bibinfo{address}{Ottawa, ON, Canada}.
\newblock
\urldef\tempurl%
\url{https://www.bsdcan.org/2019/schedule/events/1074.en.html}
\showURL{%
\tempurl}


\bibitem[Erlingsson et~al\mbox{.}(2006)]%
        {XFI:OSDI06}
\bibfield{author}{\bibinfo{person}{{\'U}lfar Erlingsson},
  \bibinfo{person}{Mart{\'\i}n Abadi}, \bibinfo{person}{Michael Vrable},
  \bibinfo{person}{Mihai Budiu}, {and} \bibinfo{person}{George~C. Necula}.}
  \bibinfo{year}{2006}\natexlab{}.
\newblock \showarticletitle{{XFI}: Software Guards for System Address Spaces}.
  In \bibinfo{booktitle}{\emph{Proceedings of the 7th USENIX Symposium on
  Operating Systems Design and Implementation}} \emph{(\bibinfo{series}{OSDI
  '06})}. \bibinfo{publisher}{USENIX Association}, \bibinfo{address}{Seattle,
  WA, USA}, \bibinfo{pages}{75--88}.
\newblock
\showISBNx{1-931971-47-1}
\urldef\tempurl%
\url{https://www.usenix.org/conference/osdi-06/xfi-software-guards-system-address-spaces}
\showURL{%
\tempurl}


\bibitem[Evans et~al\mbox{.}(2015)]%
        {CtrlJujutsu:CCS15}
\bibfield{author}{\bibinfo{person}{Isaac Evans}, \bibinfo{person}{Fan Long},
  \bibinfo{person}{Ulziibayar Otgonbaatar}, \bibinfo{person}{Howard Shrobe},
  \bibinfo{person}{Martin Rinard}, \bibinfo{person}{Hamed Okhravi}, {and}
  \bibinfo{person}{Stelios Sidiroglou-Douskos}.}
  \bibinfo{year}{2015}\natexlab{}.
\newblock \showarticletitle{{Control Jujutsu}: On the Weaknesses of
  Fine-Grained Control Flow Integrity}. In
  \bibinfo{booktitle}{\emph{Proceedings of the 22nd ACM SIGSAC Conference on
  Computer and Communications Security}} \emph{(\bibinfo{series}{CCS '15})}.
  \bibinfo{publisher}{ACM}, \bibinfo{address}{Denver, CO, USA},
  \bibinfo{pages}{901--913}.
\newblock
\showISBNx{978-1-4503-3832-5}
\urldef\tempurl%
\url{https://doi.org/10.1145/2810103.2813646}
\showDOI{\tempurl}


\bibitem[Felker et~al\mbox{.}(2021)]%
        {musl}
\bibfield{author}{\bibinfo{person}{Rich Felker} {et~al\mbox{.}}}
  \bibinfo{year}{2021}\natexlab{}.
\newblock \bibinfo{booktitle}{\emph{musl libc}}.
\newblock
\urldef\tempurl%
\url{https://musl.libc.org}
\showURL{%
\tempurl}


\bibitem[Frantzen and Shuey(2001)]%
        {StackGhost:UsenixSec01}
\bibfield{author}{\bibinfo{person}{Mike Frantzen} {and} \bibinfo{person}{Mike
  Shuey}.} \bibinfo{year}{2001}\natexlab{}.
\newblock \showarticletitle{{StackGhost}: Hardware Facilitated Stack
  Protection}. In \bibinfo{booktitle}{\emph{Proceedings of the 10th USENIX
  Security Symposium}} \emph{(\bibinfo{series}{Security '01})}.
  \bibinfo{publisher}{USENIX Association}, \bibinfo{address}{Washington, DC,
  USA}, \bibinfo{numpages}{11}~pages.
\newblock
\urldef\tempurl%
\url{https://www.usenix.org/conference/10th-usenix-security-symposium/stackghost-hardware-facilitated-stack-protection}
\showURL{%
\tempurl}


\bibitem[Gawlik et~al\mbox{.}(2016)]%
        {CROP:NDSS16}
\bibfield{author}{\bibinfo{person}{Robert Gawlik}, \bibinfo{person}{Benjamin
  Kollenda}, \bibinfo{person}{Philipp Koppe}, \bibinfo{person}{Behrad Garmany},
  {and} \bibinfo{person}{Thorsten Holz}.} \bibinfo{year}{2016}\natexlab{}.
\newblock \showarticletitle{Enabling Client-Side Crash-Resistance to Overcome
  Diversification and Information Hiding}. In
  \bibinfo{booktitle}{\emph{Proceedings of the 2016 Network and Distributed
  System Security Symposium}} \emph{(\bibinfo{series}{NDSS '16})}.
  \bibinfo{publisher}{Internet Society}, \bibinfo{address}{San Diego, CA, USA},
  \bibinfo{numpages}{15}~pages.
\newblock
\showISBNx{1-891562-41-X}
\urldef\tempurl%
\url{https://doi.org/10.14722/ndss.2016.23262}
\showDOI{\tempurl}


\bibitem[Ge et~al\mbox{.}(2017)]%
        {GRIFFIN:ASPLOS17}
\bibfield{author}{\bibinfo{person}{Xinyang Ge}, \bibinfo{person}{Weidong Cui},
  {and} \bibinfo{person}{Trent Jaeger}.} \bibinfo{year}{2017}\natexlab{}.
\newblock \showarticletitle{{\sc Griffin}: Guarding Control Flows Using {Intel}
  Processor Trace}. In \bibinfo{booktitle}{\emph{Proceedings of the 22nd
  International Conference on Architectural Support for Programming Languages
  and Operating Systems}} \emph{(\bibinfo{series}{ASPLOS '17})}.
  \bibinfo{publisher}{ACM}, \bibinfo{address}{Xi'an, China},
  \bibinfo{pages}{585--598}.
\newblock
\showISBNx{978-1-4503-4465-4}
\urldef\tempurl%
\url{https://doi.org/10.1145/3037697.3037716}
\showDOI{\tempurl}


\bibitem[Ge et~al\mbox{.}(2016)]%
        {KernelCFI:EuroSP16}
\bibfield{author}{\bibinfo{person}{Xinyang Ge}, \bibinfo{person}{Nirupama
  Talele}, \bibinfo{person}{Mathias Payer}, {and} \bibinfo{person}{Trent
  Jaeger}.} \bibinfo{year}{2016}\natexlab{}.
\newblock \showarticletitle{Fine-Grained Control-Flow Integrity for Kernel
  Software}. In \bibinfo{booktitle}{\emph{Proceedings of the 2016 IEEE European
  Symposium on Security and Privacy}} \emph{(\bibinfo{series}{EuroSP '16})}.
  \bibinfo{publisher}{IEEE Computer Society}, \bibinfo{address}{Saarbruecken,
  Germany}, \bibinfo{pages}{179--194}.
\newblock
\showISBNx{978-1-5090-1752-2}
\urldef\tempurl%
\url{https://doi.org/10.1109/EuroSP.2016.24}
\showDOI{\tempurl}


\bibitem[G\"{o}kta\c{s} et~al\mbox{.}(2014)]%
        {OutOfCtrl:Oakland14}
\bibfield{author}{\bibinfo{person}{Enes G\"{o}kta\c{s}}, \bibinfo{person}{Elias
  Athanasopoulos}, \bibinfo{person}{Herbert Bos}, {and}
  \bibinfo{person}{Georgios Portokalidis}.} \bibinfo{year}{2014}\natexlab{}.
\newblock \showarticletitle{Out of Control: Overcoming Control-Flow Integrity}.
  In \bibinfo{booktitle}{\emph{Proceedings of the 2014 IEEE Symposium on
  Security and Privacy}} \emph{(\bibinfo{series}{SP '14})}.
  \bibinfo{publisher}{IEEE Computer Society}, \bibinfo{address}{San Jose, CA,
  USA}, \bibinfo{pages}{575--589}.
\newblock
\showISBNx{978-1-4799-4686-0}
\urldef\tempurl%
\url{https://doi.org/10.1109/SP.2014.43}
\showDOI{\tempurl}


\bibitem[G\"{o}kta\c{s} et~al\mbox{.}(2016)]%
        {APM:UsenixSec16}
\bibfield{author}{\bibinfo{person}{Enes G\"{o}kta\c{s}},
  \bibinfo{person}{Robert Gawlik}, \bibinfo{person}{Benjamin Kollenda},
  \bibinfo{person}{Elias Athanasopoulos}, \bibinfo{person}{Georgios
  Portokalidis}, \bibinfo{person}{Cristiano Giuffrida}, {and}
  \bibinfo{person}{Herbert Bos}.} \bibinfo{year}{2016}\natexlab{}.
\newblock \showarticletitle{Undermining Information Hiding (and What to Do
  about It)}. In \bibinfo{booktitle}{\emph{Proceedings of the 25th USENIX
  Security Symposium}} \emph{(\bibinfo{series}{Security '16})}.
  \bibinfo{publisher}{USENIX Association}, \bibinfo{address}{Austin, TX, USA},
  \bibinfo{pages}{105--119}.
\newblock
\showISBNx{978-1-931971-32-4}
\urldef\tempurl%
\url{https://www.usenix.org/conference/usenixsecurity16/technical-sessions/presentation/goktas}
\showURL{%
\tempurl}


\bibitem[Google Cloud(2023)]%
        {Arm:GoogleCloud}
Google Cloud \bibinfo{year}{2023}\natexlab{}.
\newblock \bibinfo{booktitle}{\emph{Arm VMs on Compute}}.
\newblock
\urldef\tempurl%
\url{https://cloud.google.com/compute/docs/instances/arm-on-compute}
\showURL{%
\tempurl}


\bibitem[Gravani et~al\mbox{.}(2021)]%
        {IskiOS:RAID21}
\bibfield{author}{\bibinfo{person}{Spyridoula Gravani},
  \bibinfo{person}{Mohammad Hedayati}, \bibinfo{person}{John Criswell}, {and}
  \bibinfo{person}{Michael~L. Scott}.} \bibinfo{year}{2021}\natexlab{}.
\newblock \showarticletitle{Fast Intra-Kernel Isolation and Security with
  {IskiOS}}. In \bibinfo{booktitle}{\emph{Proceedings of the 24th International
  Symposium on Research in Attacks, Intrusions and Defenses}}
  \emph{(\bibinfo{series}{RAID '21})}. \bibinfo{publisher}{ACM},
  \bibinfo{address}{San Sebastian, Spain}, \bibinfo{pages}{119--134}.
\newblock
\showISBNx{978-1-4503-9058-3}
\urldef\tempurl%
\url{https://doi.org/10.1145/3471621.3471849}
\showDOI{\tempurl}


\bibitem[Gu et~al\mbox{.}(2022)]%
        {EPK:ATC22}
\bibfield{author}{\bibinfo{person}{Jinyu Gu}, \bibinfo{person}{Hao Li},
  \bibinfo{person}{Wentai Li}, \bibinfo{person}{Yubin Xia}, {and}
  \bibinfo{person}{Haibo Chen}.} \bibinfo{year}{2022}\natexlab{}.
\newblock \showarticletitle{{EPK}: Scalable and Efficient Memory Protection
  Keys}. In \bibinfo{booktitle}{\emph{Proceedings of the 2022 USENIX Annual
  Technical Conference}} \emph{(\bibinfo{series}{ATC '22})}.
  \bibinfo{publisher}{USENIX Association}, \bibinfo{address}{Carlsbad, CA,
  USA}, \bibinfo{pages}{609--624}.
\newblock
\showISBNx{978-1-939133-29-8}
\urldef\tempurl%
\url{https://www.usenix.org/conference/atc22/presentation/gu-jinyu}
\showURL{%
\tempurl}


\bibitem[Gu et~al\mbox{.}(2020)]%
        {UnderBridge:ATC20}
\bibfield{author}{\bibinfo{person}{Jinyu Gu}, \bibinfo{person}{Xinyue Wu},
  \bibinfo{person}{Wentai Li}, \bibinfo{person}{Nian Liu},
  \bibinfo{person}{Zeyu Mi}, \bibinfo{person}{Yubin Xia}, {and}
  \bibinfo{person}{Haibo Chen}.} \bibinfo{year}{2020}\natexlab{}.
\newblock \showarticletitle{Harmonizing Performance and Isolation in
  Microkernels with Efficient Intra-Kernel Isolation and Communication}. In
  \bibinfo{booktitle}{\emph{Proceedings of the 2020 USENIX Annual Technical
  Conference}} \emph{(\bibinfo{series}{ATC '20})}. \bibinfo{publisher}{USENIX
  Association}, \bibinfo{address}{Virtual Event}, \bibinfo{pages}{401--417}.
\newblock
\showISBNx{978-1-939133-14-4}
\urldef\tempurl%
\url{https://www.usenix.org/conference/atc20/presentation/gu}
\showURL{%
\tempurl}


\bibitem[Gu et~al\mbox{.}(2017)]%
        {PT-CFI:CODASPY17}
\bibfield{author}{\bibinfo{person}{Yufei Gu}, \bibinfo{person}{Qingchuan Zhao},
  \bibinfo{person}{Yinqian Zhang}, {and} \bibinfo{person}{Zhiqiang Lin}.}
  \bibinfo{year}{2017}\natexlab{}.
\newblock \showarticletitle{{PT-CFI}: Transparent Backward-Edge Control Flow
  Violation Detection Using {Intel} Processor Trace}. In
  \bibinfo{booktitle}{\emph{Proceedings of the 7th ACM Conference on Data and
  Application Security and Privacy}} \emph{(\bibinfo{series}{CODASPY '17})}.
  \bibinfo{publisher}{ACM}, \bibinfo{address}{Scottsdale, AZ, USA},
  \bibinfo{pages}{173--184}.
\newblock
\showISBNx{978-1-4503-4523-1}
\urldef\tempurl%
\url{https://doi.org/10.1145/3029806.3029830}
\showDOI{\tempurl}


\bibitem[Hedayati et~al\mbox{.}(2019)]%
        {Hodor:ATC19}
\bibfield{author}{\bibinfo{person}{Mohammad Hedayati},
  \bibinfo{person}{Spyridoula Gravani}, \bibinfo{person}{Ethan Johnson},
  \bibinfo{person}{John Criswell}, \bibinfo{person}{Michael~L. Scott},
  \bibinfo{person}{Kai Shen}, {and} \bibinfo{person}{Mike Marty}.}
  \bibinfo{year}{2019}\natexlab{}.
\newblock \showarticletitle{Hodor: Intra-Process Isolation for High-Throughput
  Data Plane Libraries}. In \bibinfo{booktitle}{\emph{Proceedings of the 2019
  USENIX Annual Technical Conference}} \emph{(\bibinfo{series}{ATC '19})}.
  \bibinfo{publisher}{USENIX Association}, \bibinfo{address}{Renton, WA, USA},
  \bibinfo{pages}{489--503}.
\newblock
\showISBNx{978-1-939133-03-8}
\urldef\tempurl%
\url{https://www.usenix.org/conference/atc19/presentation/hedayati-hodor}
\showURL{%
\tempurl}


\bibitem[Hu et~al\mbox{.}(2018)]%
        {uCFI:CCS18}
\bibfield{author}{\bibinfo{person}{Hong Hu}, \bibinfo{person}{Chenxiong Qian},
  \bibinfo{person}{Carter Yagemann}, \bibinfo{person}{Simon Pak~Ho Chung},
  \bibinfo{person}{William~R. Harris}, \bibinfo{person}{Taesoo Kim}, {and}
  \bibinfo{person}{Wenke Lee}.} \bibinfo{year}{2018}\natexlab{}.
\newblock \showarticletitle{Enforcing Unique Code Target Property for
  Control-Flow Integrity}. In \bibinfo{booktitle}{\emph{Proceedings of the 2018
  ACM SIGSAC Conference on Computer and Communications Security}}
  \emph{(\bibinfo{series}{CCS '18})}. \bibinfo{publisher}{ACM},
  \bibinfo{address}{Toronto, ON, Canada}, \bibinfo{pages}{1470--1486}.
\newblock
\showISBNx{978-1-4503-5693-0}
\urldef\tempurl%
\url{https://doi.org/10.1145/3243734.3243797}
\showDOI{\tempurl}


\bibitem[Hu et~al\mbox{.}(2016)]%
        {DOP:Oakland16}
\bibfield{author}{\bibinfo{person}{Hong Hu}, \bibinfo{person}{Shweta Shinde},
  \bibinfo{person}{Sendroiu Adrian}, \bibinfo{person}{Zheng~Leong Chua},
  \bibinfo{person}{Prateek Saxena}, {and} \bibinfo{person}{Zhenkai Liang}.}
  \bibinfo{year}{2016}\natexlab{}.
\newblock \showarticletitle{Data-Oriented Programming: On the Expressiveness of
  Non-Control Data Attacks}. In \bibinfo{booktitle}{\emph{Proceedings of the
  2016 IEEE Symposium on Security and Privacy}} \emph{(\bibinfo{series}{SP
  '16})}. \bibinfo{publisher}{IEEE Computer Society}, \bibinfo{address}{San
  Jose, CA, USA}, \bibinfo{pages}{969--986}.
\newblock
\showISBNx{978-1-5090-0824-7}
\urldef\tempurl%
\url{https://doi.org/10.1109/SP.2016.62}
\showDOI{\tempurl}


\bibitem[Huang et~al\mbox{.}(2016)]%
        {LMP:ACSAC16}
\bibfield{author}{\bibinfo{person}{Wei Huang}, \bibinfo{person}{Zhen Huang},
  \bibinfo{person}{Dhaval Miyani}, {and} \bibinfo{person}{David Lie}.}
  \bibinfo{year}{2016}\natexlab{}.
\newblock \showarticletitle{{LMP}: Light-Weighted Memory Protection with
  Hardware Assistance}. In \bibinfo{booktitle}{\emph{Proceedings of the 32nd
  Annual Conference on Computer Security Applications}}
  \emph{(\bibinfo{series}{ACSAC '16})}. \bibinfo{publisher}{ACM},
  \bibinfo{address}{Los Angeles, CA, USA}, \bibinfo{pages}{460--470}.
\newblock
\showISBNx{978-1-4503-4771-6}
\urldef\tempurl%
\url{https://doi.org/10.1145/2991079.2991089}
\showDOI{\tempurl}


\bibitem[Intel Corporation(2022)]%
        {X86:Intel:Manual}
Intel Corporation \bibinfo{year}{2022}\natexlab{}.
\newblock \bibinfo{booktitle}{\emph{Intel\textsuperscript{\textregistered} 64
  and {IA-32} Architectures Software Developer's Manual}}.
\newblock Intel Corporation.
\newblock
\urldef\tempurl%
\url{https://www.intel.com/content/www/us/en/developer/articles/technical/intel-sdm.html}
\showURL{%
\tempurl}
\newblock
\shownote{{Order} Number: 325462-078US}.


\bibitem[Ismail et~al\mbox{.}(2022)]%
        {PACtight:UsenixSec22}
\bibfield{author}{\bibinfo{person}{Mohannad Ismail}, \bibinfo{person}{Andrew
  Quach}, \bibinfo{person}{Christopher Jelesnianski}, \bibinfo{person}{Yeongjin
  Jang}, {and} \bibinfo{person}{Changwoo Min}.}
  \bibinfo{year}{2022}\natexlab{}.
\newblock \showarticletitle{Tightly Seal Your Sensitive Pointers with
  {PACTight}}. In \bibinfo{booktitle}{\emph{Proceedings of the 31st USENIX
  Security Symposium}} \emph{(\bibinfo{series}{Security '22})}.
  \bibinfo{publisher}{USENIX Association}, \bibinfo{address}{Boston, MA, USA},
  \bibinfo{pages}{3717--3734}.
\newblock
\showISBNx{978-1-939133-31-1}
\urldef\tempurl%
\url{https://www.usenix.org/conference/usenixsecurity22/presentation/ismail}
\showURL{%
\tempurl}


\bibitem[Ispoglou et~al\mbox{.}(2018)]%
        {BOP:CCS18}
\bibfield{author}{\bibinfo{person}{Kyriakos~K. Ispoglou},
  \bibinfo{person}{Bader AlBassam}, \bibinfo{person}{Trent Jaeger}, {and}
  \bibinfo{person}{Mathias Payer}.} \bibinfo{year}{2018}\natexlab{}.
\newblock \showarticletitle{Block Oriented Programming: Automating Data-Only
  Attacks}. In \bibinfo{booktitle}{\emph{Proceedings of the 2018 ACM SIGSAC
  Conference on Computer and Communications Security}}
  \emph{(\bibinfo{series}{CCS '18})}. \bibinfo{publisher}{ACM},
  \bibinfo{address}{Toronto, ON, Canada}, \bibinfo{pages}{1868--1882}.
\newblock
\showISBNx{978-1-4503-5693-0}
\urldef\tempurl%
\url{https://doi.org/10.1145/3243734.3243739}
\showDOI{\tempurl}


\bibitem[Jang et~al\mbox{.}(2014)]%
        {SafeDispatch:NDSS14}
\bibfield{author}{\bibinfo{person}{Dongseok Jang}, \bibinfo{person}{Zachary
  Tatlock}, {and} \bibinfo{person}{Sorin Lerner}.}
  \bibinfo{year}{2014}\natexlab{}.
\newblock \showarticletitle{{\sc SafeDispatch}: Securing {C++} Virtual Calls
  from Memory Corruption Attacks}. In \bibinfo{booktitle}{\emph{Proceedings of
  the 2014 Network and Distributed System Security Symposium}}
  \emph{(\bibinfo{series}{NDSS '14})}. \bibinfo{publisher}{Internet Society},
  \bibinfo{address}{San Diego, CA, USA}, \bibinfo{numpages}{15}~pages.
\newblock
\showISBNx{1-891562-35-5}
\urldef\tempurl%
\url{https://doi.org/10.14722/ndss.2014.23287}
\showDOI{\tempurl}


\bibitem[Johnson et~al\mbox{.}(2022)]%
        {Ombro:ATC22}
\bibfield{author}{\bibinfo{person}{Ethan Johnson}, \bibinfo{person}{Colin
  Pronovost}, {and} \bibinfo{person}{John Criswell}.}
  \bibinfo{year}{2022}\natexlab{}.
\newblock \showarticletitle{Hardening Hypervisors with {Ombro}}. In
  \bibinfo{booktitle}{\emph{Proceedings of the 2022 USENIX Annual Technical
  Conference}} \emph{(\bibinfo{series}{ATC '22})}. \bibinfo{publisher}{USENIX
  Association}, \bibinfo{address}{Carlsbad, CA, USA},
  \bibinfo{pages}{415--435}.
\newblock
\showISBNx{978-1-939133-29-8}
\urldef\tempurl%
\url{https://www.usenix.org/conference/atc22/presentation/johnson}
\showURL{%
\tempurl}


\bibitem[Kawada et~al\mbox{.}(2021)]%
        {TZmCFI:IJPP21}
\bibfield{author}{\bibinfo{person}{Tomoaki Kawada}, \bibinfo{person}{Shinya
  Honda}, \bibinfo{person}{Yutaka Matsubara}, {and} \bibinfo{person}{Hiroaki
  Takada}.} \bibinfo{year}{2021}\natexlab{}.
\newblock \showarticletitle{{TZmCFI}: {RTOS}-Aware Control-Flow Integrity Using
  {TrustZone} for {Armv8-M}}.
\newblock \bibinfo{journal}{\emph{International Journal of Parallel
  Programming}}  \bibinfo{volume}{49} (\bibinfo{date}{April}
  \bibinfo{year}{2021}), \bibinfo{pages}{216--236}.
\newblock
\showISSN{1573-7640}
\urldef\tempurl%
\url{https://doi.org/10.1007/s10766-020-00673-z}
\showDOI{\tempurl}


\bibitem[Khandaker et~al\mbox{.}(2019b)]%
        {CFI-LB:EuroSP19}
\bibfield{author}{\bibinfo{person}{Mustakimur Khandaker}, \bibinfo{person}{Abu
  Naser}, \bibinfo{person}{Wenqing Liu}, \bibinfo{person}{Zhi Wang},
  \bibinfo{person}{Yajin Zhou}, {and} \bibinfo{person}{Yueqiang Cheng}.}
  \bibinfo{year}{2019}\natexlab{b}.
\newblock \showarticletitle{Adaptive Call-Site Sensitive Control Flow
  Integrity}. In \bibinfo{booktitle}{\emph{Proceedings of the 2019 IEEE
  European Symposium on Security and Privacy}} \emph{(\bibinfo{series}{EuroSP
  '19})}. \bibinfo{publisher}{IEEE Computer Society},
  \bibinfo{address}{Stockholm, Sweden}, \bibinfo{pages}{95--110}.
\newblock
\showISBNx{978-1-7281-1148-3}
\urldef\tempurl%
\url{https://doi.org/10.1109/EuroSP.2019.00017}
\showDOI{\tempurl}


\bibitem[Khandaker et~al\mbox{.}(2019a)]%
        {OS-CFI:UsenixSec19}
\bibfield{author}{\bibinfo{person}{Mustakimur~Rahman Khandaker},
  \bibinfo{person}{Wenqing Liu}, \bibinfo{person}{Abu Naser},
  \bibinfo{person}{Zhi Wang}, {and} \bibinfo{person}{Jie Yang}.}
  \bibinfo{year}{2019}\natexlab{a}.
\newblock \showarticletitle{Origin-Sensitive Control Flow Integrity}. In
  \bibinfo{booktitle}{\emph{Proceedings of the 28th USENIX Security Symposium}}
  \emph{(\bibinfo{series}{Security '19})}. \bibinfo{publisher}{USENIX
  Association}, \bibinfo{address}{Santa Clara, CA, USA},
  \bibinfo{pages}{195--211}.
\newblock
\showISBNx{978-1-939133-06-9}
\urldef\tempurl%
\url{https://www.usenix.org/conference/usenixsecurity19/presentation/khandaker}
\showURL{%
\tempurl}


\bibitem[Koning et~al\mbox{.}(2017)]%
        {MemSentry:EuroSys17}
\bibfield{author}{\bibinfo{person}{Koen Koning}, \bibinfo{person}{Xi Chen},
  \bibinfo{person}{Herbert Bos}, \bibinfo{person}{Cristiano Giuffrida}, {and}
  \bibinfo{person}{Elias Athanasopoulos}.} \bibinfo{year}{2017}\natexlab{}.
\newblock \showarticletitle{No Need to Hide: Protecting Safe Regions on
  Commodity Hardware}. In \bibinfo{booktitle}{\emph{Proceedings of the 12th
  European Conference on Computer Systems}} \emph{(\bibinfo{series}{EuroSys
  '17})}. \bibinfo{publisher}{ACM}, \bibinfo{address}{Belgrade, Serbia},
  \bibinfo{pages}{437--452}.
\newblock
\showISBNx{978-1-4503-4938-3}
\urldef\tempurl%
\url{https://doi.org/10.1145/3064176.3064217}
\showDOI{\tempurl}


\bibitem[Kuznetsov and Morrison(2022)]%
        {Privbox:ATC22}
\bibfield{author}{\bibinfo{person}{Dmitry Kuznetsov} {and}
  \bibinfo{person}{Adam Morrison}.} \bibinfo{year}{2022}\natexlab{}.
\newblock \showarticletitle{Privbox: Faster System Calls Through Sandboxed
  Privileged Execution}. In \bibinfo{booktitle}{\emph{Proceedings of the 2022
  USENIX Annual Technical Conference}} \emph{(\bibinfo{series}{ATC '22})}.
  \bibinfo{publisher}{USENIX Association}, \bibinfo{address}{Carlsbad, CA,
  USA}, \bibinfo{pages}{233--247}.
\newblock
\showISBNx{978-1-939133-29-8}
\urldef\tempurl%
\url{https://www.usenix.org/conference/atc22/presentation/kuznetsov}
\showURL{%
\tempurl}


\bibitem[Kuznetsov et~al\mbox{.}(2014)]%
        {CPI:OSDI14}
\bibfield{author}{\bibinfo{person}{Volodymyr Kuznetsov},
  \bibinfo{person}{L\'{a}szl\'{o} Szekeres}, \bibinfo{person}{Mathias Payer},
  \bibinfo{person}{George Candea}, \bibinfo{person}{R. Sekar}, {and}
  \bibinfo{person}{Dawn Song}.} \bibinfo{year}{2014}\natexlab{}.
\newblock \showarticletitle{Code-Pointer Integrity}. In
  \bibinfo{booktitle}{\emph{Proceedings of the 11th USENIX Symposium on
  Operating Systems Design and Implementation}} \emph{(\bibinfo{series}{OSDI
  '14})}. \bibinfo{publisher}{USENIX Association},
  \bibinfo{address}{Broomfield, CO, USA}, \bibinfo{pages}{147--163}.
\newblock
\showISBNx{978-1-931971-16-4}
\urldef\tempurl%
\url{https://www.usenix.org/conference/osdi14/technical-sessions/presentation/kuznetsov}
\showURL{%
\tempurl}


\bibitem[Kwon et~al\mbox{.}(2019)]%
        {uXOM:UsenixSec19}
\bibfield{author}{\bibinfo{person}{Donghyun Kwon}, \bibinfo{person}{Jangseop
  Shin}, \bibinfo{person}{Giyeol Kim}, \bibinfo{person}{Byoungyoung Lee},
  \bibinfo{person}{Yeongpil Cho}, {and} \bibinfo{person}{Yunheung Paek}.}
  \bibinfo{year}{2019}\natexlab{}.
\newblock \showarticletitle{{uXOM}: Efficient {eXecute}-Only Memory on {ARM}
  {Cortex-M}}. In \bibinfo{booktitle}{\emph{Proceedings of the 28th USENIX
  Security Symposium}} \emph{(\bibinfo{series}{Security '19})}.
  \bibinfo{publisher}{USENIX Association}, \bibinfo{address}{Santa Clara, CA,
  USA}, \bibinfo{pages}{231--247}.
\newblock
\showISBNx{978-1-939133-06-9}
\urldef\tempurl%
\url{https://www.usenix.org/conference/usenixsecurity19/presentation/kwon}
\showURL{%
\tempurl}


\bibitem[Lattner and Adve(2004)]%
        {LLVM:CGO04}
\bibfield{author}{\bibinfo{person}{Chris Lattner} {and} \bibinfo{person}{Vikram
  Adve}.} \bibinfo{year}{2004}\natexlab{}.
\newblock \showarticletitle{{LLVM}: A Compilation Framework for Lifelong
  Program Analysis \& Transformation}. In \bibinfo{booktitle}{\emph{Proceedings
  of the 2nd International Symposium on Code Generation and Optimization}}
  \emph{(\bibinfo{series}{CGO '04})}. \bibinfo{publisher}{IEEE Computer
  Society}, \bibinfo{address}{Palo Alto, CA, USA},
  \bibinfo{numpages}{12}~pages.
\newblock
\showISBNx{0-7695-2102-9}
\urldef\tempurl%
\url{https://doi.org/10.1109/CGO.2004.1281665}
\showDOI{\tempurl}


\bibitem[Li et~al\mbox{.}(2020)]%
        {ZipperStack:ESORICS20}
\bibfield{author}{\bibinfo{person}{Jinfeng Li}, \bibinfo{person}{Liwei Chen},
  \bibinfo{person}{Qizhen Xu}, \bibinfo{person}{Linan Tian},
  \bibinfo{person}{Gang Shi}, \bibinfo{person}{Kai Chen}, {and}
  \bibinfo{person}{Dan Meng}.} \bibinfo{year}{2020}\natexlab{}.
\newblock \showarticletitle{Zipper Stack: Shadow Stacks Without Shadow}. In
  \bibinfo{booktitle}{\emph{Proceedings of the 25th European Symposium on
  Research in Computer Security}} \emph{(\bibinfo{series}{ESORICS '20})}.
  \bibinfo{publisher}{Springer-Verlag}, \bibinfo{address}{Guildford, UK},
  \bibinfo{pages}{338--358}.
\newblock
\showISBNx{978-3-030-58951-6}
\urldef\tempurl%
\url{https://doi.org/10.1007/978-3-030-58951-6_17}
\showDOI{\tempurl}


\bibitem[Liljestrand et~al\mbox{.}(2021)]%
        {PACStack:UsenixSec21}
\bibfield{author}{\bibinfo{person}{Hans Liljestrand}, \bibinfo{person}{Thomas
  Nyman}, \bibinfo{person}{Lachlan~J. Gunn}, \bibinfo{person}{Jan-Erik Ekberg},
  {and} \bibinfo{person}{N. Asokan}.} \bibinfo{year}{2021}\natexlab{}.
\newblock \showarticletitle{{PACStack}: an Authenticated Call Stack}. In
  \bibinfo{booktitle}{\emph{Proceedings of the 30th USENIX Security Symposium}}
  \emph{(\bibinfo{series}{Security '21})}. \bibinfo{publisher}{USENIX
  Association}, \bibinfo{address}{Virtual Event}, \bibinfo{pages}{357--374}.
\newblock
\showISBNx{978-1-939133-24-3}
\urldef\tempurl%
\url{https://www.usenix.org/conference/usenixsecurity21/presentation/liljestrand}
\showURL{%
\tempurl}


\bibitem[Liljestrand et~al\mbox{.}(2019)]%
        {PARTS:UsenixSec19}
\bibfield{author}{\bibinfo{person}{Hans Liljestrand}, \bibinfo{person}{Thomas
  Nyman}, \bibinfo{person}{Kui Wang}, \bibinfo{person}{Carlos~Chinea Perez},
  \bibinfo{person}{Jan-Erik Ekberg}, {and} \bibinfo{person}{N. Asokan}.}
  \bibinfo{year}{2019}\natexlab{}.
\newblock \showarticletitle{{PAC} it up: Towards Pointer Integrity using {ARM}
  Pointer Authentication}. In \bibinfo{booktitle}{\emph{Proceedings of the 28th
  USENIX Security Symposium}} \emph{(\bibinfo{series}{Security '19})}.
  \bibinfo{publisher}{USENIX Association}, \bibinfo{address}{Santa Clara, CA,
  USA}, \bibinfo{pages}{177--194}.
\newblock
\showISBNx{978-1-939133-06-9}
\urldef\tempurl%
\url{https://www.usenix.org/conference/usenixsecurity19/presentation/liljestrand}
\showURL{%
\tempurl}


\bibitem[Linux(2020)]%
        {linux-5-6-src}
Linux \bibinfo{year}{2020}\natexlab{}.
\newblock \bibinfo{booktitle}{\emph{Linux Kernel Source Tree v5.6}}.
\newblock
\urldef\tempurl%
\url{https://git.kernel.org/pub/scm/linux/kernel/git/torvalds/linux.git/tree/?h=v5.6}
\showURL{%
\tempurl}


\bibitem[Linux(2021)]%
        {linux-4-19-219-src}
Linux \bibinfo{year}{2021}\natexlab{}.
\newblock \bibinfo{booktitle}{\emph{Linux Kernel Stable Tree v4.19.219}}.
\newblock
\urldef\tempurl%
\url{https://git.kernel.org/pub/scm/linux/kernel/git/stable/linux.git/tree/?h=v4.19.219}
\showURL{%
\tempurl}


\bibitem[Lipp et~al\mbox{.}(2018)]%
        {Meltdown:UsenixSec18}
\bibfield{author}{\bibinfo{person}{Moritz Lipp}, \bibinfo{person}{Michael
  Schwarz}, \bibinfo{person}{Daniel Gruss}, \bibinfo{person}{Thomas Prescher},
  \bibinfo{person}{Werner Haas}, \bibinfo{person}{Anders Fogh},
  \bibinfo{person}{Jann Horn}, \bibinfo{person}{Stefan Mangard},
  \bibinfo{person}{Paul Kocher}, \bibinfo{person}{Daniel Genkin},
  \bibinfo{person}{Yuval Yarom}, {and} \bibinfo{person}{Mike Hamburg}.}
  \bibinfo{year}{2018}\natexlab{}.
\newblock \showarticletitle{Meltdown: Reading Kernel Memory from User Space}.
  In \bibinfo{booktitle}{\emph{Proceedings of the 27th USENIX Security
  Symposium}} \emph{(\bibinfo{series}{Security '18})}.
  \bibinfo{publisher}{USENIX Association}, \bibinfo{address}{Baltimore, MD,
  USA}, \bibinfo{pages}{973--990}.
\newblock
\showISBNx{978-1-939133-04-5}
\urldef\tempurl%
\url{https://www.usenix.org/conference/usenixsecurity18/presentation/lipp}
\showURL{%
\tempurl}


\bibitem[Liu et~al\mbox{.}(2017)]%
        {FlowGuard:HPCA17}
\bibfield{author}{\bibinfo{person}{Yutao Liu}, \bibinfo{person}{Peitao Shi},
  \bibinfo{person}{Xinran Wang}, \bibinfo{person}{Haibo Chen},
  \bibinfo{person}{Binyu Zang}, {and} \bibinfo{person}{Haibing Guan}.}
  \bibinfo{year}{2017}\natexlab{}.
\newblock \showarticletitle{Transparent and Efficient {CFI} Enforcement with
  {Intel} Processor Trace}. In \bibinfo{booktitle}{\emph{Proceedings of the
  2017 IEEE International Symposium on High Performance Computer Architecture}}
  \emph{(\bibinfo{series}{HPCA '17})}. \bibinfo{publisher}{IEEE Computer
  Society}, \bibinfo{address}{Austin, TX, USA}, \bibinfo{pages}{529--540}.
\newblock
\showISBNx{978-1-5090-4985-1}
\urldef\tempurl%
\url{https://doi.org/10.1109/HPCA.2017.18}
\showDOI{\tempurl}


\bibitem[Liu et~al\mbox{.}(2015)]%
        {SeCage:CCS15}
\bibfield{author}{\bibinfo{person}{Yutao Liu}, \bibinfo{person}{Tianyu Zhou},
  \bibinfo{person}{Kexin Chen}, \bibinfo{person}{Haibo Chen}, {and}
  \bibinfo{person}{Yubin Xia}.} \bibinfo{year}{2015}\natexlab{}.
\newblock \showarticletitle{Thwarting Memory Disclosure with Efficient
  Hypervisor-Enforced Intra-Domain Isolation}. In
  \bibinfo{booktitle}{\emph{Proceedings of the 22nd ACM SIGSAC Conference on
  Computer and Communications Security}} \emph{(\bibinfo{series}{CCS '15})}.
  \bibinfo{publisher}{ACM}, \bibinfo{address}{Denver, CO, USA},
  \bibinfo{pages}{1607--1619}.
\newblock
\showISBNx{978-1-4503-3832-5}
\urldef\tempurl%
\url{https://doi.org/10.1145/2810103.2813690}
\showDOI{\tempurl}


\bibitem[LLVM(2021a)]%
        {libc++:LLVM}
LLVM \bibinfo{year}{2021}\natexlab{a}.
\newblock \bibinfo{booktitle}{\emph{``libc++'' C++ Standard Library}}.
\newblock
\urldef\tempurl%
\url{https://libcxx.llvm.org}
\showURL{%
\tempurl}


\bibitem[LLVM(2021b)]%
        {libc++abi:LLVM}
LLVM \bibinfo{year}{2021}\natexlab{b}.
\newblock \bibinfo{booktitle}{\emph{``libc++abi'' C++ Standard Library
  Support}}.
\newblock
\urldef\tempurl%
\url{https://libcxxabi.llvm.org}
\showURL{%
\tempurl}


\bibitem[LLVM(2021c)]%
        {libunwind:LLVM}
LLVM \bibinfo{year}{2021}\natexlab{c}.
\newblock \bibinfo{booktitle}{\emph{libunwind LLVM Unwinder}}.
\newblock
\urldef\tempurl%
\url{https://bcain-llvm.readthedocs.io/projects/libunwind}
\showURL{%
\tempurl}


\bibitem[LLVM(2022)]%
        {compiler-rt:LLVM}
LLVM \bibinfo{year}{2022}\natexlab{}.
\newblock \bibinfo{booktitle}{\emph{``compiler-rt'' runtime libraries}}.
\newblock
\urldef\tempurl%
\url{https://compiler-rt.llvm.org}
\showURL{%
\tempurl}


\bibitem[LLVM(2023a)]%
        {IndirectBrExpand:LLVM}
LLVM \bibinfo{year}{2023}\natexlab{a}.
\newblock \bibinfo{booktitle}{\emph{{lib/CodeGen/IndirectBrExpandPass.cpp} File
  Reference}}.
\newblock
\urldef\tempurl%
\url{https://llvm.org/doxygen/IndirectBrExpandPass_8cpp.html}
\showURL{%
\tempurl}


\bibitem[LLVM(2023b)]%
        {LLD:LLVM}
LLVM \bibinfo{year}{2023}\natexlab{b}.
\newblock \bibinfo{booktitle}{\emph{LLD - The LLVM Linker}}.
\newblock
\urldef\tempurl%
\url{https://lld.llvm.org}
\showURL{%
\tempurl}


\bibitem[Mashtizadeh et~al\mbox{.}(2015)]%
        {CCFI:CCS15}
\bibfield{author}{\bibinfo{person}{Ali~Jose Mashtizadeh},
  \bibinfo{person}{Andrea Bittau}, \bibinfo{person}{Dan Boneh}, {and}
  \bibinfo{person}{David Mazi\`{e}res}.} \bibinfo{year}{2015}\natexlab{}.
\newblock \showarticletitle{{CCFI}: Cryptographically Enforced Control Flow
  Integrity}. In \bibinfo{booktitle}{\emph{Proceedings of the 22nd ACM SIGSAC
  Conference on Computer and Communications Security}}
  \emph{(\bibinfo{series}{CCS '15})}. \bibinfo{publisher}{ACM},
  \bibinfo{address}{Denver, CO, USA}, \bibinfo{pages}{941--951}.
\newblock
\showISBNx{978-1-4503-3832-5}
\urldef\tempurl%
\url{https://doi.org/10.1145/2810103.2813676}
\showDOI{\tempurl}


\bibitem[McCamant and Morrisett(2006)]%
        {PittSFIeld:UsenixSec06}
\bibfield{author}{\bibinfo{person}{Stephen McCamant} {and}
  \bibinfo{person}{Greg Morrisett}.} \bibinfo{year}{2006}\natexlab{}.
\newblock \showarticletitle{Evaluating {SFI} for a {CISC} Architecture}. In
  \bibinfo{booktitle}{\emph{Proceedings of the 15th USENIX Security Symposium}}
  \emph{(\bibinfo{series}{Security '06})}. \bibinfo{publisher}{USENIX
  Association}, \bibinfo{address}{Vancouver, BC. Canada},
  \bibinfo{pages}{209--224}.
\newblock
\urldef\tempurl%
\url{https://www.usenix.org/conference/15th-usenix-security-symposium/evaluating-sfi-cisc-architecture}
\showURL{%
\tempurl}


\bibitem[McVoy and Staelin(1996)]%
        {lmbench:ATC96}
\bibfield{author}{\bibinfo{person}{Larry McVoy} {and} \bibinfo{person}{Carl
  Staelin}.} \bibinfo{year}{1996}\natexlab{}.
\newblock \showarticletitle{{lmbench}: Portable Tools for Performance
  Analysis}. In \bibinfo{booktitle}{\emph{Proceedings of the 1996 USENIX Annual
  Technical Conference}} \emph{(\bibinfo{series}{ATC '96})}.
  \bibinfo{publisher}{USENIX Association}, \bibinfo{address}{San Diego, CA,
  USA}, \bibinfo{numpages}{16}~pages.
\newblock
\urldef\tempurl%
\url{https://www.usenix.org/legacy/publications/library/proceedings/sd96/mcvoy.html}
\showURL{%
\tempurl}


\bibitem[Mi et~al\mbox{.}(2019)]%
        {SkyBridge:EuroSys19}
\bibfield{author}{\bibinfo{person}{Zeyu Mi}, \bibinfo{person}{Dingji Li},
  \bibinfo{person}{Zihan Yang}, \bibinfo{person}{Xinran Wang}, {and}
  \bibinfo{person}{Haibo Chen}.} \bibinfo{year}{2019}\natexlab{}.
\newblock \showarticletitle{{SkyBridge}: Fast and Secure Inter-Process
  Communication for Microkernels}. In \bibinfo{booktitle}{\emph{Proceedings of
  the 14th European Conference on Computer Systems}}
  \emph{(\bibinfo{series}{EuroSys '19})}. \bibinfo{publisher}{ACM},
  \bibinfo{address}{Dresden, Germany}, Article \bibinfo{articleno}{9},
  \bibinfo{numpages}{15}~pages.
\newblock
\showISBNx{978-1-4503-6281-8}
\urldef\tempurl%
\url{https://doi.org/10.1145/3302424.3303946}
\showDOI{\tempurl}


\bibitem[Microsoft Azure(2022)]%
        {Arm:MicrosoftAzure}
Microsoft Azure \bibinfo{year}{2022}\natexlab{}.
\newblock \bibinfo{booktitle}{\emph{Azure Virtual Machines with Ampere Altra
  Arm–based processors---generally available}}.
\newblock
\urldef\tempurl%
\url{https://azure.microsoft.com/en-us/blog/azure-virtual-machines-with-ampere-altra-arm-based-processors-generally-available}
\showURL{%
\tempurl}


\bibitem[Mohan et~al\mbox{.}(2015)]%
        {O-CFI:NDSS15}
\bibfield{author}{\bibinfo{person}{Vishwath Mohan}, \bibinfo{person}{Per
  Larsen}, \bibinfo{person}{Stefan Brunthaler}, \bibinfo{person}{Kevin~W.
  Hamlen}, {and} \bibinfo{person}{Michael Franz}.}
  \bibinfo{year}{2015}\natexlab{}.
\newblock \showarticletitle{Opaque Control-Flow Integrity}. In
  \bibinfo{booktitle}{\emph{Proceedings of the 2015 Network and Distributed
  System Security Symposium}} \emph{(\bibinfo{series}{NDSS '15})}.
  \bibinfo{publisher}{Internet Society}, \bibinfo{address}{San Diego, CA, USA},
  \bibinfo{numpages}{15}~pages.
\newblock
\showISBNx{1-891562-38-X}
\urldef\tempurl%
\url{https://doi.org/10.14722/ndss.2015.23271}
\showDOI{\tempurl}


\bibitem[Nagarakatte et~al\mbox{.}(2009)]%
        {SoftBound:PLDI09}
\bibfield{author}{\bibinfo{person}{Santosh Nagarakatte},
  \bibinfo{person}{Jianzhou Zhao}, \bibinfo{person}{Milo~M.K. Martin}, {and}
  \bibinfo{person}{Steve Zdancewic}.} \bibinfo{year}{2009}\natexlab{}.
\newblock \showarticletitle{{SoftBound}: Highly Compatible and Complete Spatial
  Memory Safety for {C}}. In \bibinfo{booktitle}{\emph{Proceedings of the 2009
  ACM SIGPLAN Conference on Programming Language Design and Implementation}}
  \emph{(\bibinfo{series}{PLDI '09})}. \bibinfo{publisher}{ACM},
  \bibinfo{address}{Dublin, Ireland}, \bibinfo{pages}{245--258}.
\newblock
\showISBNx{978-1-60558-392-1}
\urldef\tempurl%
\url{https://doi.org/10.1145/1542476.1542504}
\showDOI{\tempurl}


\bibitem[Nagarakatte et~al\mbox{.}(2010)]%
        {CETS:ISMM10}
\bibfield{author}{\bibinfo{person}{Santosh Nagarakatte},
  \bibinfo{person}{Jianzhou Zhao}, \bibinfo{person}{Milo~M.K. Martin}, {and}
  \bibinfo{person}{Steve Zdancewic}.} \bibinfo{year}{2010}\natexlab{}.
\newblock \showarticletitle{{CETS}: Compiler Enforced Temporal Safety for {C}}.
  In \bibinfo{booktitle}{\emph{Proceedings of the 2010 International Symposium
  on Memory Management}} \emph{(\bibinfo{series}{ISMM '10})}.
  \bibinfo{publisher}{ACM}, \bibinfo{address}{Toronto, ON, Canada},
  \bibinfo{pages}{31--40}.
\newblock
\showISBNx{978-1-4503-0054-4}
\urldef\tempurl%
\url{https://doi.org/10.1145/1806651.1806657}
\showDOI{\tempurl}


\bibitem[Narayanan et~al\mbox{.}(2020)]%
        {LVDs:VEE20}
\bibfield{author}{\bibinfo{person}{Vikram Narayanan}, \bibinfo{person}{Yongzhe
  Huang}, \bibinfo{person}{Gang Tan}, \bibinfo{person}{Trent Jaeger}, {and}
  \bibinfo{person}{Anton Burtsev}.} \bibinfo{year}{2020}\natexlab{}.
\newblock \showarticletitle{Lightweight Kernel Isolation with Virtualization
  and {VM} Functions}. In \bibinfo{booktitle}{\emph{Proceedings of the 16th ACM
  SIGPLAN/SIGOPS International Conference on Virtual Execution Environments}}
  \emph{(\bibinfo{series}{VEE '20})}. \bibinfo{publisher}{ACM},
  \bibinfo{address}{Lausanne, Switzerland}, \bibinfo{pages}{157--171}.
\newblock
\showISBNx{978-1-4503-7554-2}
\urldef\tempurl%
\url{https://doi.org/10.1145/3381052.3381328}
\showDOI{\tempurl}


\bibitem[Niu and Tan(2013)]%
        {MIP:CCS13}
\bibfield{author}{\bibinfo{person}{Ben Niu} {and} \bibinfo{person}{Gang Tan}.}
  \bibinfo{year}{2013}\natexlab{}.
\newblock \showarticletitle{Monitor Integrity Protection with Space Efficiency
  and Separate Compilation}. In \bibinfo{booktitle}{\emph{Proceedings of the
  2013 ACM SIGSAC Conference on Computer and Communications Security}}
  \emph{(\bibinfo{series}{CCS '13})}. \bibinfo{publisher}{ACM},
  \bibinfo{address}{Berlin, Germany}, \bibinfo{pages}{199--210}.
\newblock
\showISBNx{978-1-4503-2477-9}
\urldef\tempurl%
\url{https://doi.org/10.1145/2508859.2516649}
\showDOI{\tempurl}


\bibitem[Niu and Tan(2014)]%
        {MCFI:PLDI14}
\bibfield{author}{\bibinfo{person}{Ben Niu} {and} \bibinfo{person}{Gang Tan}.}
  \bibinfo{year}{2014}\natexlab{}.
\newblock \showarticletitle{Modular Control-Flow Integrity}. In
  \bibinfo{booktitle}{\emph{Proceedings of the 35th ACM SIGPLAN Conference on
  Programming Language Design and Implementation}} \emph{(\bibinfo{series}{PLDI
  '14})}. \bibinfo{publisher}{ACM}, \bibinfo{address}{Edinburgh, UK},
  \bibinfo{pages}{577--587}.
\newblock
\showISBNx{978-1-4503-2784-8}
\urldef\tempurl%
\url{https://doi.org/10.1145/2594291.2594295}
\showDOI{\tempurl}


\bibitem[Niu and Tan(2015)]%
        {PICFI:CCS15}
\bibfield{author}{\bibinfo{person}{Ben Niu} {and} \bibinfo{person}{Gang Tan}.}
  \bibinfo{year}{2015}\natexlab{}.
\newblock \showarticletitle{Per-Input Control-Flow Integrity}. In
  \bibinfo{booktitle}{\emph{Proceedings of the 22nd ACM SIGSAC Conference on
  Computer and Communications Security}} \emph{(\bibinfo{series}{CCS '15})}.
  \bibinfo{publisher}{ACM}, \bibinfo{address}{Denver, CO, USA},
  \bibinfo{pages}{914--926}.
\newblock
\showISBNx{978-1-4503-3832-5}
\urldef\tempurl%
\url{https://doi.org/10.1145/2810103.2813644}
\showDOI{\tempurl}


\bibitem[Nyman et~al\mbox{.}(2017)]%
        {CaRE:RAID17}
\bibfield{author}{\bibinfo{person}{Thomas Nyman}, \bibinfo{person}{Jan-Erik
  Ekberg}, \bibinfo{person}{Lucas Davi}, {and} \bibinfo{person}{N. Asokan}.}
  \bibinfo{year}{2017}\natexlab{}.
\newblock \showarticletitle{{CFI CaRE}: Hardware-Supported Call and Return
  Enforcement for Commercial Microcontrollers}. In
  \bibinfo{booktitle}{\emph{Proceedings of the 20th International Symposium on
  Research in Attacks, Intrusions, and Defenses}} \emph{(\bibinfo{series}{RAID
  '17})}. \bibinfo{publisher}{Springer-Verlag}, \bibinfo{address}{Atlanta, GA,
  USA}, \bibinfo{pages}{259--284}.
\newblock
\showISBNx{978-3-319-66332-6}
\urldef\tempurl%
\url{https://doi.org/10.1007/978-3-319-66332-6_12}
\showDOI{\tempurl}


\bibitem[Oikonomopoulos et~al\mbox{.}(2016)]%
        {AllocOracle:UsenixSec16}
\bibfield{author}{\bibinfo{person}{Angelos Oikonomopoulos},
  \bibinfo{person}{Elias Athanasopoulos}, \bibinfo{person}{Herbert Bos}, {and}
  \bibinfo{person}{Cristiano Giuffrida}.} \bibinfo{year}{2016}\natexlab{}.
\newblock \showarticletitle{Poking Holes in Information Hiding}. In
  \bibinfo{booktitle}{\emph{Proceedings of the 25th USENIX Security Symposium}}
  \emph{(\bibinfo{series}{Security '16})}. \bibinfo{publisher}{USENIX
  Association}, \bibinfo{address}{Austin, TX, USA}, \bibinfo{pages}{121--138}.
\newblock
\showISBNx{978-1-931971-32-4}
\urldef\tempurl%
\url{https://www.usenix.org/conference/usenixsecurity16/technical-sessions/presentation/oikonomopoulos}
\showURL{%
\tempurl}


\bibitem[Oracle Cloud Infrastructure(2023)]%
        {AmpereA1:OCI}
Oracle Cloud Infrastructure \bibinfo{year}{2023}\natexlab{}.
\newblock \bibinfo{booktitle}{\emph{Ampere A1 Compute}}.
\newblock
\urldef\tempurl%
\url{https://www.oracle.com/cloud/compute/arm}
\showURL{%
\tempurl}


\bibitem[Pappas et~al\mbox{.}(2013)]%
        {kBouncer:UsenixSec13}
\bibfield{author}{\bibinfo{person}{Vasilis Pappas}, \bibinfo{person}{Michalis
  Polychronakis}, {and} \bibinfo{person}{Angelos~D. Keromytis}.}
  \bibinfo{year}{2013}\natexlab{}.
\newblock \showarticletitle{Transparent {ROP} Exploit Mitigation Using Indirect
  Branch Tracing}. In \bibinfo{booktitle}{\emph{Proceedings of the 22nd USENIX
  Security Symposium}} \emph{(\bibinfo{series}{Security '13})}.
  \bibinfo{publisher}{USENIX Association}, \bibinfo{address}{Washington, DC,
  USA}, \bibinfo{pages}{447--462}.
\newblock
\showISBNx{978-1-931971-03-4}
\urldef\tempurl%
\url{https://www.usenix.org/conference/usenixsecurity13/technical-sessions/paper/pappas}
\showURL{%
\tempurl}


\bibitem[Park et~al\mbox{.}(2019)]%
        {libmpk:ATC19}
\bibfield{author}{\bibinfo{person}{Soyeon Park}, \bibinfo{person}{Sangho Lee},
  \bibinfo{person}{Wen Xu}, \bibinfo{person}{HyunGon Moon}, {and}
  \bibinfo{person}{Taesoo Kim}.} \bibinfo{year}{2019}\natexlab{}.
\newblock \showarticletitle{{libmpk}: Software Abstraction for {Intel} Memory
  Protection Keys ({Intel MPK})}. In \bibinfo{booktitle}{\emph{Proceedings of
  the 2019 USENIX Annual Technical Conference}} \emph{(\bibinfo{series}{ATC
  '19})}. \bibinfo{publisher}{USENIX Association}, \bibinfo{address}{Renton,
  WA, USA}, \bibinfo{pages}{241--254}.
\newblock
\showISBNx{978-1-939133-03-8}
\urldef\tempurl%
\url{https://www.usenix.org/conference/atc19/presentation/park-soyeon}
\showURL{%
\tempurl}


\bibitem[PaX Team(2000)]%
        {NoExec:PaX00}
PaX Team \bibinfo{year}{2000}\natexlab{}.
\newblock \bibinfo{booktitle}{\emph{Non-Executable Pages Design \&
  Implementation}}.
\newblock
\urldef\tempurl%
\url{https://pax.grsecurity.net/docs/noexec.txt}
\showURL{%
\tempurl}


\bibitem[PaX Team(2001)]%
        {ASLR:PaX01}
PaX Team \bibinfo{year}{2001}\natexlab{}.
\newblock \bibinfo{booktitle}{\emph{Address Space Layout Randomization}}.
\newblock
\urldef\tempurl%
\url{https://pax.grsecurity.net/docs/aslr.txt}
\showURL{%
\tempurl}


\bibitem[Payer et~al\mbox{.}(2015)]%
        {Lockdown:DIMVA15}
\bibfield{author}{\bibinfo{person}{Mathias Payer}, \bibinfo{person}{Antonio
  Barresi}, {and} \bibinfo{person}{Thomas~R. Gross}.}
  \bibinfo{year}{2015}\natexlab{}.
\newblock \showarticletitle{Fine-Grained Control-Flow Integrity Through Binary
  Hardening}. In \bibinfo{booktitle}{\emph{Proceedings of the 12th
  International Conference on Detection of Intrusions and Malware, and
  Vulnerability Assessment}} \emph{(\bibinfo{series}{DIMVA '15})}.
  \bibinfo{publisher}{Springer-Verlag}, \bibinfo{address}{Milan, Italy},
  \bibinfo{pages}{144--164}.
\newblock
\showISBNx{978-3-319-20550-2}
\urldef\tempurl%
\url{https://doi.org/10.1007/978-3-319-20550-2_8}
\showDOI{\tempurl}


\bibitem[Pewny and Holz(2013)]%
        {CFR:ACSAC13}
\bibfield{author}{\bibinfo{person}{Jannik Pewny} {and}
  \bibinfo{person}{Thorsten Holz}.} \bibinfo{year}{2013}\natexlab{}.
\newblock \showarticletitle{Control-Flow Restrictor: Compiler-Based {CFI} for
  {iOS}}. In \bibinfo{booktitle}{\emph{Proceedings of the 29th Annual Computer
  Security Applications Conference}} \emph{(\bibinfo{series}{ACSAC '13})}.
  \bibinfo{publisher}{ACM}, \bibinfo{address}{New Orleans, LA, USA},
  \bibinfo{pages}{309--318}.
\newblock
\showISBNx{978-1-4503-2015-3}
\urldef\tempurl%
\url{https://doi.org/10.1145/2523649.2523674}
\showDOI{\tempurl}


\bibitem[Proskurin et~al\mbox{.}(2020)]%
        {xMP:Oakland20}
\bibfield{author}{\bibinfo{person}{Sergej Proskurin}, \bibinfo{person}{Marius
  Momeu}, \bibinfo{person}{Seyedhamed Ghavamnia}, \bibinfo{person}{Vasileios~P.
  Kemerlis}, {and} \bibinfo{person}{Michalis Polychronakis}.}
  \bibinfo{year}{2020}\natexlab{}.
\newblock \showarticletitle{{xMP}: Selective Memory Protection for Kernel and
  User Space}. In \bibinfo{booktitle}{\emph{Proceedings of the 2020 IEEE
  Symposium on Security and Privacy}} \emph{(\bibinfo{series}{SP '20})}.
  \bibinfo{publisher}{IEEE Computer Society}, \bibinfo{address}{San Francisco,
  CA, USA}, \bibinfo{pages}{563--577}.
\newblock
\showISBNx{978-1-7281-3497-0}
\urldef\tempurl%
\url{https://doi.org/10.1109/SP40000.2020.00041}
\showDOI{\tempurl}


\bibitem[Qualcomm(2017)]%
        {PAuth:Qualcomm-WP17}
Qualcomm \bibinfo{year}{2017}\natexlab{}.
\newblock \bibinfo{booktitle}{\emph{Pointer Authentication on {ARMv8.3}: Design
  and Analysis of the New Software Security Instructions}}.
\newblock \bibinfo{type}{White Paper}. \bibinfo{institution}{Qualcomm
  Technologies, Inc.}
\newblock
\urldef\tempurl%
\url{https://www.qualcomm.com/content/dam/qcomm-martech/dm-assets/documents/pointer-auth-v7.pdf}
\showURL{%
\tempurl}


\bibitem[Roemer et~al\mbox{.}(2012)]%
        {ROP:TISSEC12}
\bibfield{author}{\bibinfo{person}{Ryan Roemer}, \bibinfo{person}{Erik
  Buchanan}, \bibinfo{person}{Hovav Shacham}, {and} \bibinfo{person}{Stefan
  Savage}.} \bibinfo{year}{2012}\natexlab{}.
\newblock \showarticletitle{Return-Oriented Programming: Systems, Languages,
  and Applications}.
\newblock \bibinfo{journal}{\emph{ACM Transactions on Information and System
  Security}} \bibinfo{volume}{15}, \bibinfo{number}{1}, Article
  \bibinfo{articleno}{2} (\bibinfo{date}{March} \bibinfo{year}{2012}),
  \bibinfo{numpages}{34}~pages.
\newblock
\showISSN{1094-9224}
\urldef\tempurl%
\url{https://doi.org/10.1145/2133375.2133377}
\showDOI{\tempurl}


\bibitem[Sartakov et~al\mbox{.}(2021)]%
        {CubicleOS:ASPLOS21}
\bibfield{author}{\bibinfo{person}{Vasily~A. Sartakov},
  \bibinfo{person}{Llu\'{\i}s Vilanova}, {and} \bibinfo{person}{Peter
  Pietzuch}.} \bibinfo{year}{2021}\natexlab{}.
\newblock \showarticletitle{{CubicleOS}: A Library {OS} with Software
  Componentisation for Practical Isolation}. In
  \bibinfo{booktitle}{\emph{Proceedings of the 26th ACM International
  Conference on Architectural Support for Programming Languages and Operating
  Systems}} \emph{(\bibinfo{series}{ASPLOS '21})}. \bibinfo{publisher}{ACM},
  \bibinfo{address}{Virtual Event}, \bibinfo{pages}{546--558}.
\newblock
\showISBNx{978-1-4503-8317-2}
\urldef\tempurl%
\url{https://doi.org/10.1145/3445814.3446731}
\showDOI{\tempurl}


\bibitem[Schrammel et~al\mbox{.}(2020)]%
        {Donky:UsenixSec20}
\bibfield{author}{\bibinfo{person}{David Schrammel}, \bibinfo{person}{Samuel
  Weiser}, \bibinfo{person}{Stefan Steinegger}, \bibinfo{person}{Martin
  Schwarzl}, \bibinfo{person}{Michael Schwarz}, \bibinfo{person}{Stefan
  Mangard}, {and} \bibinfo{person}{Daniel Gruss}.}
  \bibinfo{year}{2020}\natexlab{}.
\newblock \showarticletitle{{Donky}: Domain Keys – Efficient In-Process
  Isolation for {RISC-V} and x86}. In \bibinfo{booktitle}{\emph{Proceedings of
  the 29th USENIX Security Symposium}} \emph{(\bibinfo{series}{Security '20})}.
  \bibinfo{publisher}{USENIX Association}, \bibinfo{address}{Boston, MA, USA},
  \bibinfo{pages}{1677--1694}.
\newblock
\showISBNx{978-1-939133-17-5}
\urldef\tempurl%
\url{https://www.usenix.org/conference/usenixsecurity20/presentation/schrammel}
\showURL{%
\tempurl}


\bibitem[Schuster et~al\mbox{.}(2015)]%
        {COOP:Oakland15}
\bibfield{author}{\bibinfo{person}{Felix Schuster}, \bibinfo{person}{Thomas
  Tendyck}, \bibinfo{person}{Christopher Liebchen}, \bibinfo{person}{Lucas
  Davi}, \bibinfo{person}{Ahmad-Reza Sadeghi}, {and} \bibinfo{person}{Thorsten
  Holz}.} \bibinfo{year}{2015}\natexlab{}.
\newblock \showarticletitle{Counterfeit Object-oriented Programming: On the
  Difficulty of Preventing Code Reuse Attacks in {C++} Applications}. In
  \bibinfo{booktitle}{\emph{Proceedings of the 2015 IEEE Symposium on Security
  and Privacy}} \emph{(\bibinfo{series}{SP '15})}. \bibinfo{publisher}{IEEE
  Computer Society}, \bibinfo{address}{San Jose, CA, USA},
  \bibinfo{pages}{745--762}.
\newblock
\showISBNx{978-1-4673-6949-7}
\urldef\tempurl%
\url{https://doi.org/10.1109/SP.2015.51}
\showDOI{\tempurl}


\bibitem[Sehr et~al\mbox{.}(2010)]%
        {PNaCl:UsenixSec10}
\bibfield{author}{\bibinfo{person}{David Sehr}, \bibinfo{person}{Robert Muth},
  \bibinfo{person}{Cliff Biffle}, \bibinfo{person}{Victor Khimenko},
  \bibinfo{person}{Egor Pasko}, \bibinfo{person}{Karl Schimpf},
  \bibinfo{person}{Bennet Yee}, {and} \bibinfo{person}{Brad Chen}.}
  \bibinfo{year}{2010}\natexlab{}.
\newblock \showarticletitle{Adapting Software Fault Isolation to Contemporary
  {CPU} Architectures}. In \bibinfo{booktitle}{\emph{Proceedings of the 19th
  USENIX Security Symposium}} \emph{(\bibinfo{series}{Security '10})}.
  \bibinfo{publisher}{USENIX Association}, \bibinfo{address}{Washington, DC,
  USA}, \bibinfo{numpages}{11}~pages.
\newblock
\urldef\tempurl%
\url{https://www.usenix.org/conference/usenixsecurity10/adapting-software-fault-isolation-contemporary-cpu-architectures}
\showURL{%
\tempurl}


\bibitem[Shacham(2007)]%
        {ROP:CCS07}
\bibfield{author}{\bibinfo{person}{Hovav Shacham}.}
  \bibinfo{year}{2007}\natexlab{}.
\newblock \showarticletitle{The Geometry of Innocent Flesh on the Bone:
  Return-into-libc Without Function Calls (on the x86)}. In
  \bibinfo{booktitle}{\emph{Proceedings of the 14th ACM Conference on Computer
  and Communications Security}} \emph{(\bibinfo{series}{CCS '07})}.
  \bibinfo{publisher}{ACM}, \bibinfo{address}{Alexandria, VA, USA},
  \bibinfo{pages}{552--561}.
\newblock
\showISBNx{978-1-59593-703-2}
\urldef\tempurl%
\url{https://doi.org/10.1145/1315245.1315313}
\showDOI{\tempurl}


\bibitem[Shanbhogue et~al\mbox{.}(2019)]%
        {CET:HASP19}
\bibfield{author}{\bibinfo{person}{Vedvyas Shanbhogue}, \bibinfo{person}{Deepak
  Gupta}, {and} \bibinfo{person}{Ravi Sahita}.}
  \bibinfo{year}{2019}\natexlab{}.
\newblock \showarticletitle{Security Analysis of Processor Instruction Set
  Architecture for Enforcing Control-Flow Integrity}. In
  \bibinfo{booktitle}{\emph{Proceedings of the 8th International Workshop on
  Hardware and Architectural Support for Security and Privacy}}
  \emph{(\bibinfo{series}{HASP '19})}. \bibinfo{publisher}{ACM},
  \bibinfo{address}{Phoenix, AZ, USA}, Article \bibinfo{articleno}{8},
  \bibinfo{numpages}{11}~pages.
\newblock
\showISBNx{978-1-4503-7226-8}
\urldef\tempurl%
\url{https://doi.org/10.1145/3337167.3337175}
\showDOI{\tempurl}


\bibitem[Shen et~al\mbox{.}(2020)]%
        {PicoXOM:SecDev20}
\bibfield{author}{\bibinfo{person}{Zhuojia Shen}, \bibinfo{person}{Komail
  Dharsee}, {and} \bibinfo{person}{John Criswell}.}
  \bibinfo{year}{2020}\natexlab{}.
\newblock \showarticletitle{Fast Execute-Only Memory for Embedded Systems}. In
  \bibinfo{booktitle}{\emph{Proceedings of the 2020 IEEE Secure Development
  Conference}} \emph{(\bibinfo{series}{SecDev '20})}. \bibinfo{publisher}{IEEE
  Computer Society}, \bibinfo{address}{Atlanta, GA, USA},
  \bibinfo{pages}{7--14}.
\newblock
\showISBNx{978-1-7281-8388-6}
\urldef\tempurl%
\url{https://doi.org/10.1109/SecDev45635.2020.00017}
\showDOI{\tempurl}


\bibitem[Shen et~al\mbox{.}(2022)]%
        {Randezvous:ACSAC22}
\bibfield{author}{\bibinfo{person}{Zhuojia Shen}, \bibinfo{person}{Komail
  Dharsee}, {and} \bibinfo{person}{John Criswell}.}
  \bibinfo{year}{2022}\natexlab{}.
\newblock \showarticletitle{Randezvous: Making Randomization Effective on
  {MCUs}}. In \bibinfo{booktitle}{\emph{Proceedings of the 38th Annual Computer
  Security Applications Conference}} \emph{(\bibinfo{series}{ACSAC '22})}.
  \bibinfo{publisher}{ACM}, \bibinfo{address}{Austin, TX, USA},
  \bibinfo{pages}{28--41}.
\newblock
\showISBNx{978-1-4503-9759-9}
\urldef\tempurl%
\url{https://doi.org/10.1145/3564625.3567970}
\showDOI{\tempurl}


\bibitem[Snow et~al\mbox{.}(2013)]%
        {JIT-ROP:Oakland13}
\bibfield{author}{\bibinfo{person}{Kevin~Z. Snow}, \bibinfo{person}{Fabian
  Monrose}, \bibinfo{person}{Lucas Davi}, \bibinfo{person}{Alexandra
  Dmitrienko}, \bibinfo{person}{Christopher Liebchen}, {and}
  \bibinfo{person}{Ahmad-Reza Sadeghi}.} \bibinfo{year}{2013}\natexlab{}.
\newblock \showarticletitle{Just-In-Time Code Reuse: On the Effectiveness of
  Fine-Grained Address Space Layout Randomization}. In
  \bibinfo{booktitle}{\emph{Proceedings of the 2013 IEEE Symposium on Security
  and Privacy}} \emph{(\bibinfo{series}{SP '13})}. \bibinfo{publisher}{IEEE
  Computer Society}, \bibinfo{address}{San Francisco, CA, USA},
  \bibinfo{pages}{574--588}.
\newblock
\showISBNx{978-0-7695-4977-4}
\urldef\tempurl%
\url{https://doi.org/10.1109/SP.2013.45}
\showDOI{\tempurl}


\bibitem[Standard Performance Evaluation Corporation(2022)]%
        {CPU2017:SPEC}
Standard Performance Evaluation Corporation \bibinfo{year}{2022}\natexlab{}.
\newblock \bibinfo{booktitle}{\emph{SPEC CPU\textregistered 2017}}.
\newblock
\urldef\tempurl%
\url{https://www.spec.org/cpu2017}
\showURL{%
\tempurl}


\bibitem[Strackx et~al\mbox{.}(2009)]%
        {Overread:EuroSec09}
\bibfield{author}{\bibinfo{person}{Raoul Strackx}, \bibinfo{person}{Yves
  Younan}, \bibinfo{person}{Pieter Philippaerts}, \bibinfo{person}{Frank
  Piessens}, \bibinfo{person}{Sven Lachmund}, {and} \bibinfo{person}{Thomas
  Walter}.} \bibinfo{year}{2009}\natexlab{}.
\newblock \showarticletitle{Breaking the Memory Secrecy Assumption}. In
  \bibinfo{booktitle}{\emph{Proceedings of the 2nd European Workshop on System
  Security}} \emph{(\bibinfo{series}{EuroSec '09})}. \bibinfo{publisher}{ACM},
  \bibinfo{address}{Nuremburg, Germany}, \bibinfo{pages}{1--8}.
\newblock
\showISBNx{978-1-60558-472-0}
\urldef\tempurl%
\url{https://doi.org/10.1145/1519144.1519145}
\showDOI{\tempurl}


\bibitem[Sung et~al\mbox{.}(2020)]%
        {libhermitMPK:VEE20}
\bibfield{author}{\bibinfo{person}{Mincheol Sung}, \bibinfo{person}{Pierre
  Olivier}, \bibinfo{person}{Stefan Lankes}, {and} \bibinfo{person}{Binoy
  Ravindran}.} \bibinfo{year}{2020}\natexlab{}.
\newblock \showarticletitle{Intra-Unikernel Isolation with Intel Memory
  Protection Keys}. In \bibinfo{booktitle}{\emph{Proceedings of the 16th ACM
  SIGPLAN/SIGOPS International Conference on Virtual Execution Environments}}
  \emph{(\bibinfo{series}{VEE '20})}. \bibinfo{publisher}{ACM},
  \bibinfo{address}{Lausanne, Switzerland}, \bibinfo{pages}{143--156}.
\newblock
\showISBNx{978-1-4503-7554-2}
\urldef\tempurl%
\url{https://doi.org/10.1145/3381052.3381326}
\showDOI{\tempurl}


\bibitem[Sysoev et~al\mbox{.}(2022)]%
        {Nginx}
\bibfield{author}{\bibinfo{person}{Igor Sysoev} {et~al\mbox{.}}}
  \bibinfo{year}{2022}\natexlab{}.
\newblock \bibinfo{booktitle}{\emph{{nginx}}}.
\newblock
\urldef\tempurl%
\url{https://nginx.org/en}
\showURL{%
\tempurl}


\bibitem[Tice et~al\mbox{.}(2014)]%
        {IFCC:UsenixSec14}
\bibfield{author}{\bibinfo{person}{Caroline Tice}, \bibinfo{person}{Tom
  Roeder}, \bibinfo{person}{Peter Collingbourne}, \bibinfo{person}{Stephen
  Checkoway}, \bibinfo{person}{\'{U}lfar Erlingsson}, \bibinfo{person}{Luis
  Lozano}, {and} \bibinfo{person}{Geoff Pike}.}
  \bibinfo{year}{2014}\natexlab{}.
\newblock \showarticletitle{Enforcing Forward-Edge Control-Flow Integrity in
  {GCC} \& {LLVM}}. In \bibinfo{booktitle}{\emph{Proceedings of the 23rd USENIX
  Security Symposium}} \emph{(\bibinfo{series}{Security '14})}.
  \bibinfo{publisher}{USENIX Association}, \bibinfo{address}{San Diego, CA,
  USA}, \bibinfo{pages}{941--955}.
\newblock
\showISBNx{978-1-931971-15-7}
\urldef\tempurl%
\url{https://www.usenix.org/conference/usenixsecurity14/technical-sessions/presentation/tice}
\showURL{%
\tempurl}


\bibitem[Tran et~al\mbox{.}(2011)]%
        {Ret2Libc:RAID11}
\bibfield{author}{\bibinfo{person}{Minh Tran}, \bibinfo{person}{Mark
  Etheridge}, \bibinfo{person}{Tyler Bletsch}, \bibinfo{person}{Xuxian Jiang},
  \bibinfo{person}{Vincent Freeh}, {and} \bibinfo{person}{Peng Ning}.}
  \bibinfo{year}{2011}\natexlab{}.
\newblock \showarticletitle{On the Expressiveness of Return-into-libc Attacks}.
  In \bibinfo{booktitle}{\emph{Proceedings of the 14th International Symposium
  on Recent Advances in Intrusion Detection}} \emph{(\bibinfo{series}{RAID
  '11})}. \bibinfo{publisher}{Springer-Verlag}, \bibinfo{address}{Menlo Park,
  CA, USA}, \bibinfo{pages}{121--141}.
\newblock
\showISBNx{978-3-642-23643-3}
\urldef\tempurl%
\url{https://doi.org/10.1007/978-3-642-23644-0_7}
\showDOI{\tempurl}


\bibitem[Vahldiek-Oberwagner et~al\mbox{.}(2019)]%
        {ERIM:UsenixSec19}
\bibfield{author}{\bibinfo{person}{Anjo Vahldiek-Oberwagner},
  \bibinfo{person}{Eslam Elnikety}, \bibinfo{person}{Nuno~O. Duarte},
  \bibinfo{person}{Michael Sammler}, \bibinfo{person}{Peter Druschel}, {and}
  \bibinfo{person}{Deepak Garg}.} \bibinfo{year}{2019}\natexlab{}.
\newblock \showarticletitle{{ERIM}: Secure, Efficient In-process Isolation with
  Protection Keys ({MPK})}. In \bibinfo{booktitle}{\emph{Proceedings of the
  28th USENIX Security Symposium}} \emph{(\bibinfo{series}{Security '19})}.
  \bibinfo{publisher}{USENIX Association}, \bibinfo{address}{Santa Clara, CA,
  USA}, \bibinfo{pages}{1221--1238}.
\newblock
\showISBNx{978-1-939133-06-9}
\urldef\tempurl%
\url{https://www.usenix.org/conference/usenixsecurity19/presentation/vahldiek-oberwagner}
\showURL{%
\tempurl}


\bibitem[van~der Veen et~al\mbox{.}(2015)]%
        {PathArmor:CCS15}
\bibfield{author}{\bibinfo{person}{Victor van~der Veen},
  \bibinfo{person}{Dennis Andriesse}, \bibinfo{person}{Enes G\"{o}kta\c{s}},
  \bibinfo{person}{Ben Gras}, \bibinfo{person}{Lionel Sambuc},
  \bibinfo{person}{Asia Slowinska}, \bibinfo{person}{Herbert Bos}, {and}
  \bibinfo{person}{Cristiano Giuffrida}.} \bibinfo{year}{2015}\natexlab{}.
\newblock \showarticletitle{Practical Context-Sensitive {CFI}}. In
  \bibinfo{booktitle}{\emph{Proceedings of the 22nd ACM SIGSAC Conference on
  Computer and Communications Security}} \emph{(\bibinfo{series}{CCS '15})}.
  \bibinfo{publisher}{ACM}, \bibinfo{address}{Denver, CO, USA},
  \bibinfo{pages}{927--940}.
\newblock
\showISBNx{978-1-4503-3832-5}
\urldef\tempurl%
\url{https://doi.org/10.1145/2810103.2813673}
\showDOI{\tempurl}


\bibitem[van~der Veen et~al\mbox{.}(2016)]%
        {TypeArmor:Oakland16}
\bibfield{author}{\bibinfo{person}{Victor van~der Veen}, \bibinfo{person}{Enes
  G\"{o}kta\c{s}}, \bibinfo{person}{Moritz Contag}, \bibinfo{person}{Andre
  Pawoloski}, \bibinfo{person}{Xi Chen}, \bibinfo{person}{Sanjay Rawat},
  \bibinfo{person}{Herbert Bos}, \bibinfo{person}{Thorsten Holz},
  \bibinfo{person}{Elias Athanasopoulos}, {and} \bibinfo{person}{Cristiano
  Giuffrida}.} \bibinfo{year}{2016}\natexlab{}.
\newblock \showarticletitle{A Tough {\tt call}: Mitigating Advanced Code-Reuse
  Attacks at the Binary Level}. In \bibinfo{booktitle}{\emph{Proceedings of the
  2016 IEEE Symposium on Security and Privacy}} \emph{(\bibinfo{series}{SP
  '16})}. \bibinfo{publisher}{IEEE Computer Society}, \bibinfo{address}{San
  Jose, CA, USA}, \bibinfo{pages}{934--953}.
\newblock
\showISBNx{978-1-5090-0824-7}
\urldef\tempurl%
\url{https://doi.org/10.1109/SP.2016.60}
\showDOI{\tempurl}


\bibitem[Vilanova et~al\mbox{.}(2014)]%
        {CODOMs:ISCA14}
\bibfield{author}{\bibinfo{person}{Llu\"{\i}s Vilanova}, \bibinfo{person}{Muli
  Ben-Yehuda}, \bibinfo{person}{Nacho Navarro}, \bibinfo{person}{Yoav Etsion},
  {and} \bibinfo{person}{Mateo Valero}.} \bibinfo{year}{2014}\natexlab{}.
\newblock \showarticletitle{{CODOMs}: Protecting Software with Code-Centric
  Memory Domains}. In \bibinfo{booktitle}{\emph{Proceeding of the 41st Annual
  International Symposium on Computer Architecuture}}
  \emph{(\bibinfo{series}{ISCA '14})}. \bibinfo{publisher}{IEEE Computer
  Society}, \bibinfo{address}{Minneapolis, MN, USA}, \bibinfo{pages}{469--480}.
\newblock
\showISBNx{978-1-4799-4394-4}
\urldef\tempurl%
\url{https://doi.org/10.1109/ISCA.2014.6853202}
\showDOI{\tempurl}


\bibitem[Voulimeneas et~al\mbox{.}(2022)]%
        {Cerberus:EuroSys22}
\bibfield{author}{\bibinfo{person}{Alexios Voulimeneas}, \bibinfo{person}{Jonas
  Vinck}, \bibinfo{person}{Ruben Mechelinck}, {and} \bibinfo{person}{Stijn
  Volckaert}.} \bibinfo{year}{2022}\natexlab{}.
\newblock \showarticletitle{You Shall Not (by)Pass! Practical, Secure, and Fast
  {PKU}-Based Sandboxing}. In \bibinfo{booktitle}{\emph{Proceedings of the 17th
  European Conference on Computer Systems}} \emph{(\bibinfo{series}{EuroSys
  '22})}. \bibinfo{publisher}{ACM}, \bibinfo{address}{Rennes, France},
  \bibinfo{pages}{266--282}.
\newblock
\showISBNx{978-1-4503-9162-7}
\urldef\tempurl%
\url{https://doi.org/10.1145/3492321.3519560}
\showDOI{\tempurl}


\bibitem[Wahbe et~al\mbox{.}(1993)]%
        {SFI:SOSP93}
\bibfield{author}{\bibinfo{person}{Robert Wahbe}, \bibinfo{person}{Steven
  Lucco}, \bibinfo{person}{Thomas~E. Anderson}, {and} \bibinfo{person}{Susan~L.
  Graham}.} \bibinfo{year}{1993}\natexlab{}.
\newblock \showarticletitle{Efficient Software-Based Fault Isolation}. In
  \bibinfo{booktitle}{\emph{Proceedings of the 14th ACM Symposium on Operating
  Systems Principles}} \emph{(\bibinfo{series}{SOSP '93})}.
  \bibinfo{publisher}{ACM}, \bibinfo{address}{Asheville, NC, USA},
  \bibinfo{pages}{203--216}.
\newblock
\showISBNx{0-89791-632-8}
\urldef\tempurl%
\url{https://doi.org/10.1145/168619.168635}
\showDOI{\tempurl}


\bibitem[Walls et~al\mbox{.}(2019)]%
        {RECFISH:ECRTS19}
\bibfield{author}{\bibinfo{person}{Robert~J. Walls},
  \bibinfo{person}{Nicholas~F. Brown}, \bibinfo{person}{Thomas Le~Baron},
  \bibinfo{person}{Craig~A. Shue}, \bibinfo{person}{Hamed Okhravi}, {and}
  \bibinfo{person}{Bryan~C. Ward}.} \bibinfo{year}{2019}\natexlab{}.
\newblock \showarticletitle{Control-Flow Integrity for Real-Time Embedded
  Systems}. In \bibinfo{booktitle}{\emph{Proceedings of the 31st Euromicro
  Conference on Real-Time Systems}} \emph{(\bibinfo{series}{ECRTS '19})}.
  \bibinfo{publisher}{Schloss Dagstuhl--Leibniz-Zentrum f\"{u}er Informatik},
  \bibinfo{address}{Stuttgart, Germany}, \bibinfo{pages}{2:1--2:24}.
\newblock
\showISBNx{978-3-95977-110-8}
\urldef\tempurl%
\url{https://doi.org/10.4230/LIPIcs.ECRTS.2019.2}
\showDOI{\tempurl}


\bibitem[Wang et~al\mbox{.}(2015)]%
        {BinCC:ACSAC15}
\bibfield{author}{\bibinfo{person}{Minghua Wang}, \bibinfo{person}{Heng Yin},
  \bibinfo{person}{Abhishek~Vasisht Bhaskar}, \bibinfo{person}{Purui Su}, {and}
  \bibinfo{person}{Dengguo Feng}.} \bibinfo{year}{2015}\natexlab{}.
\newblock \showarticletitle{Binary Code Continent: Finer-Grained Control Flow
  Integrity for Stripped Binaries}. In \bibinfo{booktitle}{\emph{Proceedings of
  the 31st Annual Computer Security Applications Conference}}
  \emph{(\bibinfo{series}{ACSAC '15})}. \bibinfo{publisher}{ACM},
  \bibinfo{address}{Los Angeles, CA, USA}, \bibinfo{pages}{331--340}.
\newblock
\showISBNx{978-1-4503-3682-6}
\urldef\tempurl%
\url{https://doi.org/10.1145/2818000.2818017}
\showDOI{\tempurl}


\bibitem[Wang et~al\mbox{.}(2020b)]%
        {MonGuard:EuroSec20}
\bibfield{author}{\bibinfo{person}{Xiaoguang Wang}, \bibinfo{person}{SengMing
  Yeoh}, \bibinfo{person}{Pierre Olivier}, {and} \bibinfo{person}{Binoy
  Ravindran}.} \bibinfo{year}{2020}\natexlab{b}.
\newblock \showarticletitle{Secure and Efficient In-Process Monitor (and
  Library) Protection with {Intel MPK}}. In
  \bibinfo{booktitle}{\emph{Proceedings of the 13th European Workshop on
  Systems Security}} \emph{(\bibinfo{series}{EuroSec '20})}.
  \bibinfo{publisher}{ACM}, \bibinfo{address}{Heraklion, Greece},
  \bibinfo{pages}{7--12}.
\newblock
\showISBNx{978-1-4503-7523-8}
\urldef\tempurl%
\url{https://doi.org/10.1145/3380786.3391398}
\showDOI{\tempurl}


\bibitem[Wang et~al\mbox{.}(2022)]%
        {RetTag:EuroSec22}
\bibfield{author}{\bibinfo{person}{Yu Wang}, \bibinfo{person}{Jinting Wu},
  \bibinfo{person}{Tai Yue}, \bibinfo{person}{Zhenyu Ning}, {and}
  \bibinfo{person}{Fengwei Zhang}.} \bibinfo{year}{2022}\natexlab{}.
\newblock \showarticletitle{{RetTag}: Hardware-Assisted Return Address
  Integrity on {RISC-V}}. In \bibinfo{booktitle}{\emph{Proceedings of the 15th
  European Workshop on Systems Security}} \emph{(\bibinfo{series}{EuroSec
  '22})}. \bibinfo{publisher}{ACM}, \bibinfo{address}{Rennes, France},
  \bibinfo{pages}{50--56}.
\newblock
\showISBNx{978-1-4503-9255-6}
\urldef\tempurl%
\url{https://doi.org/10.1145/3517208.3523758}
\showDOI{\tempurl}


\bibitem[Wang and Jiang(2010)]%
        {HyperSafe:Oakland10}
\bibfield{author}{\bibinfo{person}{Zhi Wang} {and} \bibinfo{person}{Xuxian
  Jiang}.} \bibinfo{year}{2010}\natexlab{}.
\newblock \showarticletitle{{HyperSafe}: A Lightweight Approach to Provide
  Lifetime Hypervisor Control-Flow Integrity}. In
  \bibinfo{booktitle}{\emph{Proceedings of the 2010 IEEE Symposium on Security
  and Privacy}} \emph{(\bibinfo{series}{SP '10})}. \bibinfo{publisher}{IEEE
  Computer Society}, \bibinfo{address}{Oakland, CA, USA},
  \bibinfo{pages}{380--395}.
\newblock
\showISBNx{978-1-4244-6895-9}
\urldef\tempurl%
\url{https://doi.org/10.1109/SP.2010.30}
\showDOI{\tempurl}


\bibitem[Wang et~al\mbox{.}(2020a)]%
        {SEIMI:Oakland20}
\bibfield{author}{\bibinfo{person}{Zhe Wang}, \bibinfo{person}{Chenggang Wu},
  \bibinfo{person}{Mengyao Xie}, \bibinfo{person}{Yinqian Zhang},
  \bibinfo{person}{Kangjie Lu}, \bibinfo{person}{Xiaofeng Zhang},
  \bibinfo{person}{Yuanming Lai}, \bibinfo{person}{Yan Kang}, {and}
  \bibinfo{person}{Min Yang}.} \bibinfo{year}{2020}\natexlab{a}.
\newblock \showarticletitle{{SEIMI}: Efficient and Secure {SMAP}-Enabled
  Intra-process Memory Isolation}. In \bibinfo{booktitle}{\emph{Proceedins of
  the 2020 IEEE Symposium on Security and Privacy}} \emph{(\bibinfo{series}{SP
  '20})}. \bibinfo{publisher}{IEEE Computer Society}, \bibinfo{address}{San
  Francisco, CA, USA}, \bibinfo{pages}{592--607}.
\newblock
\showISBNx{978-1-7281-3497-0}
\urldef\tempurl%
\url{https://doi.org/10.1109/SP40000.2020.00087}
\showDOI{\tempurl}


\bibitem[Wikipedia(2023)]%
        {ARMCoreList:Wikipedia}
Wikipedia \bibinfo{year}{2023}\natexlab{}.
\newblock \bibinfo{booktitle}{\emph{Comparison of ARM processors}}.
\newblock
\urldef\tempurl%
\url{https://en.wikipedia.org/wiki/Comparison_of_ARM_processors#ARMv8-A}
\showURL{%
\tempurl}


\bibitem[Witchel et~al\mbox{.}(2002)]%
        {Mondrian:ASPLOS02}
\bibfield{author}{\bibinfo{person}{Emmett Witchel}, \bibinfo{person}{Josh
  Cates}, {and} \bibinfo{person}{Krste Asanovi\'{c}}.}
  \bibinfo{year}{2002}\natexlab{}.
\newblock \showarticletitle{Mondrian Memory Protection}. In
  \bibinfo{booktitle}{\emph{Proceedings of the 10th International Conference on
  Architectural Support for Programming Languages and Operating Systems}}
  \emph{(\bibinfo{series}{ASPLOS '02})}. \bibinfo{publisher}{ACM},
  \bibinfo{address}{San Jose, CA, USA}, \bibinfo{pages}{304--316}.
\newblock
\showISBNx{1-58113-574-2}
\urldef\tempurl%
\url{https://doi.org/10.1145/605397.605429}
\showDOI{\tempurl}


\bibitem[Witchel et~al\mbox{.}(2005)]%
        {Mondrix:SOSP05}
\bibfield{author}{\bibinfo{person}{Emmett Witchel}, \bibinfo{person}{Junghwan
  Rhee}, {and} \bibinfo{person}{Krste Asanovi\'{c}}.}
  \bibinfo{year}{2005}\natexlab{}.
\newblock \showarticletitle{Mondrix: Memory Isolation for {Linux} Using
  {Mondriaan} Memory Protection}. In \bibinfo{booktitle}{\emph{Proceedings of
  the 20th ACM Symposium on Operating Systems Principles}}
  \emph{(\bibinfo{series}{SOSP '05})}. \bibinfo{publisher}{ACM},
  \bibinfo{address}{Brighton, UK}, \bibinfo{pages}{31--44}.
\newblock
\showISBNx{1-59593-079-5}
\urldef\tempurl%
\url{https://doi.org/10.1145/1095810.1095814}
\showDOI{\tempurl}


\bibitem[{XAMPPRocky} et~al\mbox{.}(2021)]%
        {Tokei:GitHub}
\bibfield{author}{\bibinfo{person}{{XAMPPRocky}} {et~al\mbox{.}}}
  \bibinfo{year}{2021}\natexlab{}.
\newblock \bibinfo{booktitle}{\emph{Tokei: Count your code, quickly}}.
\newblock
\urldef\tempurl%
\url{https://github.com/XAMPPRocky/tokei}
\showURL{%
\tempurl}


\bibitem[Xia et~al\mbox{.}(2012)]%
        {CFIMon:DSN12}
\bibfield{author}{\bibinfo{person}{Yubin Xia}, \bibinfo{person}{Yutao Liu},
  \bibinfo{person}{Haibo Chen}, {and} \bibinfo{person}{Binyu Zang}.}
  \bibinfo{year}{2012}\natexlab{}.
\newblock \showarticletitle{{CFIMon}: Detecting Violation of Control Flow
  Integrity using Performance Counters}. In
  \bibinfo{booktitle}{\emph{Proceedings of the 42nd Annual IEEE/IFIP
  International Conference on Dependable Systems and Networks}}
  \emph{(\bibinfo{series}{DSN '12})}. \bibinfo{publisher}{IEEE Computer
  Society}, \bibinfo{address}{Boston, MA, USA}, \bibinfo{numpages}{12}~pages.
\newblock
\showISBNx{978-1-4673-1625-5}
\urldef\tempurl%
\url{https://doi.org/10.1109/DSN.2012.6263958}
\showDOI{\tempurl}


\bibitem[Xie et~al\mbox{.}(2022)]%
        {CETIS:CCS22}
\bibfield{author}{\bibinfo{person}{Mengyao Xie}, \bibinfo{person}{Chenggang
  Wu}, \bibinfo{person}{Yinqian Zhang}, \bibinfo{person}{Jiali Xu},
  \bibinfo{person}{Yuanming Lai}, \bibinfo{person}{Yan Kang},
  \bibinfo{person}{Wei Wang}, {and} \bibinfo{person}{Zhe Wang}.}
  \bibinfo{year}{2022}\natexlab{}.
\newblock \showarticletitle{{CETIS}: Retrofitting {Intel CET} for Generic and
  Efficient Intra-Process Memory Isolation}. In
  \bibinfo{booktitle}{\emph{Proceedings of the 2022 ACM SIGSAC Conference on
  Computer and Communications Security}} \emph{(\bibinfo{series}{CCS '22})}.
  \bibinfo{publisher}{ACM}, \bibinfo{address}{Los Angeles, CA, USA},
  \bibinfo{pages}{2989--3002}.
\newblock
\showISBNx{978-1-4503-9450-5}
\urldef\tempurl%
\url{https://doi.org/10.1145/3548606.3559344}
\showDOI{\tempurl}


\bibitem[Yee et~al\mbox{.}(2009)]%
        {NaCl:Oakland09}
\bibfield{author}{\bibinfo{person}{Bennet Yee}, \bibinfo{person}{David Sehr},
  \bibinfo{person}{Gregory Dardyk}, \bibinfo{person}{J.~Bradley Chen},
  \bibinfo{person}{Robert Muth}, \bibinfo{person}{Tavis Ormandy},
  \bibinfo{person}{Shiki Okasaka}, \bibinfo{person}{Neha Narula}, {and}
  \bibinfo{person}{Nicholas Fullagar}.} \bibinfo{year}{2009}\natexlab{}.
\newblock \showarticletitle{Native Client: A Sandbox for Portable, Untrusted
  x86 Native Code}. In \bibinfo{booktitle}{\emph{Proceedings of the 2009 IEEE
  Symposium on Security and Privacy}} \emph{(\bibinfo{series}{SP '09})}.
  \bibinfo{publisher}{IEEE Computer Society}, \bibinfo{address}{Oakland, CA,
  USA}, \bibinfo{pages}{79--93}.
\newblock
\showISBNx{978-0-7695-3633-0}
\urldef\tempurl%
\url{https://doi.org/10.1109/SP.2009.25}
\showDOI{\tempurl}


\bibitem[Yoo et~al\mbox{.}(2022)]%
        {PAL:UsenixSec22}
\bibfield{author}{\bibinfo{person}{Sungbae Yoo}, \bibinfo{person}{Jinbum Park},
  \bibinfo{person}{Seolheui Kim}, \bibinfo{person}{Yeji Kim}, {and}
  \bibinfo{person}{Taesoo Kim}.} \bibinfo{year}{2022}\natexlab{}.
\newblock \showarticletitle{In-Kernel Control-Flow Integrity on Commodity
  {OSes} using {ARM} Pointer Authentication}. In
  \bibinfo{booktitle}{\emph{Proceedings of the 31st USENIX Security Symposium}}
  \emph{(\bibinfo{series}{Security '22})}. \bibinfo{publisher}{USENIX
  Association}, \bibinfo{address}{Boston, MA, USA}, \bibinfo{pages}{89--106}.
\newblock
\showISBNx{978-1-939133-31-1}
\urldef\tempurl%
\url{https://www.usenix.org/conference/usenixsecurity22/presentation/yoo}
\showURL{%
\tempurl}


\bibitem[Yuan et~al\mbox{.}(2015)]%
        {CFIGuard:RAID15}
\bibfield{author}{\bibinfo{person}{Pinghai Yuan}, \bibinfo{person}{Qingkai
  Zeng}, {and} \bibinfo{person}{Xuhua Ding}.} \bibinfo{year}{2015}\natexlab{}.
\newblock \showarticletitle{Hardware-Assisted Fine-Grained Code-Reuse Attack
  Detection}. In \bibinfo{booktitle}{\emph{Proceedings of the 18th
  International Symposium on Research in Attacks, Intrusions, and Defenses}}
  \emph{(\bibinfo{series}{RAID '15})}. \bibinfo{publisher}{Springer-Verlag},
  \bibinfo{address}{Kyoto, Japan}, \bibinfo{pages}{66--85}.
\newblock
\showISBNx{978-3-319-26362-5}
\urldef\tempurl%
\url{https://doi.org/10.1007/978-3-319-26362-5_4}
\showDOI{\tempurl}


\bibitem[Zeng et~al\mbox{.}(2013)]%
        {Strato:UsenixSec13}
\bibfield{author}{\bibinfo{person}{Bin Zeng}, \bibinfo{person}{Gang Tan}, {and}
  \bibinfo{person}{{\'U}lfar Erlingsson}.} \bibinfo{year}{2013}\natexlab{}.
\newblock \showarticletitle{Strato: A Retargetable Framework for Low-Level
  Inlined-Reference Monitors}. In \bibinfo{booktitle}{\emph{Proceedings of the
  22nd USENIX Security Symposium}} \emph{(\bibinfo{series}{Security '13})}.
  \bibinfo{publisher}{USENIX Association}, \bibinfo{address}{Washington, DC,
  USA}, \bibinfo{pages}{369--382}.
\newblock
\showISBNx{978-1-931971-03-4}
\urldef\tempurl%
\url{https://www.usenix.org/conference/usenixsecurity13/technical-sessions/presentation/zeng}
\showURL{%
\tempurl}


\bibitem[Zeng et~al\mbox{.}(2011)]%
        {Zeng:CCS11}
\bibfield{author}{\bibinfo{person}{Bin Zeng}, \bibinfo{person}{Gang Tan}, {and}
  \bibinfo{person}{Greg Morrisett}.} \bibinfo{year}{2011}\natexlab{}.
\newblock \showarticletitle{Combining Control-Flow Integrity and Static
  Analysis for Efficient and Validated Data Sandboxing}. In
  \bibinfo{booktitle}{\emph{Proceedings of the 18th ACM Conference on Computer
  and Communications Security}} \emph{(\bibinfo{series}{CCS '11})}.
  \bibinfo{publisher}{ACM}, \bibinfo{address}{Chicago, IL, USA},
  \bibinfo{pages}{29--40}.
\newblock
\showISBNx{978-1-4503-0948-6}
\urldef\tempurl%
\url{https://doi.org/10.1145/2046707.2046713}
\showDOI{\tempurl}


\bibitem[Zhang et~al\mbox{.}(2013)]%
        {CCFIR:Oakland13}
\bibfield{author}{\bibinfo{person}{Chao Zhang}, \bibinfo{person}{Tao Wei},
  \bibinfo{person}{Zhaofeng Chen}, \bibinfo{person}{Lei Duan},
  \bibinfo{person}{Laszlo Szekeres}, \bibinfo{person}{Stephen McCamant},
  \bibinfo{person}{Dawn Song}, {and} \bibinfo{person}{Wei Zou}.}
  \bibinfo{year}{2013}\natexlab{}.
\newblock \showarticletitle{Practical Control Flow Integrity and Randomization
  for Binary Executables}. In \bibinfo{booktitle}{\emph{Proceedings of the 2013
  IEEE Symposium on Security and Privacy}} \emph{(\bibinfo{series}{SP '13})}.
  \bibinfo{publisher}{IEEE Computer Society}, \bibinfo{address}{San Francisco,
  CA, USA}, \bibinfo{pages}{559--573}.
\newblock
\showISBNx{978-0-7695-4977-4}
\urldef\tempurl%
\url{https://doi.org/10.1109/SP.2013.44}
\showDOI{\tempurl}


\bibitem[Zhang and Sekar(2013)]%
        {Bin-CFI:UsenixSec13}
\bibfield{author}{\bibinfo{person}{Mingwei Zhang} {and} \bibinfo{person}{R.
  Sekar}.} \bibinfo{year}{2013}\natexlab{}.
\newblock \showarticletitle{Control Flow Integrity for {COTS} Binaries}. In
  \bibinfo{booktitle}{\emph{Proceedings of the 22nd USENIX Security Symposium}}
  \emph{(\bibinfo{series}{Security '13})}. \bibinfo{publisher}{USENIX
  Association}, \bibinfo{address}{Washington, DC, USA},
  \bibinfo{pages}{337--352}.
\newblock
\showISBNx{978-1-931971-03-4}
\urldef\tempurl%
\url{https://www.usenix.org/conference/usenixsecurity13/technical-sessions/presentation/Zhang}
\showURL{%
\tempurl}


\bibitem[Zhang et~al\mbox{.}(2019)]%
        {BOGO:ASPLOS19}
\bibfield{author}{\bibinfo{person}{Tong Zhang}, \bibinfo{person}{Dongyoon Lee},
  {and} \bibinfo{person}{Changhee Jung}.} \bibinfo{year}{2019}\natexlab{}.
\newblock \showarticletitle{{BOGO}: Buy Spatial Memory Safety, Get Temporal
  Memory Safety (Almost) Free}. In \bibinfo{booktitle}{\emph{Proceedings of the
  24th International Conference on Architectural Support for Programming
  Languages and Operating Systems}} \emph{(\bibinfo{series}{ASPLOS '19})}.
  \bibinfo{publisher}{ACM}, \bibinfo{address}{Providence, RI, USA},
  \bibinfo{pages}{631--644}.
\newblock
\showISBNx{978-1-4503-6240-5}
\urldef\tempurl%
\url{https://doi.org/10.1145/3297858.3304017}
\showDOI{\tempurl}


\bibitem[Zhou et~al\mbox{.}(2020)]%
        {Silhouette:UsenixSec20}
\bibfield{author}{\bibinfo{person}{Jie Zhou}, \bibinfo{person}{Yufei Du},
  \bibinfo{person}{Zhuojia Shen}, \bibinfo{person}{Lele Ma},
  \bibinfo{person}{John Criswell}, {and} \bibinfo{person}{Robert~J. Walls}.}
  \bibinfo{year}{2020}\natexlab{}.
\newblock \showarticletitle{Silhouette: Efficient Protected Shadow Stacks for
  Embedded Systems}. In \bibinfo{booktitle}{\emph{Proceedings of the 29th
  USENIX Security Symposium}} \emph{(\bibinfo{series}{Security '20})}.
  \bibinfo{publisher}{USENIX Association}, \bibinfo{address}{Boston, MA, USA},
  \bibinfo{pages}{1219--1236}.
\newblock
\showISBNx{978-1-939133-17-5}
\urldef\tempurl%
\url{https://www.usenix.org/conference/usenixsecurity20/presentation/zhou-jie}
\showURL{%
\tempurl}


\bibitem[Zhou et~al\mbox{.}(2014)]%
        {ARMlock:CCS14}
\bibfield{author}{\bibinfo{person}{Yajin Zhou}, \bibinfo{person}{Xiaoguang
  Wang}, \bibinfo{person}{Yue Chen}, {and} \bibinfo{person}{Zhi Wang}.}
  \bibinfo{year}{2014}\natexlab{}.
\newblock \showarticletitle{{ARMlock}: Hardware-Based Fault Isolation for
  {ARM}}. In \bibinfo{booktitle}{\emph{Proceedings of the 21st ACM Conference
  on Computer and Communications Security}} \emph{(\bibinfo{series}{CCS '14})}.
  \bibinfo{publisher}{ACM}, \bibinfo{address}{Scottsdale, AZ, USA},
  \bibinfo{pages}{558--569}.
\newblock
\showISBNx{978-1-4503-2957-6}
\urldef\tempurl%
\url{https://doi.org/10.1145/2660267.2660344}
\showDOI{\tempurl}


\bibitem[Zieris and Horsch(2018)]%
        {Zieris:ASIACCS18}
\bibfield{author}{\bibinfo{person}{Philipp Zieris} {and}
  \bibinfo{person}{Julian Horsch}.} \bibinfo{year}{2018}\natexlab{}.
\newblock \showarticletitle{A Leak-Resilient Dual Stack Scheme for
  Backward-Edge Control-Flow Integrity}. In
  \bibinfo{booktitle}{\emph{Proceedings of the 2018 ACM Asia Conference on
  Computer and Communications Security}} \emph{(\bibinfo{series}{ASIACCS
  '18})}. \bibinfo{publisher}{ACM}, \bibinfo{address}{Incheon, Republic of
  Korea}, \bibinfo{pages}{369--380}.
\newblock
\showISBNx{978-1-4503-5576-6}
\urldef\tempurl%
\url{https://doi.org/10.1145/3196494.3196531}
\showDOI{\tempurl}


\end{thebibliography}


\end{document}